\documentclass[11pt]{article}
\pdfoutput=1
\usepackage[utf8]{inputenc}
\usepackage{dsfont}
\usepackage{diagbox}
\usepackage{amsfonts}
\usepackage{mathtools}
\allowdisplaybreaks[4]        
\usepackage{amssymb}
\usepackage{lipsum}
\usepackage{euscript}     
\usepackage{braket}
\usepackage{starfont}
\usepackage{color,soul}         
\usepackage{braket}
\usepackage{tensor}        
\usepackage{amsthm}
\usepackage{graphicx}
\usepackage{slashed}
\usepackage{leftidx}
\usepackage{subfigure}
\usepackage{tikz}
\usepackage{bbm}
\definecolor{outerspace}{rgb}{0.25, 0.29, 0.3}
\definecolor{scarlet}{rgb}{1.0, 0.13, 0.0}
\usepackage[header,title,page,titletoc]{appendix}  
\definecolor{princetonorange}{rgb}{1.0, 0.56, 0.0}
\definecolor{WildStrawberry}{rgb}{1.0, 0.26, 0.64}
\definecolor{rossocorsa}{rgb}{0.83, 0.0, 0.0}
\definecolor{navyblue}{rgb}{0.0, 0.0, 0.5}
\usepackage[numbers,sort&compress]{natbib}  
\usepackage{float}
\usepackage[paper=letterpaper,margin=1in]{geometry}
\newcommand{\req}[1]{(\ref{#1})} 

\newcommand{\bea}{\begin{eqnarray}}

\newcommand{\eea}{\end{eqnarray}}
\newcommand{\ba}{\begin{eqnarray}}
\newcommand{\ea}{\end{eqnarray}}

\newcommand{\be}{\begin{equation}}
\newcommand{\ee}{\end{equation} }
\newcommand{\beqa}{\begin{eqnarray}}
\newcommand{\eeqa}{\end{eqnarray}}
\newcommand{\beqar}{\begin{eqnarray*}}
\newcommand{\eeqar}{\end{eqnarray*}}

\renewcommand{\req}[1]{eq.~(\ref{#1})}

\DeclareMathOperator{\arccot}{arccot}

\newcommand{\ssc}{\scriptscriptstyle}
\newcommand{\eg}{{\it e.g.,}\ }
\newcommand{\ie}{{\it i.e.,}\ }

\newcommand{\D}{\mathcal{D}}

\newcommand{\cO}{\mathcal{O}}

\DeclareMathOperator{\tr}{tr}

\def\({\left(}
\def\){\right)}
\def\[{\left[}
\def\]{\right]}

\usepackage{hyperref}
\hypersetup{
    colorlinks,
    citecolor=rossocorsa,
    filecolor=navyblue,
    linkcolor=navyblue,
    urlcolor=navyblue
}

\begin{document}

\begin{titlepage}

\begin{center}

\phantom{ }
\vspace{2cm}

{\bf \huge{Aspects of $N$-partite information\\ in conformal field theories}}
\vskip 0.7cm
C\'esar A. Ag\'on${}^{\text{\Hades}}$, Pablo Bueno${}^{\text{\Zeus,\Cupido}}$, Oscar Lasso Andino${}^{\text{\Poseidon}}$ and Alejandro Vilar L\'opez${}^{\text{\Apollon}}$
\vskip 0.05in

\small{${}^{\text{\Hades}}$ \textit{Instituto Balseiro, Centro At\'omico Bariloche}}
\vskip -.4cm
\small{\textit{ 8400-S.C. de Bariloche, R\'io Negro, Argentina}}

\small{${}^{\text{\Zeus}}$ \textit{CERN, Theoretical Physics Department,}}
\vskip -.4cm
\small{\textit{CH-1211 Geneva 23, Switzerland}}

\small{${}^{\text{\Cupido}}$ \textit{Departament de F\'isica Qu\`antica i Astrof\'isica, Institut de Ci\`encies del Cosmos, Universitat de Barcelona,}}
\vskip -.4cm
\small{\textit{Mart\'i i Franqu\`es 1, E-08028 Barcelona, Spain}}

\small{${}^{\text{\Poseidon}}$\textit{Escuela de Ciencias F\'isicas y Matem\'aticas, Universidad de Las Am\'ericas,
}}
\vskip -.4cm
\small{\textit{Redondel del Ciclista, Antigua V\'ia a Nay\'on, C.P.170124, Quito, Ecuador}}

\small{${}^{\text{\Apollon}}$ \textit{Physique Th\'eorique et Math\'ematique and International Solvay Institutes, Universit\'e Libre
de Bruxelles,}}
\vskip -.4cm
\small{\textit{C.P. 231, B-1050 Brussels, Belgium
}}

\begin{abstract}
We present several new results for the $N$-partite information, $I_N$, of spatial regions in the ground state of $d$-dimensional conformal field theories. First, we show that $I_N$ can be written in terms of a single $N$-point function of twist operators. Using this, we argue that in the limit in which all mutual separations are much  greater than the regions sizes, the $N$-partite information scales as $I_N \sim r^{-2N\Delta}$, where $r$ is the typical distance between pairs of regions and $\Delta$ is the lowest primary scaling dimension. In the case of spherical entangling surfaces, we obtain a completely explicit formula for the $I_4$ in terms of 2-, 3- and 4-point functions of the lowest-dimensional primary. Then, we consider a three-dimensional scalar field in the lattice. We verify the predicted long-distance scaling and provide strong evidence that $I_N$ is always positive for general regions and arbitrary $N$ for that theory. For the $I_4$, we find excellent numerical agreement between our general formula and the lattice result for disk regions. We also perform lattice calculations of the mutual information for more general regions and general separations both for a free scalar and a free fermion, and conjecture that, normalized by the corresponding disk entanglement entropy coefficients, the scalar result is always greater than the fermion one. Finally, we verify explicitly the equality between the $N$-partite information of bulk and boundary fields in holographic theories for spherical entangling surfaces in general dimensions. 
\end{abstract}

\end{center}
\end{titlepage}

\newpage

\tableofcontents


\section{Introduction}
In quantum field theory (QFT), vacuum expectation values are statistical measures of the vacuum state fluctuations in the quantum fields, and a complete knowledge of such expectation values can be used to fully characterize the theory \cite{Wightman:1956zz}. An alternative formulation of QFT can be performed in terms of algebras associated to spacetime regions \cite{Haag:1963dh} ---prototypically, causal diamonds. In this, the r\^ole played by the vacuum expectation values of the quantum fields in the usual formulation should be played by statistical measures of the vacuum state in the corresponding algebras. The natural quantities that one can associate to such algebras are entanglement measures such as entanglement entropy (EE). 

As a matter of fact, the EE of any pair of regions is ill-defined in the continuum, although one can often construct well-defined regularized versions of it which capture universal information about the continuum theory.\footnote{A few examples in the context of conformal field theories (CFTs) are: the trace-anomaly coefficients for even dimensional theories \cite{Calabrese:2004eu,Calabrese:2009qy,Solodukhin:2008dh,Fursaev:2012mp,Safdi:2012sn,Miao:2015iba}, the Euclidean sphere partition function in odd dimensions \cite{Casini:2011kv,Dowker:2010yj}, the stress-tensor two-point  function charge $C_T$ in general dimensions \cite{Perlmutter:2013gua,Faulkner:2015csl,Bueno:2015rda}, the thermal entropy coefficient $C_S$ in general  dimensions \cite{Swingle:2013hga,Bueno:2015qya,Bueno:2018xqc}, and others \cite{Lee:2014xwa,Lewkowycz:2014jia,Miao:2015dua,Anastasiou:2022pzm,Bueno:2022jbl,Baiguera:2022sao}.} On the other hand, an entanglement-based axiomatic approach to QFT would naturally require the use of measures which can be defined properly in the continuum theory. A natural choice is the mutual information (MI) which, given a pair of disjoint regions $A_1,A_2$ can be defined in terms of the EE as
\begin{equation}
I_2(A_1,A_2)\equiv S(A_1)+S(A_2)-S(A_1 A_2)\, .
\end{equation}
Strictly speaking, the RHS of this formula is only defined in the continuum as a limit of regulated EEs, but there exists an alternative equivalent definition which makes sense in the continuum theory in terms of relative entropy ---see \eg \cite{Witten:2018zxz}. The mutual information is positive semi-definite, $I_2\geq 0$, monotonous under inclusion, and symmetric in its arguments, among other properties ---see \eg \cite{Agon:2021zvp}. It has been suggested that the long-distance expansion of the mutual information of pairs of regions could be used to systematically extract the operatorial content of the theory \cite{Agon:2021zvp,Agon:2021lus}. Indeed, the exponents of the inverse distance powers appearing in such expansion are linear combinations of the conformal dimensions  of the theory \cite{Calabrese:2010he} and, in the case of spherical regions, the expansion can be organized as a sum of conformal blocks associated to each primary field \cite{Long:2016vkg, Chen:2016mya, Chen:2017hbk}. In that case, the leading term is given by \cite{Agon:2015ftl}
\begin{equation}\label{i2l}
I_2=c(2\Delta) \frac{R_{1}^{2\Delta}R_2^{2\Delta}}{r_{12}^{4\Delta}}\, ,\quad \text{where} \quad c(N\Delta)\equiv \frac{\sqrt{\pi}}{4} \frac{ \Gamma[N\Delta+1]}{\Gamma[N\Delta+\frac{3}{2}]}\, ,\quad (R_{1,2}\ll r_{12})
\end{equation}
In this expression, which is valid in general dimensions, $R_1,R_2$ are the radii of the spheres, $r_{12}$ is the distance between them, and $\Delta$ is the lowest scaling dimension of the theory, which is assumed to correspond to a scalar field.\footnote{Analogous formulas for the case in which the lowest primary has an arbitrary spin have also been obtained \cite{Chen:2017hbk,Casini:2021raa}. In the case of a free fermion, the subleading term has been computed in \cite{Agon:2021zvp}.} For more general regions, $c(2\Delta)$ is replaced by a complicated function of the spacetime dimension, the geometry of the regions, and $\Delta$. 

Subleading terms in the mutual information long-distance expansion are expected to include information also about the Operator Product Expansion (OPE) coefficients. In fact, one such coefficient appears already in the analogous leading long-distance term in the case of the tripartite information, $I_3$. Given three entangling regions $A_{1},A_{2},A_3$, this is defined as
\begin{equation}
I_3(A_1,A_2,A_3)\equiv S(A_1)+S(A_2)+S(A_3)-S(A_1A_2)-S(A_1A_3)-S(A_2A_3)+S(A_1A_2 A_3)\, .
\end{equation}
As opposed to the mutual information, the tripartite information can be both positive and negative for different theories \cite{Lieb:1974qr,Vedral:2002zz}. When a given theory satisfies $I_3 \leq 0$ for arbitrary regions, the mutual information is said to be ``monogamous'', which is the case, for instance, of holographic theories \cite{Hayden:2011ag,Cui:2018dyq} ---see also \cite{Rangamani:2015qwa,Rota:2015wge,Czech:2022fzb}. On the other hand, examples of theories which exhibit non-monogamous mutual informations include free fields \cite{Casini:2008wt}. Finally, as argued in \cite{Agon:2021lus}, a CFT with $I_3=0$ for general regions is only possible in $d=2$ ---such theory corresponding to a free fermion \cite{Casini:2008wt}. In the long-distance regime, it has been shown that the tripartite information of three spherical regions is given, for a general $d$-dimensional CFT, by\footnote{Observe also that, in the case of a lowest-dimensional fermionic primary of dimension $\Delta_f$, the leading power in $I_3$ is no longer $\sim (R/r)^{6\Delta_f}$, but rather $\sim (R/r)^{6\tilde \Delta_f}$, where $\Delta_f < \tilde \Delta_f\leq \Delta_f+\frac{1}{6}$, the saturation of the inequality occurring at least for a free fermion \cite{Agon:2021lus}.} \cite{Agon:2021lus}
\begin{equation}\label{tripi}
I_3=\left[\frac{2^{6\Delta} \Gamma(\Delta+\frac{1}{2})^3}{2\pi \Gamma(3\Delta+\frac{3}{2})}-c(3\Delta)(C_{\mathcal{O}\mathcal{O}\mathcal{O}})^2 \right] \frac{R_1^{2\Delta} R_2^{2\Delta} R_3^{2\Delta} }{r_{12}^{2\Delta}r_{13}^{2\Delta}r_{23}^{2\Delta}}\, , \quad (R_{1,2,3}\ll r_{12},r_{23},r_{13})
\end{equation}
where $\Delta$ is again the scaling dimension of the lowest scaling-dimension primary of the theory, and $C_{\mathcal{O}\mathcal{O}\mathcal{O}}$ its three-point structure constant.

 In view of the above results, it is conceivable to imagine a procedure which would bootstrap the CFT data from long-distance expansions of this kind. In order to do this, one could either go to higher orders in the corresponding long-distance expansion or, complementarily, consider the leading terms  in higher-$N$ generalizations of the mutual information. Given $N$ disjoint regions $A_1,\dots,A_N\equiv \{ A_{i}\}$, a somewhat canonical choice is the $N$-partite information, which can be defined as
\bea\label{IN1}
I_N( A_1,\dots,A_N)\equiv -\sum_{\sigma}(-)^{|\sigma|} S(\sigma)\, ,
\eea
where the sum is over all the subsets $\sigma \subset \{A_1,\dots,A_N\}$ and $|\sigma|$ is the number of elements in the subset. 
$I_N$ can be alternatively written in terms of lower-partite informations, \eg as
\begin{equation}\label{IN2}
I_N(\boldsymbol{\cdot},A_{N-1},A_{N})=I_{N-1}(\boldsymbol{\cdot},A_{N-1})+I_{N-1}(\boldsymbol{\cdot},A_{N})-I_{N-1}(\boldsymbol{\cdot},A_{N-1}A_{N})\, ,
\end{equation}
where $\boldsymbol{\cdot}\equiv A_1,\dots,A_{N-2}$, which makes manifest both the fact that $I_N$ is a well-defined quantity in QFT and also that it can be interpreted as a measure of the non-extensivity of $I_{N-1}$. Not much is known about $I_N$ for general QFTs. In fact, studies of  $I_N$ for $N\geq 4$ in QFT have been mostly limited to holographic theories. In that context, it has been argued that  $I_N$ does not have a definite sign beyond $N=3$ \cite{Hayden:2011ag,Mirabi:2016elb}. The $N$-partite information also plays an important r\^ole in various systematic approaches aimed at a full characterization of various multipartite entanglement measures and their properties for holography theories ---see \eg \cite{Bao:2015bfa,Hubeny:2018trv,Hubeny:2018ijt,He:2019ttu,HernandezCuenca:2019wgh,Fadel:2021urx}.  Additional holographic explorations involving the $N$-partite information include \eg \cite{Alishahiha:2014jxa,Mahapatra:2019uql}.  In the context of two-dimensional CFTs, the $N$-partite information has also been studied using replica trick techniques in \cite{Coser:2013qda,DeNobili:2015dla}. 


In this paper we present several new results involving the $N$-partite information of spatial regions in CFTs. First of all, we show that for general CFTs in arbitrary dimensions, the $N$-partite information can be written as a single $N$-point correlator of the form
\bea\label{IN44}
I_N(\{A_i\})=\lim_{n\to 1}\frac{\(-\)^{N+1}}{1-n}\langle \tilde{\Sigma}_{A_1}^{(n)}\tilde{\Sigma}_{A_2}^{(n)} \cdots \tilde{\Sigma}_{A_N}^{(n)}\rangle\, ,
\eea
where $\tilde{\Sigma}_{A_i}$ is the ``twist operator'' with support on $A_i$ which implements the identification between copies when computing the R\'enyi entropy $S^{(n)}(A_i)$ using the replica trick.  Using this, we show that the leading term in the long-distance expansion of the $N$-partite information behaves as
\begin{equation}\label{longdi}
I_N(\{A_i\}) \propto \left[ \frac{R}{r}\right]^{2N \Delta}\, , \quad (R\ll r)
\end{equation}
for general theories and dimensions. Here, $\Delta$ is the lowest scaling dimension of the theory and, for simplicity, we assumed all region sizes to be equal to $R$ and the same for the relative distances (up to $\mathcal{O}(1)$ factors), which we denoted $r$. Generalizing the $N=2,3$ cases, we observe that the leading term in $I_N$ generically contains $N$-,$(N-1)$-,$\dots$ and 2-point correlators of the smallest-dimension primary operator. In particular, we obtain the following general formula for the long-distance limit of the four-partite information in the case of spherical regions
\begin{align}\label{I4-final0}
\frac{I_4}{R^{8\Delta}}=&+\left[\langle \cO_1 \cO_2\cO_3 \cO_4\rangle - \frac{1}{r_{12}^{2\Delta}r_{34}^{2\Delta}} - \frac{1}{r_{13}^{2\Delta}r_{24}^{2\Delta}} - \frac{1}{r_{14}^{2\Delta}r_{23}^{2\Delta}} \right]^2 c(4\Delta)\\  \nonumber
& - \left[\frac{1}{r_{13}^{\Delta}r_{14}^{\Delta}r_{34}^{2\Delta}r_{12}^{2\Delta}r_{23}^{\Delta}r_{24}^{\Delta}} +\frac{1}{r_{12}^{\Delta}r_{14}^{\Delta}r_{24}^{2\Delta}r_{13}^{2\Delta}r_{23}^{\Delta}r_{34}^{\Delta}}+\frac{1}{r_{12}^{\Delta}r_{13}^{\Delta}r_{23}^{2\Delta}r_{14}^{2\Delta}r_{24}^{\Delta}r_{34}^{\Delta}}    \right]  \frac{9\sqrt{\pi} \Gamma\(3\Delta\)^2 (C_{\mathcal{O}\mathcal{O}\mathcal{O}})^2}{\Gamma\(\Delta\)^2 \Gamma\(4\Delta+\frac32\)} \\ \nonumber &+\left[\frac{1}{r_{12}^{2\Delta}r_{23}^{2\Delta}r_{34}^{2\Delta}r_{14}^{2\Delta}}+\frac{1}{r_{13}^{2\Delta}r_{23}^{2\Delta}r_{24}^{2\Delta}r_{14}^{2\Delta}}+\frac{1}{r_{12}^{2\Delta}r_{13}^{2\Delta}r_{24}^{2\Delta}r_{34}^{2\Delta}} \right] \[\,8c_{4:4}^{(1,1,1,1)}(\Delta)-c(4\Delta)\] \, .
\end{align}
In this formula, $R\ll r_{ij}$  $\forall i\neq j$ and for simplicity we considered equal-radii spheres, $c(4\Delta)$ was defined in \req{i2l} and $c_{4:4}^{(1,1,1,1)}(\Delta)$ is a complicated integral which appears explicitly in \req{c44:Splitting} and which we have not been able to integrate analytically, with the exception of the $\Delta=1/2$ case, for which we find\footnote{Additional numerical values are presented in Table \ref{tablaC44}.} 
\begin{equation}
c_{4:4}^{(1,1,1,1)}(\Delta=1/2)=\frac{4}{45}+\frac{2}{3\pi^2}\, .
\end{equation}
Observe that the above formula for the $I_4$ contains a four-point function of the leading primary which will be a theory-dependent  function of the  two invariant cross-ratios.  

We perform a highly non-trivial check of \req{I4-final0} by computing $I_4$ in the case of four identical disk regions of diameter $R$ centered at the corners of a square of side length $\sqrt{2}r$ for a three-dimensional free scalar in the lattice. The results read, respectively, 
\begin{equation}
\left. I_4\right|^{\rm eq.\,(\ref{I4-final0})}_{d=3\, {\rm free\, scalar}}=\left[\frac{1}{180}+\frac{1}{6\pi^2}\right]\frac{R^4}{r^4} \simeq 0.0224 \frac{R^4}{r^4}\, , \quad \left. I_4\right|^{\rm lattice}_{d=3\, {\rm free\, scalar}} \simeq 0.0207 \frac{R^4}{r^4}\, , \quad (R\ll r)
\end{equation}
which, given the limitations of the lattice, is an excellent agreement. We carry out additional lattice calculations for the free scalar for $N=2,3,4,5,6$. We show that the scaling given by \req{longdi} is respected in all cases for different shapes of the entangling regions. Our results suggest that the $N$-partite information of a three-dimensional free scalar is positive for general regions and arbitrary $N$, \ie we provide evidence that
\begin{equation}
\left. I_N(\{A_i\})\right|_{d=3\, {\rm free\, scalar}} \overset{\rm (?)}{\geq }0\quad \forall \,\{A_i\}, N \,,
\end{equation}
a result which most likely holds in general dimensions.

In the particular case of the mutual information, we perform additional lattice calculations for different pairs of regions at arbitrary distances. The dependence of mutual information on the shape of the entangling regions has been previously studied \eg in \cite{Fonda:2014cca,Casini:2015woa,MohammadiMozaffar:2015wnx,Nakaguchi:2014pha,Agon:2015mja} for holographic theories and \eg in \cite{Casini:2009sr,Bueno:2019mex,Bueno:2021fxb,Agon:2021zvp} for free fields in the lattice and in the so-called ``Extensive Mutual Information'' model. Here we consider both the free scalar as well as a free fermion in the lattice and compute the mutual information for pairs of identical regions with different shapes as a function of the separation. We observe that, conveniently normalized by the corresponding disk EE universal coefficient, $F_0$, the scalar results are systematically larger than the fermion ones, which leads us to conjecture that
\begin{equation}
\left.\frac{I_2(A_1,A_2)}{F_0}\right|_{d=3\, {\rm free\, scalar}}  \overset{(?)}{> } \left.\frac{I_2(A_1,A_2)}{F_0}\right|_{d=3\, {\rm free\, fermion}} \quad \forall \, A_1,A_2 \,,
\end{equation}
holds for arbitrary pairs of regions.

Finally, we close with a discussion on holographic conformal field theories. We first argue that at long distances the $N$-partite information of a holographic theory equals the $N$-partite information of the dual bulk theory in the gravitational background dual to the ground state of the CFT \cite{Faulkner:2013ana}. Then we prove that, at leading order in the long distance expansion, both boundary and bulk twist operators equal each other. This is done by explicit computation using the universality of the modular flow for boundary spheres and bulk hemispheres, as well as the extrapolate dictionary. The whole result implies the equality between boundary and bulk $N$-partite information for the cases in which each individual boundary region has spherical boundary.

The remainder of the paper is organized as follows. In Section \ref{Npar} we present a derivation of the $N$-partite information formula (\ref{IN44}). We then study the long-distance expansion of the twist operator and keep only the leading long-distance term. A new method for determining the OPE coefficients of such term is also presented and, in the particular case of a spherical boundary region, this
is evaluated explicitly in the $n\to 1$ limit, leading to a new explicit formula ---see \req{Cij-coeffs-eq} below. We then present a diagrammatic method to organize the computation of the leading contribution to the $N$-partite information at long distances. The method is illustrated via examples, through the explicit analysis of the $I_2$, $I_3$ and $I_4$ cases. We present closed formulas for each of these quantities in the aforementioned regime. In Section \ref{Free-CFT}
we apply the general formulas derived in the previous section to the particular case of a free scalar in arbitrary dimensions. We then consider particular geometric arrangements of the regions in three spacetime dimensions and determine the associated coefficients. 
In Section \ref{sec:Lattice} 
we compute the $N$-partite information (for $N=2,3,4,5,6$)   for various entangling regions for a three-dimensional free scalar in the lattice. We verify the leading scaling predicted by our general formula and perform a highly non-trivial check of \req{I4-final0} in the case of disk regions, finding excellent agreement. In all cases, we find $I_N$ to be positive for the free scalar. Additionally, also in the lattice, we compute the mutual information for pairs of regions both for the scalar and for a free fermion for arbitrary separations, verifying the expected scalings at short and long separations. In the case of disk regions, we provide analytic approximations valid for most separations. Based on the results obtained, we conjecture the free scalar mutual information to be greater than the free fermion one for general configurations.
 In Section \ref{holographysec}, we reinterpret our previous results from the perspective of holography. In that context, we argue that, at leading order in the long-distance expansion, the CFT twist operator associated to a spherical region equals the bulk twist operator associated to the dual hemispherical region, establishing the equality between boundary and bulk $N$-partite informations, as expected from \cite{Faulkner:2013ana}.  Appendix \ref{app-coeffs} presents an explicit computation of the various coefficients appearing in the long-distance terms of  $I_2$, $I_3$ and $I_4$. Those coefficients are given in terms of sums of products of the twist OPE coefficients $C_{ij}$. In order to carry out the various required computations of this Appendix we present a collection of formulas (and in some cases its derivations) in  Appendix \ref{App-integrals}. Appendix \ref{subsec:ScalarNumerics}  contains a table where we compare various possible fits to the lattice data obtained in the long-distance regime for the free scalar and verify that the scaling predicted by \req{longdi} is always preferred by the data. Appendix \ref{EMIMEMI} contains analytic results for the mutual information of the pairs of regions considered in Section  \ref{sec:Lattice}  for two toy models which possess, respectively, the same scaling as the free scalar and the free fermion for  long separations.

\section{$N$-partite information  for general CFTs}\label{Npar}
In this Section we provide new general formulas for the $N$-partite information of spatial regions in general CFTs and arbitrary dimensions. We start by proving that $I_N$ can be written in terms of a single $N$-point correlator of twist operators. Then, we obtain the general scaling of $I_N$ for long separations and obtain a new explicit formula for the OPE coefficients of the leading long-distance term in the case of spherical entangling surfaces.


\subsection{General formula in terms of twist operators}
As explained in the introduction, given $N$ non-empty disjoint regions we can define the $N$-partite information as in \req{IN1} or \req{IN2}.
Explicitly, \req{IN1} takes the form
\bea\label{IN-explicit}
I_N(\{ A_i\})&=&\sum_{i} S(A_i)-\sum_{i<j}S(A_iA_j)+\sum_{i<j<k}S(A_i A_jA_k)-\cdots \nonumber  +(-)^{N+1} S(A_1A_2  \cdots  A_N)\,.
\eea
This formula has a natural separation in terms of the number of regions involved in each of the entropies. We can thus use the following compact notation to write the $N$-partite information
\bea\label{IN-2}
I_N(\{A_i\})&=&-\sum_{\alpha=1}^N(-)^\alpha \sum_{i_1<\cdots<i_{\alpha}} S(A_{i_1} \cdots  A_{i_\alpha} )\,.
\eea
We will use this expression later on. In order to proceed, let us consider the replica trick, which entails the evaluation of the R\'enyi entropies for integer $n$ in the form
 \bea
S^{(n)}(A)=\frac{1}{1-n}\log\left[\frac{Z({\cal C}^{(n)}_A)}{Z^n}\right]\,,
\eea
so that analytically continuing for real $n$ and taking the $n \to 1$ limit yields the EE of region $A$,
\bea
S(A)=\lim_{n\rightarrow 1} S^{(n)}(A)\, .
\eea
In the above formula, ${\cal C}^{(n)}_A$ represents  the replica manifold for the $n$ copies of the original spacetime geometry, after suitably identifying the region $A$ of the $i$-th copy with the $(i+1)$-th one, and $n+1\leftrightarrow 1$. $Z(X)$ is the partition function of the theory defined on the manifold $X$ (for simplicity we use $Z$ when the manifold is a single copy of the original spacetime). Observe that replacing all EEs in the above definitions of the $N$-partite information by R\'enyi entropies, it is straightforward to define the $n$-th R\'enyi $N$-partite  information, $I_N^{(n)}$, which reduces to $I_N$ for $n\rightarrow 1$.

The identification process required to define the manifold ${\cal C}^{(n)}_A$ can be implemented by introducing a twist operator $\Sigma^{(n)}_A$ with support on the entangling region  \cite{Calabrese:2004eu,Calabrese:2005zw,Hung:2014npa}. This operator identifies the operators of different replicas in the region $A$, namely, $\phi_i\to \phi_{i+1}$ when $x\in A$. $\Sigma^{(n)}_A$ can be normalized such that we extract the contribution from the identity operator and write\footnote{Notice that this definition implies $\langle \tilde{\Sigma}_{A}^{(n)} \rangle=0$.}
\bea
\label{SigmaA1}
\Sigma_{A}^{(n)}= \langle \Sigma_{A}^{(n)} \rangle(1+\tilde{\Sigma}_{A}^{(n)})\, \quad { \rm with} \quad  \langle \Sigma_{A}^{(n)} \rangle= \frac{Z({\cal C}^{(n)}_A)}{Z^n}\,.
\eea
We would like to write down a closed formula for $I^{(n)}_N$ in terms of correlators of $\tilde{\Sigma}_{A}^{(n)}$. Let us start by translating equation (\ref{IN-2}) in the language of twist operators. More concretely, we take the terms of order $\alpha$ which have the form
\bea
\sum_{i_1<\cdots<i_{\alpha}} S(A_{i_1} \cdots  A_{i_\alpha} )=\lim_{n\to 1}\sum_{i_1<\cdots<i_{\alpha}}\frac{1}{1-n}\log\langle \Sigma_{A_{i_1}}^{(n)} \cdots \Sigma_{A_{i_\alpha}}^{(n)} \rangle\,.
\eea
Now, let us apply relation (\ref{SigmaA1}) to further massage the order-$\alpha$ correlators,
\bea\label{alpha-order-term-2}
\log\langle \Sigma_{A_{i_1}}^{(n)} \cdots \Sigma_{A_{i_\alpha}}^{(n)} \rangle&=&\log\left[\langle \Sigma_{A_{i_1}}^{(n)}\rangle \cdots \langle \Sigma_{A_{i_\alpha}}^{(n)} \rangle  \right] +
\log\Big[\Big\langle \(1+\tilde{\Sigma}_{A_{i_1}}^{(n)}\) \cdots \(1+\tilde{\Sigma}_{A_{i_\alpha}}^{(n)}\) \Big\rangle \Big]\, ,
 \eea
where we separated the expectation values of the individual twist operators from their tilded parts. The part that contains the products of tilded twist operators can be expanded and it becomes
\bea\label{alpha-order-term-3}
&\log\Big[\Big\langle \(1+\tilde{\Sigma}_{A_{i_1}}^{(n)}\) \cdots \(1+\tilde{\Sigma}_{A_{i_\alpha}}^{(n)}\) \Big\rangle \Big]=\\  \nonumber 
&\log\Big[1+\langle \tilde{\Sigma}_{A_{i_1}}^{(n)}\tilde{\Sigma}_{A_{i_2}}^{(n)} \rangle +\underbrace{\dots}_{\rm other\,\, pairings}+\langle \tilde{\Sigma}_{A_{i_1}}^{(n)}\tilde{\Sigma}_{A_{i_2}}^{(n)}\tilde{\Sigma}_{A_{i_3}}^{(n)}  \rangle+\underbrace{\dots}_{\rm other\,\, triplets}+\cdots + \langle \tilde{\Sigma}_{A_{i_1}}^{(n)}\tilde{\Sigma}_{A_{i_2}}^{(n)}  \cdots \tilde{\Sigma}_{A_{i_\alpha}}^{(n)}  \rangle\Big]\, ,
\eea
where, as previously noted, there are not terms proportinal to $\langle \tilde{\Sigma}_{A_i}^{(n)}\rangle$ as those are zero.

We start by analyzing the contribution of the twist normalization factors to the 
$N$-partite information. This requires studying the quantity
\bea\label{Normalization-term}
\sum_{\alpha=1}^N\(-\)^{\alpha}\left[\sum_{i_1<\cdots<i_\alpha} \frac{1}{1-n}\left[\log\langle \Sigma_{A_{i_1}}^{(n)}\rangle+ \cdots +\log\langle \Sigma_{A_{i_\alpha}}^{(n)}\rangle  \right] \right]\,.
\eea 
Notice that we have separated the logarithm of the product of expectations values of twist operators appearing in (\ref{alpha-order-term-2}) into the sum of logarithms of each individual expectation value. 
For each order $\alpha$, the number of ordered $\alpha$ indices is simply given by the number of combinations of $\alpha$ elements one can take from a set of $N$ objects, \ie by the binomial coefficient $\binom{N}{\alpha}$. For each choice of ordered indices we have the sum of $\alpha$ different terms, therefore, the total number of terms of the form $\log\langle \Sigma^{(n)}_{A_j} \rangle $  is $\alpha\cdot \binom{N}{\alpha}$ for each $\alpha$. On the other hand, among all these terms the logarithm of the twist operator of a particular region should appear the same number of times regardless of the chosen region, since the formula treats all regions on an equal footing. Thus, we conclude that for each $\alpha$ we have
\bea
\sum_{i_1<\cdots<i_\alpha} \frac{1}{1-n}\left[\log\langle \Sigma_{A_{i_1}}^{(n)}\rangle+ \cdots \log\langle \Sigma_{A_{i_\alpha}}^{(n)}\rangle  \right] =\frac{\alpha }{N}\binom{N}{\alpha}\sum_{i=1}^N\frac{1}{1-n}\log\langle \Sigma_{A_{i}}^{(n)}\rangle\, .
\eea
Summing the above equation over $\alpha$ with the appropriate signs one gets that \req{Normalization-term} equals
\bea 
\(\sum_{\alpha=1}^N\(-\)^{\alpha} \binom{N-1}{\alpha-1}\)\sum_{i=1}^N\frac{1}{1-n}\log\langle \Sigma_{A_{i}}^{(n)}\rangle\,.
\eea
Shifting the summation index $\alpha$ by $1$, the term in parenthesis can be identified with the binomial expansion of $(1-1)^{N-1}$, and therefore, vanishes identically for $N>1$. This shows that $I_N$ for $N>1$ does not depend on the expectation value of the individual twist operators. Notice that for $N=1$, $I_1$ reduces to the usual entanglement entropy and in that case the  normalization factor is the answer. This result is true even for the Renyi $I^{(n)}_N$, as we did not need to take the $n\to 1$ limit to arrive at this conclusion.

Now, we analyze the contribution of the second term in the right hand side of (\ref{alpha-order-term-2}) to $I_N$. The first thing to notice is that to simplify those contributions, it is convenient to take the $n\to 1$ at this stage. The correlator of an arbitrary number of products of $\tilde{\Sigma}^{(n)}$'s vanishes as $(n-1)$ in the $n\to 1$ limit, therefore, the linear approximation of the logarithm becomes exact in such a limit. This means that the contribution to $I_N$ we are interested in is 
\begin{align}\label{alpha-order-1}
I_N\supset  \lim_{n\to 1} \frac{1}{1-n}\sum_{\alpha=1}^N\(-\)^{\alpha}\!\!\! \sum_{i_1<\cdots <i_\alpha}\Big[ &+\langle \tilde{\Sigma}_{A_{i_1}}^{(n)}\tilde{\Sigma}_{A_{i_2}}^{(n)} \rangle +\underbrace{\dots}_{\rm other\,\, pairings}  + \langle \tilde{\Sigma}_{A_{i_1}}^{(n)}\tilde{\Sigma}_{A_{i_2}}^{(n)}\tilde{\Sigma}_{A_{i_3}}^{(n)}  \rangle+\underbrace{\dots}_{\rm other\,\, triplets} \\
&+\cdots+ \langle \tilde{\Sigma}_{A_{i_1}}^{(n)}\tilde{\Sigma}_{A_{i_2}}^{(n)}  \cdots \tilde{\Sigma}_{A_{i_\alpha}}^{(n)}  \rangle\Big]\,.
\end{align}
For each $\alpha$ there would be contributions coming from correlators of each order $K$ with $\alpha \geq K$. Thus we will study the contribution to $I_N$ from all correlators of  order $K$ for each $K$. These correlators will only appear for $\alpha \geq K$. Now, for each $\alpha \geq K$ the number of ordered $\alpha$ indices appearing in \req{alpha-order-1} is given by $\binom{N}{\alpha}$ and each such term will contain $\binom{\alpha}{K}$ correlators of $K$ tilded twist operators, since that is the number of groups of $K$ elements one can made out of $\alpha$ operators. Finally, the number of times a given correlator of $K$ operators appears in such term will be independent of the choice of correlator and therefore we have that the contribution to $I_N$ from the order $K$ correlators is given by 
\bea\label{cont-K-corr}
 \lim_{n\to 1} \(\sum_{\alpha=K}^N (-)^\alpha \frac{\binom{N}{\alpha}\binom{\alpha}{K}}{\binom{N}{K}} \) \sum_{i_1<\cdots <i_K}\langle \tilde{\Sigma}_{A_{i_1}}^{(n)} \cdots \tilde{\Sigma}_{A_{i_K}}^{(n)}\rangle
\eea
where the factor of $\binom{N}{K}$ in the denominator is the total number of distinct $K$ correlators we can form out of $N$ twist operators, and thus, it coincides with the number of terms appearing in the above sum over $K$ indices.  Using the identity
\bea
\frac{\binom{N}{\alpha}\binom{\alpha}{K}}{\binom{N}{K}}=\binom{N-K}{\alpha-K}\,,
\eea
and shifting the summation index in \req{cont-K-corr}, we find that the term in parenthesis is nothing but the binomial expansion of $(1-1)^{N-K}$, and thus, it identically vanishes for $N>K$. This means that the only correlator of twist operators that can contribute to the $N$-partite information is precisely the correlator of $N$ twist operators. Hence, we find that the $N$-partite information is given by 
\bea\label{IN4}
I_N(\{A_i\})=\lim_{n\to 1}\frac{\(-\)^{N+1}}{1-n}\langle \tilde{\Sigma}_{A_1}^{(n)}\tilde{\Sigma}_{A_2}^{(n)} \cdots \tilde{\Sigma}_{A_N}^{(n)}\rangle\,.
\eea

This result can be alternatively proved by induction. This method requires guessing \req{IN4}, but such equation is the natural generalization of the $I_2$ and $I_3$ cases worked out in detailed for example in \cite{Agon:2021lus}. Therefore, we start assuming that \req{IN4} holds for $N=K-1$. Then, we use the recursive definition of $I_N$ presented in \req{IN2} to write down a formula for $I_K$, namely,
\bea\label{IK}
I_K(\{A_i\})=\lim_{n\to 1}\frac{\(-\)^{K}}{1-n}\Big\langle \tilde{\Sigma}_{A_1}^{(n)}\tilde{\Sigma}_{A_2}^{(n)} \cdots  \tilde{\Sigma}_{A_{K-2}}^{(n)} \(\tilde{\Sigma}_{A_{K-1}}^{(n)}+\tilde{\Sigma}_{A_{K}}^{(n)}-\tilde{\Sigma}_{A_{K-1} A_K}^{(n)} \)\Big\rangle\,.
\eea
Using \req{SigmaA1} for the operator $\tilde{\Sigma}_{A_{K-1} A_K}^{(n)}$ we have
\bea\label{tildeSigma-prod}
\tilde{\Sigma}_{A_{K-1} A_K}^{(n)}&=&\frac{\Sigma_{A_{K-1}}^{(n)}\Sigma_{A_K}^{(n)}}{\langle \Sigma_{A_{K-1}}^{(n)}\Sigma_{A_K}^{(n)} \rangle }-1
=\frac{1+\tilde{\Sigma}_{A_{K-1}}^{(n)}+\tilde{\Sigma}_{A_{K}}^{(n)}+\tilde{\Sigma}_{A_{K-1}}^{(n)}\tilde{\Sigma}_{A_{K}}^{(n)}}{1+\langle \tilde{\Sigma}_{A_{K-1}}^{(n)}\tilde{\Sigma}_{A_{K}}^{(n)} \rangle }-1 \, ,
\eea
where in the first equality we used $\Sigma_{A_{K-1} A_K}^{(n)}=\Sigma_{A_{K-1}}^{(n)}\Sigma_{A_K}^{(n)}$ while in the second we used \req{SigmaA1}. Taking the $n\to 1$ limit, the above equation reduces to 
\bea\label{tildeSigma-prod-2}
\tilde{\Sigma}_{A_{K-1}A_K}^{(n)}&\approx& \tilde{\Sigma}_{A_{K-1}}^{(n)}+\tilde{\Sigma}_{A_{K}}^{(n)}+\tilde{\Sigma}_{A_{K-1}}^{(n)}\tilde{\Sigma}_{A_{K}}^{(n)}\,.
\eea
Notice that we have ignored the extra term in the denominator.  This is possible because in the $n\to 1$ limit, the correlator of an arbitrary number of twist operators goes as $(n-1)$ and therefore, any extra $(n-1)$ factor in \req{tildeSigma-prod} would not contribute to the $N$-partite information. Plugging \req{tildeSigma-prod-2} into \req{IK} leads to \req{IN4} for $N=K$,
and therefore it completes the proof.

\subsection{Long-distance behavior of $I_N$\label{ldb}}
With \req{IN4} at hand, let us now study the long-distance behavior of the $N$-partite information.
In this regime, we can approximate each twist operator by the product of local operators located at a convenient location inside the region $X$ as \cite{Headrick:2010zt,Cardy:2013nua} 
\bea\label{OPE-primary-expansion}
\tilde{\Sigma}_{X}^{(n)}=\sum_{\{k_j\}\neq \mathbb{I}}C_{\{k_j\}}^{X} \prod_{j=0}^{n-1} \Phi^{(j)}_{k_j}(x_X)\, .
\eea
For the $j$-th copy, the $k_j$ label a complete  set of operators (with the identity removed).
The leading contribution at long distances will come from the lowest scaling dimension operator of the theory. Assuming for simplicity that such operator is a scalar of conformal dimension $\Delta$, which we denote by $\cO$, then we can simply keep the term 
\bea\label{Sigmatilde}
\tilde{\Sigma}^{(n)}_X\approx \sum_{i<j}C^X_{ij} \cO^{i}(x_X)\cO^{j}(x_X) \, .
\eea
Observe that in principle there is also a contribution of the form $\sum_i C^X_i\, {\cal O}^i(x_X) $, but such single-sheet operators do not contribute in the $n\to 1$ limit.\footnote{The argument is the following: the coefficient $C_i$ has support on a single sheet and by the replica symmetry it cannot depend on $i$, so it should be a constant. On the other hand, since $\tilde{\Sigma}^{(n)}_X \to 1$ as $n\to 1$, then, one concludes that $C_i\sim \mathcal{O}(n-1)$. In the computation of the $I_{N>1}$ we would always have products of more than one of those coefficients. Therefore, they cannot contribute to $I_N$ in the $n\to 1$ limit.} We abbreviate $\cO^i(x_X)$ with $\cO^i_X$. We apply this formalism to the particular case when $X$ is a ball of radius $R_X$. In that situation, it is convenient to use the new coefficients $C_{ij}\equiv C^X_{ij}R_X^{-2\Delta}$ which are scale independent and thus are the same for all the ball-like regions involved. 

The coefficients $C_{ij}$ can be computed by studying the effect of a twist operator in the correlator of test operators far away from the region $X$, and located on different sheets. This is, one considers 
\bea\label{twist-test}
\langle \tilde{\Sigma}^{(n)}_X  \mathcal{O}^i (x) \mathcal{O}^j (x)\rangle \, ,
\eea
with $ |x-x_X|\gg R_X$. Using the above equation, \req{Sigmatilde} and the two-point function
\begin{equation}
\braket{\mathcal{O}^i(x) \cO^{j}(x_X)}=\frac{\delta^{ij}}{|x-x_X|^{2\Delta}}\, ,
\end{equation}
one finds the following formula for $C_{ij}$
\bea\label{Cij-r-limit}
C_{ij}=\lim_{x\to \infty} \frac{|x-x_X|^{4\Delta}}{R_X^{2\Delta}}\langle \tilde{\Sigma}^{(n)}_X  \mathcal{O}^i (x) \mathcal{O}^j (x)\rangle\,.
\eea
The $C_{ij}$ coefficients are determined by the two-point function in replica space. Earlier works on the type of computations presented in this paper did not obtain explicitly the form of those coefficients but instead they managed to compute sums over these coefficients in the $n\to 1$ limit. We devote subsection \ref{Cij-coeffs} to study the $C_{ij}$ coefficients and argue that 
\bea\label{Cij-coeffs-eq}
C_{ij}=\frac{1}{\sin^{2\Delta}\[\frac{\pi (i-j)}n\]} \, ,
\eea
in the $n\to 1$ limit. Namely, the above expression is valid for the purposes of computing entanglement related quantities, so after adding over all the indices we need to keep only the piece proportional to $(n-1)$ in replica computations. 

Now, let us write down the correlator in \req{IN4} by keeping only the leading long distance piece appearing in \req{Sigmatilde}, 
\bea\label{correlator-IN}
\langle \prod_{\alpha=1}^N \tilde{\Sigma}_{A_\alpha}^{(n)}\rangle\,=\prod_{\alpha=1}^N R^{2\Delta}_{A_\alpha}\sum_{i_\alpha<j_\alpha}C_{i_\alpha j_\alpha}\langle \prod_{\beta=1}^N \cO^{i_\beta}_{A_\beta}\cO^{j_\beta}_{A_\beta}  \rangle  \, .
\eea
The case in which all the indices appearing in the above correlator are different reduces to a product of the expectation values of the individual operators, and since $\langle \cO^i_{A_{J}} \rangle=0 $ such configuration does not contribute. 
On the other hand, the configurations that use the minimum number of sheets simply put all the twist operators on the same pair of sheets. Following this philosophy, we will separate the possible configurations in the correlator (\ref{correlator-IN}) in terms of the number of sheets $\cal N$ with non-trivial operator insertions. Below, we will represent those configurations using matrices where the columns represent the different sheets while rows represent the different regions. 

Before moving on, observe that \req{correlator-IN} makes  evident the leading scaling of the $N$-partite information at long distances. Indeed, if we assume for simplicity that all ball regions are characterized by the same radius $R$ and that each pair is separated a distance $ r$, up to order-$1$ factors, we have 
\begin{equation}\label{iNN}
I_N(\{A_i \}) \propto \left[ \frac{R}{r}\right]^{2 N \Delta} \,,
\end{equation}
where the $r$-dependent piece comes from the $N$-point correlator of the operators $\cO^{i_\beta}_{A_\beta}\cO^{j_\beta}_{A_\beta}$ which, as we show explicitly below for the $N=2,3,4$ cases, gives rise to various products of correlators of same-sheet different-region operators involving a total of $2N$ operators each. Schematically, for each pair of operators we get a contribution $\sim r^{-2\Delta}$, and so $I_N(\{A_i \}) \propto  r^{-2N\Delta}$. While this result is derived assuming the entangling surfaces to be spheres, the scaling in \req{iNN} holds for arbitrary regions as long as all relevant scales characterizing them are much smaller than the mutual separations.  

For the analysis of the correlator appearing in \req{correlator-IN} it is useful to introduce the following graph representation of the twist operator,
 \begin{figure}[h!]
\centering
\begin{tikzpicture}
  \draw [thick] (0,0) -- (2,0);
\draw [fill] (0,0) circle[radius=0.06]; 
\draw [fill] (2,0) circle[radius=0.06]; 
\node [below, left] at (-0.5,0.1) {$\tilde{\Sigma}_A\approx \sum_{i<j}C_{ij}\mathcal{O}^i_A\mathcal{O}^j_A\approx$};
\node [below, left] at (-0.05,-0.2) {$i$};
\node [below, right] at (2.05,-0.2) {$j$};
\node [above, right] at (0.7,0.3) {$A$};
\end{tikzpicture}
\label{fig-0}
\end{figure}
where vertices represent the different sheets on which local operators are supported, and the lines represent the spatial region on which the operator is located.  We also introduce some notation for the various coefficients appearing in the $I_N$. We will denote them by  $c_{N:K}^{(i_1,\cdots,i_\alpha)}$, where the subindex $N$ stands for the fact we are considering the $N$-partite information while the index $K$ means that such coefficient is given in terms of a sum involving $K$ different sheets. The super index $(i_1,\cdots, i_\alpha)$ characterizes the various powers of $C_{ij}$ appearing in the expression for that coefficient. Notice that the $c_{N:K}^{(i_1,\cdots,i_\alpha)}$ are symmetric under permutations of the indices $\{i_1,\cdots,i_\alpha\}$ and thus, we can chose to denote them in increasing order.

\subsubsection{The $C_{ij}$ coefficients \label{Cij-coeffs}}
Let us now return to the problem of determining the functional form of the $C_{ij}$ coefficients. In order to evaluate \req{Cij-r-limit}, one needs a formula for the correlator (\ref{twist-test}). Recently, one such formula was derived in  \cite{Casini:2021raa}. For the special case in which the operators in question are scalars, this takes the form 
\bea \label{twist-correlator}
\langle \Omega |\Sigma^{(n)}_A {\cal O}^{l}(x) {\cal O}^{k}(x)|\Omega\rangle &=&\frac{\tr\left\{{\cal O}(x)\rho_A^{n-(l-k)}\,{\cal O}(x) \rho_A^{(l-k)} \right\}}{\tr\rho^n_A}\,.
\eea
The above result can be put in a more familiar form in terms of the density operator $\tilde{\rho}_A(n):=\rho_A^n/\tr{\rho_A^n}$. Since $\rho_A$ is a density operator, and therefore Hermitian, completely positive and satisfying $\tr{\rho_A}=1$; the same holds for $\tilde{\rho}_A(n)$ for $n>0$. Thus \req{twist-correlator} becomes
\bea \label{twist-correlator-2}
\langle \Omega |\Sigma^{(n)}_A {\cal O}^{l}(x) {\cal O}^{k}(x)|\Omega\rangle &=&\tr\left\{\tilde{\rho}_A(n)\,{\cal O}^{(n)}_{A}\[x,i\frac{\tau_{kl}}{\pi}\]{ \cal O}(x) \right\}\, ,
\eea
with $\tau_{kl}\equiv \pi(k-l)/n$. Notice that the operators on the RHS are defined in the original CFT (a single sheet of the replica space), and 
\bea
{\cal O}^{(n)}_{A}\[x,s\]\equiv \tilde{\rho}_A(n)^{-i s} {\cal O}(x)\tilde{\rho}_A(n)^{i s}
\eea
is the operator transformed by the modular flow induced by $\tilde{\rho}_A(n)$. On the RHS of \req{twist-correlator-2} we have the correlation function of a field with a modular-evolved field. Such correlator is analytic in the strip with Im$\,s \in \(0,1\)$, and obeys the KMS periodicity between the boundaries of the strip \cite{Haag:1992hx}. This implies that the correlator is well defined on $0\leq \tau_{kl}\leq 1$. It follows from its definition that $1/n<\tau_{kl}<1-1/n$ provided we ordered the indices such that $k>l$. We will do so here. 

The correlator on the  RHS of \req{twist-correlator-2} can be studied for any density operator $\rho$. This observation suggests to approximate the operator 
$\tilde{\rho}_A(n)$ by $\rho_A$ in the $n\to 1$ limit. Any correction to this approximation will be of order $(n-1)$ and thus it will not contribute to the entropies.
Taking the $n\to 1$ limit on the state while keeping the $n$ dependence on the modular parameters is the only consistent approximation of that correlator. Thus, in this regime we have
\bea \label{twist-correlator-3}
\langle \Omega |\Sigma^{(n)}_A {\cal O}^{l}(x) {\cal O}^{k}(x)|\Omega\rangle &\approx &\tr\left\{\rho_A\,{\cal O}_{A}\[x,i\frac{\tau_{kl}}{\pi}\]{ \cal O}(x) \right\}=\Big\langle{\cal O}_{A}\[x,i\frac{\tau_{kl}}{\pi}\]{ \cal O}(x)  \Big\rangle \,.
\eea
This relation coincides with the so called $1/n$ prescription derived in \cite{Chen:2017hbk}. Such prescription has been sucessfully applied in similar entanglement computations in the literature, see for example \cite{Chen:2017hbk, Chen:2016mya} and references therein.

Now, we are ready to compute the $C_{ij}$ coefficients starting from \req{Cij-r-limit} and the large $r$ limit of \req{twist-correlator-3}. For a sphere of radius $R$, the modular flow acts geometrically on local operators via 
\bea\label{sphere-modular}
r^{\pm}[s]=R\frac{\(R+r^{\pm}\)-e^{\mp 2\pi s} \(R-r^{\pm}\) }{\(R+r^{\pm}\)+e^{\mp 2\pi s} \(R-r^{\pm}\)}\, ,
\eea
where $r^\pm=r\pm t$ are standard null coordinates \cite{Casini:2011kv}. This flow can be interpreted as a conformal transformation, which helps us derive the form in which the unitary acts on the operator, this is 
\bea
{\cal O}_{A}[x,s]=U_A(s){\cal O}(x)U_A^{\dagger}(s)=\Omega^\Delta[x,s] {\cal O}(x[s]) \, ,
\eea 
where $U_A(s)$ is the unitary that implements the modular evolution, while $\Omega[x,s]$ is the conformal factor associated to the coordinate transformation. This can be derived from \req{sphere-modular} and the relation $d x[s]^2=\Omega^2[x,s]d x^2$\,. When $r\to \infty$ we find 
\bea
\langle{\cal O}_{A}\[x,s\]{ \cal O}(x)\rangle=\frac{1}{\(-\sinh^2[\pi s]\)^{\Delta}}\frac{R^{2\Delta}}{r^{4\Delta}}\,.
\eea
Replacing $s\to i \tau_{kl}/\pi$ in the above formula, and plugging the result into \req{Cij-r-limit} leads to our final formula for the $C_{kl}$, which reads 
\bea\label{Ckl-coeff}
C_{kl} =\frac{1}{\sin^{2\Delta} \tau_{kl}} \quad {\rm with}\quad \tau_{kl}= \frac{\pi (k-l)}{n}\,.
\eea
In Appendix \ref{app-coeffs} we use the above expression for the $C_{kl}$ to compute the relevant coefficients appearing in the mutual and tripartite information at long distances, this is done in sections \ref{App-mutual} and \ref{App-tripartite} respectively. We reproduce results for these quantities previously appeared in the literature, which provides strong support for the validity of the above formula. In the same appendix we also compute the coefficients appearing in the four-partite information at long distances (which are all new in the literature). 

\subsubsection{Explicit formulas for $I_2$}
Completely explicit formulas for $I_N$ in the long-distance regime can in principle be obtained from \req{IN4} on a case-by-case basis for arbitrary values of $N$. In the following subsections we present such formulas for $N=2,3,4$ in the case of spherical entangling surfaces. Starting with the mutual information case, we have
\bea\label{I2}
I_2(A_1, A_2)=\lim_{n\to 1}\frac{1}{n-1}\langle \tilde{\Sigma}_{A_1}^{(n)}\tilde{\Sigma}_{A_2}^{(n)}\rangle\,.
\eea
As we have seen, in the long-distance regime the correlator can be written as
\bea\label{correlator-I2}
\langle \tilde{\Sigma}_{A_1}^{(n)}\tilde{\Sigma}_{A_2}^{(n)}\rangle\,=\(R_{1}R_{2}\)^{2\Delta}\sum_{i<j}\sum_{k<l}C_{ij}C_{kl}\langle \cO^i_{A_1}\cO^j_{A_1} \cO^k_{A_2}\cO^l_{A_2}   \rangle \, ,  
\eea
where $R_{i}$ are the radii of the spheres. The various contributions can be organized in terms of the number of sheets with non-trivial operator insertions, \vspace{0.4cm} \\ 
 \noindent\begin{minipage}{0.65\linewidth}
 \begin{equation}  \notag
           \text{${\cal N}=2,$} \quad 
\displaystyle \begin{pmatrix}
1 &\cdots \cO^i_{A_1}& \cdots \cO^j_{A_1} & \cdots 1 \\
1 &\cdots \cO^i_{A_2} & \cdots \cO^j_{A_2} &\cdots 1 \\
\end{pmatrix}  \, ,    \end{equation}
    \end{minipage} \hfill
    \begin{minipage}{0.5\linewidth}
       \begin{tikzpicture}
 \draw [thick] (0,0) to [out=30, in=150] (2,0);
  \draw [thick] (0,0) to [out=-30, in=-150] (2,0);
\draw [fill] (0,0) circle[radius=0.06]; 
\draw [fill] (2,0) circle[radius=0.06]; 
\node [below, left] at (-0.05,-0.2) {$i$};
\node [below, right] at (2.05,-0.2) {$j$};
\end{tikzpicture} 
    \end{minipage}  \vspace{0.05cm}  

\noindent
as previously described. We define the following coefficient 
\bea
c_{2:2}^{(2)}\equiv \lim_{n\to 1}\frac{1}{n-1}\sum_{i<j}C^2_{ij}\, , 
\eea
which we compute explicitly in \req{C2:2(2)} below. Plugging the result into the mutual information we get
\bea\label{I2-2}
I_2(A_1, A_2)=\frac{\sqrt{\pi}}{4}\frac{\Gamma\(2\Delta+1\)}{\Gamma\(2\Delta+\frac{3}{2}\)}\langle \cO_{A_1} \cO_{A_2} \rangle^2 R_1^{2\Delta}R_2^{2\Delta}\,.
\eea
In Fig.\,\ref{I2-space-rep} we provide a graph representation of the long distance mutual information in terms of standard spatial correlators. 
\begin{figure}[t!]
\centering
\begin{tikzpicture}

  \draw [thick]  (0,0.15) to (1.7,0.15);
  \draw [thick]  (0,-0.15) to  (1.7,-0.15);      

  \draw [fill] (0,0.15) circle[radius=0.06]; 
   \draw [fill] (0,-0.15) circle[radius=0.06]; 
\draw [thick] (0,0) circle[radius=0.35]; 
\draw [fill] (1.7,0.15) circle[radius=0.06]; 
\draw [fill] (1.7,-0.15) circle[radius=0.06]; 
\draw [thick] (1.7,0) circle[radius=0.35]; 

\node [below, left] at (-0.2,-0.3) {$A_1$};
\node [below, right] at (1.9,-0.3) {$A_2$};
\end{tikzpicture}
\caption{Graph representation of the mutual information. The result involves the square of the correlator of the primary operator with the lowest conformal dimension of the theory. In each correlator, the primary is defined at two points lying at the (centers of each of the) spheres.  \label{I2-space-rep}}
\end{figure}
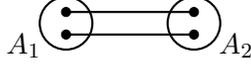 The above formula can be written more explicitly using the expression for the correlator of the leading primary operator,
\begin{equation}\label{2pf}
\langle \cO(x_{A_1}) \cO(x_{A_2}) \rangle =\frac{1}{r_{12}^{2\Delta}}\, ,
\end{equation}
where $r_{12}\equiv |x_{A_1}-x_{A_2}|$ is the distance between the spheres centers. Then we get
\bea\label{I2-2}
I_2(A_1,A_2)=\frac{\sqrt{\pi}}{4}\frac{\Gamma\(2\Delta+1\)}{\Gamma\(2\Delta+\frac{3}{2}\)} \frac{R_1^{2\Delta} R_{2}^{2\Delta}}{r_{12}^{4\Delta}}\,,
\eea
in agreement with previous results \cite{Calabrese:2010he,Agon:2015ftl,Chen:2017hbk,Casini:2021raa}.

\subsubsection{Explicit formulas for $I_3$}
Let us now consider the tripartite information. In that case, we have
\bea\label{I3}
I_3(A_1,A_2, A_3)=-\lim_{n\to 1}\frac{1}{n-1}\langle \tilde{\Sigma}_{A_1}^{(n)}\tilde{\Sigma}_{A_2}^{(n)}\tilde{\Sigma}_{A_3}^{(n)}\rangle\,.
\eea
The leading contribution to $I_3$ comes from replacing the twist operators by a  product of two operators of lowest scaling dimension as in \req{Sigmatilde}, which leads to the following expression for the correlator in \req{I3},
\bea\label{correlator-I3}
\langle \tilde{\Sigma}_{A_1}^{(n)}\tilde{\Sigma}_{A_2}^{(n)}\tilde{\Sigma}_{A_3}^{(n)}\rangle\,=\(R_{1} R_{2} R_{3}\)^{2\Delta}\sum_{i<j}\sum_{k<l}\sum_{m<n}C_{ij}C_{kl}C_{mn}\langle \cO^i_{A_1}\cO^j_{A_1} \cO^k_{A_2}\cO^l_{A_2}   \cO^m_{A_3}\cO^n_{A_3} \rangle  \,.
\eea
For ${\cal N}=2$ non-trivial sheets there is one only possible configuration which has the following structure\vspace{0.2cm} \\
 \noindent\begin{minipage}{0.65\linewidth}
 \begin{equation}  \notag
           \text{${\cal N}=2,$} \quad 
\displaystyle \begin{pmatrix}
1 &\cdots \cO^i_{A_1}& \cdots \cO^j_{A_1} & \cdots 1 \\
1 &\cdots \cO^i_{A_2} & \cdots \cO^j_{A_2} &\cdots 1 \\
1 &\cdots \cO^i_{A_3}& \cdots \cO^j_{A_3} &\cdots 1 \\
\end{pmatrix}  \, ,    \end{equation}
    \end{minipage} \hfill
    \begin{minipage}{0.5\linewidth}
        \begin{tikzpicture}
 \draw [thick] (0,0) to [out=30, in=150] (2,0);
  \draw [thick] (0,0) -- (2,0);
  \draw [thick] (0,0) to [out=-30, in=-150] (2,0);
\draw [fill] (0,0) circle[radius=0.06]; 
\draw [fill] (2,0) circle[radius=0.06]; 
\node [below, left] at (-0.05,-0.2) {$i$};
\node [below, right] at (2.05,-0.2) {$j$};
\end{tikzpicture}
    \end{minipage}  \vspace{0.05cm}  
    
 \noindent as previously described. The contribution of such term to \req{correlator-I3} is
\bea\label{I3-3pt}
\sum_{i<j}C^3_{ij}\langle \cO^i_{A_1} \cO^i_{A_2} \cO^i_{A_3} \rangle \langle \cO^j_{A_1} \cO^j_{A_2}\cO^j_{A_3} \rangle\,.
\eea
Notice in the above graph that each vertex represents a correlator. The number of legs attached to it will correspond to the number of operators, thus, we have a product of two three point functions, one at sheet $i$ and the other one at sheet $j$.  

The other non-zero configurations will necesarilly have two operators on each row (sheet), one such example will be \\
 \noindent\begin{minipage}{0.65\linewidth}
 \begin{equation}  \notag
           \text{${\cal N}=3,$} \quad 
\displaystyle \begin{pmatrix}
1 &\cdots \cO^i_{A_1} & \cdots \cO^j_{A_1} & \cdots 1 &\cdots 1  \\
1 &\cdots 1 & \cdots \cO^j_{A_2}& \cdots  \cO^k_{A_2} &\cdots 1\\
1 &\cdots \cO^i_{A_3}& \cdots 1 & \cdots  \cO^k_{A_3} &\cdots  1 \\
\end{pmatrix}  \, ,    \end{equation}
    \end{minipage} \hfill
    \begin{minipage}{0.5\linewidth}
         \begin{tikzpicture}
\draw [thick] (0,0) --(1,1.7);
 \draw [thick] (1,1.7)-- (2,0); 
 \draw [thick] (0,0) -- (2,0); 
\draw [fill] (0,0) circle[radius=0.06]; 
\draw [fill] (1,1.7) circle[radius=0.06];
\draw [fill] (2,0) circle[radius=0.06]; 
\node [below, left] at (-0.05,-0.2) {$i$};
\node [below, right] at (2.05,-0.2) {$j$};
\node [above] at (1,1.8) {$k$};
\end{tikzpicture}
    \end{minipage} \vspace{0.05cm}

    \noindent
This graph is obtained by taking three vertices (corresponding to three different sheets), and three lines (corresponding to the three twist operators) and constructing  a graph with the condition that on each vertex there must be at least two lines connected to it. This leads to the single graph above. The terms contributing to the correlator with such a graph are 
\bea
\sum_{i<j<k}C_{ij}C_{ik}C_{jk}\(\langle \cO^i_{A_1} \cO^i_{A_3} \rangle \langle \cO^j_{A_1} \cO^j_{A_2}  \rangle\langle \cO^k_{A_2} \cO^k_{A_3} \rangle +{\rm permutations \,\, of }\,\, \{A_1,A_2,A_3\} \)\,.
\eea
The number of terms in the parenthesis with the exact same contribution is $3!$, which equals the number of different ways we can label the sides of the triangle with labels $A_1, A_2,$ and $A_3$.
Thus, we conclude that the $I_3$, is given by 
\begin{align}\notag
I_3=-\lim_{n\to 1}\frac{\(R_{1} R_{2} R_{3}\)^{2\Delta}}{n-1}  \Big[  &+\sum_{i<j}C^3_{ij}\langle \cO^i_{A_1} \cO^i_{A_2}\cO^i_{A_3} \rangle \langle \cO^j_{A_1} \cO^j_{A_2}\cO^j_{A_3} \rangle  \\ & + 6\sum_{i<j<k}C_{ij}C_{jk}C_{ki}\langle \cO^i_{A_1} \cO^i_{A_3} \rangle \langle \cO^j_{A_1} \cO^j_{A_2}  \rangle\langle \cO^k_{A_2} \cO^k_{A_3} \rangle\Big]\, .
\end{align}
Let us now define the following coefficients
\bea
c_{3:2}^{(3)}\equiv \lim_{n\to 1}\frac{1}{n-1}\sum_{i<j}C^3_{ij}\qquad {\rm and }\qquad c_{3:3}^{(1,1,1)}\equiv \lim_{n\to 1}\frac{1}{n-1}\sum_{i<j<k}C_{ij}C_{jk}C_{ki}\, ,
\eea 
which we explicitly compute in detail in Appendix \ref{App-tripartite}. The results appear in \req{C3:2(3)} and \req{C3:3(1,1,1)}, respectively. In terms of these expressions the tripartite information adopts the final form \cite{Agon:2021lus}
\bea\label{I3-final}
I_3=\Bigg[ \frac{2^{6 \Delta} \Gamma\(\Delta+\frac12\)^3}{2 \pi \Gamma\( 3\Delta+\frac32\)}\langle \cO_{1} \cO_{3} \rangle \langle \cO_{1} \cO_{2}  \rangle\langle \cO_{2} \cO_{3} \rangle- \frac{2^{6\Delta}\(\Gamma\(3\Delta+1\)\)^2}{\Gamma\(6\Delta+2\)}\langle \cO_{1} \cO_{2}\cO_{3} \rangle^2\, \,\Bigg] \(R_{1} R_{2} R_{3}\)^{2\Delta} \, ,
\eea
where we used the notation $\cO_{i}\equiv \cO_{A_i}$. Notice that in cases in which $\langle  \cO \cO \cO\rangle =0$, we have $I_3>0$. For those, the mutual information is non-monogamous at long distances.

The above formula can also be illustrated graphically in the usual space representation of correlators as shown in Fig.\,\ref{I3-space-rep}. 
\begin{figure}[t]
\centering
\begin{tikzpicture}
  \draw [thick]  (-0.1,0.1) to (0.65,0.85);
    \draw [thick]  (0.1,-0.1) to  (1.05,0.85);      
  \draw [thick] (1.05,0.85) to (1.8,0.1);
   \draw [thick] (1.6,-0.1) to (0.65,0.85);
       \draw [thick]  (1.05,0.85) to (1.05, 1.7);   
       \draw [thick]  (0.65,0.85) to (0.65, 1.7);   
   
  \draw [fill] (0.65, 1.7) circle[radius=0.06]; 
   \draw [fill] (1.05, 1.7) circle[radius=0.06]; 
\draw [thick] (0,0) circle[radius=0.35]; 
\draw [fill] (-0.1,0.1) circle[radius=0.06]; 
\draw [fill] (0.1,-0.1) circle[radius=0.06]; 
\draw [thick] (1.7,0) circle[radius=0.35]; 
\draw [fill] (1.6,-0.1) circle[radius=0.06]; 
\draw [fill] (1.8,0.1) circle[radius=0.06]; 
\draw [thick] (0.85,1.7) circle[radius=0.35]; 

 \draw [fill] (0.58,0.78)-- (0.72,0.78)--(0.65,0.95)--(0.57,0.78); 
 
 \draw [fill] (0.98,0.78)--  (1.12,0.78)--(1.05,0.95)-- (0.98,0.78); 

\node [below, left] at (-0.2,-0.3) {$A_1$};
\node [below, right] at (0.57,2.35) {$A_3$};
\node [below, right] at (1.9,-0.3) {$A_2$};
\end{tikzpicture}
\begin{tikzpicture}

  \draw [thick]  (-0.1,0.1) to (0.65,1.7);
  \draw [thick]  (0.1,-0.1) to  (1.6,-0.1);      
   \draw [thick]  (1.8,0.1) to (1.05, 1.7);

  \draw [fill] (0.65, 1.7) circle[radius=0.06]; 
   \draw [fill] (1.05, 1.7) circle[radius=0.06]; 
\draw [thick] (0,0) circle[radius=0.35]; 
\draw [fill] (-0.1,0.1) circle[radius=0.06]; 
\draw [fill] (0.1,-0.1) circle[radius=0.06]; 
\draw [thick] (1.7,0) circle[radius=0.35]; 
\draw [fill] (1.6,-0.1) circle[radius=0.06]; 
\draw [fill] (1.8,0.1) circle[radius=0.06]; 
\draw [thick] (0.85,1.7) circle[radius=0.35]; 

\node [below, left] at (-0.2,-0.3) {$A_1$};
\node [below, right] at (0.57,2.35) {$A_3$};
\node [below, right] at (1.9,-0.3) {$A_2$};
\end{tikzpicture}
\caption{Graph representation of the two types of terms contributing to the $I_3$ at long distances. Both contributions are represented via totally connected diagrams, as expected from clustering.\label{I3-space-rep}}
\end{figure}
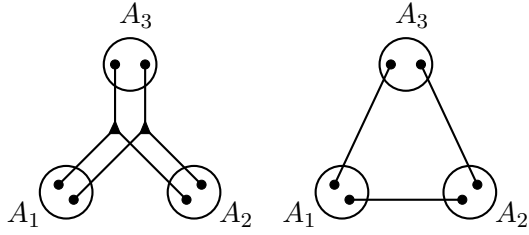There, simple lines represent correlators, while lines joint by a triangle vertex represents three-point functions. Thus, diagrammatically, the tripartite information is determined by two contributions: one made out of products of two-point functions connecting the three regions and another one which is the square of a three-point function.  

The final formula for the tripartite information can be written in terms of the relative separations of the spheres using \req{2pf} and the three-point function formula
\begin{equation}
\langle \cO(x_{A_1}) \cO(x_{A_2}) \cO(x_{A_3}) \rangle = \frac{C_{\cO\cO\cO}}{r_{12}^{\Delta}r_{23}^{\Delta}r_{13}^{\Delta}}\, ,
\end{equation}
as in \req{tripi}.

\subsubsection{Explicit formulas for $I_4$}
\begin{figure}[h!]
\centering
\begin{tikzpicture}
 \draw [thick] (0,0) to [out=20, in=160] (2,0);
  \draw [thick] (0,0) to [out=50, in=130] (2,0);
  \draw [thick] (0,0) to [out=-20, in=-160] (2,0);
   \draw [thick] (0,0) to [out=-50, in=-130] (2,0);
\draw [fill] (0,0) circle[radius=0.06]; 
\draw [fill] (2,0) circle[radius=0.06]; 
\node [below, left] at (-0.05,-0.2) {$i$};
\node [below, right] at (2.05,-0.2) {$j$};
\end{tikzpicture}
\caption{Graph representation of the contribution of the correlator of four twist operators to $I_4$ when the twist operators have non-trivial support on only two sheets. In the above graph vertices represent sheets while lines represents operators on particular regions.  Observe that the number of lines ending at a vertex gives the order of the associated correlator. }
\label{Figure-C(4)}
\end{figure}
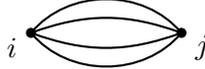

Let us now move to the $N=4$ case. We have
\bea\label{I4}
I_4(A_1,A_2,A_3,A_4)=\lim_{n\to 1}\frac{1}{n-1}\langle \tilde{\Sigma}_{{A_1}}^{(n)}\tilde{\Sigma}_{{A_2}}^{(n)}\tilde{\Sigma}_{{A_3}}^{(n)}\tilde{\Sigma}_{{A_4}}^{(n)}\rangle\,.
\eea
Here again the leading contribution to $I_4$ comes from replacing the twist operators by a  product of two operators of lowest scaling dimension as in (\ref{Sigmatilde}), which leads to the following expression for the correlator in (\ref{I4})
\bea\label{correlator-I4}
\langle \tilde{\Sigma}_{1}^{(n)}\tilde{\Sigma}_{2}^{(n)}\tilde{\Sigma}_{3}^{(n)}\tilde{\Sigma}_{4}^{(n)}\rangle\,=\(R_1R_2R_3 R_4\)^{2\Delta}\sum_{i<j}\sum_{k<l}\sum_{m<n}\sum_{p<q}C_{ij}C_{kl}C_{mn}C_{pq}\langle \cO^i_1\cO^j_1 \cO^k_2\cO^l_2   \cO^m_3\cO^n_3 \cO^p_4\cO^q_4\rangle  \, ,
\eea
where throughout the subsection we use the notation $\tilde{\Sigma}_{{A_i}}\equiv \tilde{\Sigma}_{{i}}$ and $\cO_i\equiv \cO_{A_i}$.
Organizing the contributions to \req{correlator-I4} in terms of the number ${\cal N}$ of sheets with non-trivial operator insertions, one finds that for ${\cal N}=2$ there is one only  possible configuration, which has the following structure
\begin{eqnarray}\label{conf1}
\text{${\cal N}=2,$} \quad 
\left(\begin{array}{ccccc}
1 &\cdots \cO^i_1& \cdots \cO^j_1 & \cdots 1 \\
1 &\cdots \cO^i_2 & \cdots \cO^j_2 &\cdots 1 \\
1 &\cdots \cO^i_3& \cdots \cO^j_3  &\cdots 1 \\
1 &\cdots \cO^i_4 & \cdots \cO^j_4 &\cdots 1 \\
\end{array}
\right)\, .
\end{eqnarray}
\,The contribution of such term to \req{correlator-I4} is 
\bea\label{C(4:4)}
\sum_{i<j}C^4_{ij}\langle \cO_1 \cO_2\cO_3 \cO_4\rangle^2 \quad{\rm thus}\quad I_4\supset c_{4:2}^{(4)} \langle \cO_1 \cO_2\cO_3 \cO_4\rangle^2 R^{8\Delta} \, ,
\eea
where
\bea
c_{4:2}^{(4)}\equiv \lim_{n\to 1}\frac{1}{n-1}\sum_{j<k}C^4_{ij}\,,
\eea
and we can associate to it the graph presented in Fig.\,\ref{Figure-C(4)}.

The next configurations we study are the ones that are defined on three different sheets, \ie ${\cal N}=3$. Let us illustrate one of such configurations in matrix form,
\begin{eqnarray}\label{conf2-I4}
\text{${\cal N}=3,$} \quad 
\left(\begin{array}{cccccc}
1 &\cdots \cO^i_1 & \cdots 1 &\cdots \cO^k_1 & \cdots 1  \\
1 &\cdots 1 & \cdots  \cO^j_2 &\cdots \cO^k_2& \cdots 1\\
1 &\cdots \cO^i_3&\cdots  \cO^j_3 & \cdots 1 & \cdots  1 \\
1 &\cdots \cO^i_4&\cdots \cO^j_4 & \cdots 1 & \cdots 1  \\
\end{array}
\right)\, ,
\end{eqnarray}
which corresponds to the left graph in Fig.\,\ref{Figure-C(1,1,2)}. The other two graphs in that figure have similar matrix representations obtained from the above by permutation of columns. In the graphs of Fig.\,\ref{Figure-C(1,1,2)}, the indices $\{i,j,k\}$ have been ordered and, therefore, they represent different contributions to $I_4$. 
 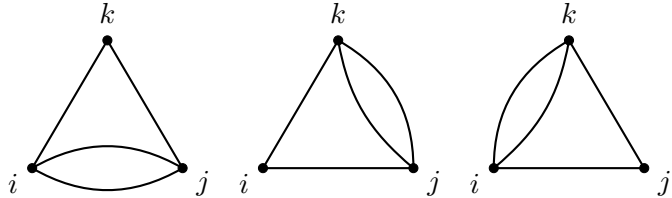
\begin{figure}[h!]
\centering
\begin{tikzpicture}
\draw [thick] (0,0) --(1,1.7);
 \draw [thick] (1,1.7)-- (2,0); 
 \draw [thick] (0,0) to [out=30, in=150] (2,0);
  \draw [thick] (0,0) to [out=-30, in=-150] (2,0);
\draw [fill] (0,0) circle[radius=0.06]; 
\draw [fill] (1,1.7) circle[radius=0.06];
\draw [fill] (2,0) circle[radius=0.06]; 
\node [below, left] at (-0.05,-0.2) {$i$};
\node [below, right] at (2.05,-0.2) {$j$};
\node [above] at (1,1.8) {$k$};
\end{tikzpicture}
\begin{tikzpicture}
\draw [thick] (0,0) --(1,1.7);
\draw [thick] (2,0) to [out=90, in=-30] (1,1.7);
\draw [thick] (2,0) to [out=140, in=-80] (1,1.7);
 \draw [thick] (2,0) --(0,0);
\draw [fill] (0,0) circle[radius=0.06]; 
\draw [fill] (1,1.7) circle[radius=0.06];
\draw [fill] (2,0) circle[radius=0.06]; 
\node [below, left] at (-0.05,-0.2) {$i$};
\node [below, right] at (2.05,-0.2) {$j$};
\node [above] at (1,1.8) {$k$};
\end{tikzpicture}
\begin{tikzpicture}
\draw [thick] (0,0) to [out=90, in=210] (1,1.7);
\draw [thick] (0,0) to [out=40, in=260] (1,1.7);
 \draw [thick] (1,1.7)-- (2,0); 
 \draw [thick] (2,0) --(0,0);
\draw [fill] (0,0) circle[radius=0.06]; 
\draw [fill] (1,1.7) circle[radius=0.06];
\draw [fill] (2,0) circle[radius=0.06]; 
\node [below, left] at (-0.05,-0.2) {$i$};
\node [below, right] at (2.05,-0.2) {$j$};
\node [above] at (1,1.8) {$k$};
\end{tikzpicture}
\caption{Graph representation of one class of contributions of the correlator of four twist operators to $I_4$ when the twist operators have non-trivial support on three sheets. In the above graph, vertices represent sheets while lines represents operators on particular regions.  
 }
\label{Figure-C(1,1,2)}
\end{figure}

\noindent
However, all of them will be proportional to the product of the following correlators
\bea \label{I4-term-1}
\(\langle \cO_1 \cO_3 \cO_4 \rangle\langle \cO_1 \cO_2  \rangle\langle \cO_2 \cO_3 \cO_4 \rangle +{\rm permutations \,\, of }\,\, \{1,2,3,4\}/ {\rm \sim equivalence} \)\, ,
\eea
where we include only the permutations that lead to independent configurations. For instance, in the matrix representation of \req{conf2-I4} the interchange $3\leftrightarrow 4$ leads to the same configuration and thus, it should not be included into the permutations of \req{I4-term-1}. Thus, the number of terms in \req{I4-term-1} equals the different orderings of $4$ regions where two of these are indistinguishable. This corresponds to ${4!}/{2!}=12$  configurations. 
The coefficient can be obtained by summing over the contributions from all the graphs presented in Fig.\,\ref{Figure-C(1,1,2)}, which can be grouped together into the single coefficient
\bea
c_{4:3}^{(1,1,2)}\equiv \lim_{n\to 1}\frac{1}{n-1}\sum_{i<j<k}\[C^2_{ij}C_{jk}C_{ki}+C_{ij}C^2_{jk}C_{ki}+C_{ij}C_{jk}C^2_{ki}\] \,.
\eea
In the above formula each term represents the coefficient associated to the contribution of each graph from left to right. 
Thus, the contribution to $I_4$ coming from those graphs is given by 
\bea\label{C:43:112}
I_4\supset C_{4:3}^{(1,1,2)}\(\langle \cO_1 \cO_3 \cO_4 \rangle\langle \cO_1 \cO_2  \rangle\langle \cO_2 \cO_3 \cO_4 \rangle +{\rm perms. \,\, of }\,\, \{1,2,3,4\}/ {\rm \sim equivalence} \)\,R^{8\Delta} \,.
\eea
For ${\cal N}=3$, there is another set of configurations contributing to $I_4$ which are represented graphically in Fig.\,\ref{Figure-C(2,2:3)}.
\begin{figure}[h!]
\centering
\begin{tikzpicture}
\draw [thick] (0,0) to [out=90, in=210] (1,1.7);
\draw [thick] (0,0) to [out=40, in=260] (1,1.7);
 \draw [thick] (0,0) to [out=30, in=150] (2,0);
  \draw [thick] (0,0) to [out=-30, in=-150] (2,0);
\draw [fill] (0,0) circle[radius=0.06]; 
\draw [fill] (1,1.7) circle[radius=0.06];
\draw [fill] (2,0) circle[radius=0.06]; 
\node [below, left] at (-0.05,-0.2) {$i$};
\node [below, right] at (2.05,-0.2) {$j$};
\node [above] at (1,1.8) {$k$};
\end{tikzpicture}
\begin{tikzpicture}
 \draw [thick] (0,0) to [out=30, in=150] (2,0);
  \draw [thick] (0,0) to [out=-30, in=-150] (2,0);
\draw [thick] (2,0) to [out=90, in=-30] (1,1.7);
\draw [thick] (2,0) to [out=140, in=-80] (1,1.7);
\draw [fill] (0,0) circle[radius=0.06]; 
\draw [fill] (1,1.7) circle[radius=0.06];
\draw [fill] (2,0) circle[radius=0.06]; 
\node [below, left] at (-0.05,-0.2) {$i$};
\node [below, right] at (2.05,-0.2) {$j$};
\node [above] at (1,1.8) {$k$};
\end{tikzpicture}
\begin{tikzpicture}
\draw [thick] (0,0) to [out=90, in=210] (1,1.7);
\draw [thick] (0,0) to [out=40, in=260] (1,1.7);
\draw [thick] (2,0) to [out=90, in=-30] (1,1.7);
\draw [thick] (2,0) to [out=140, in=-80] (1,1.7);
\draw [fill] (0,0) circle[radius=0.06]; 
\draw [fill] (1,1.7) circle[radius=0.06];
\draw [fill] (2,0) circle[radius=0.06]; 
\node [below, left] at (-0.05,-0.2) {$i$};
\node [below, right] at (2.05,-0.2) {$j$};
\node [above] at (1,1.8) {$k$};
\end{tikzpicture}
\caption{Graph representation of another class of contributions of the correlator of four twist operators to $I_4$ when the twist operators have non-trivial support on three sheets.}
\label{Figure-C(2,2:3)}
\end{figure}
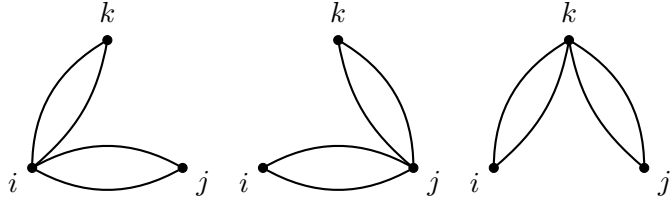

\noindent
The contributions of each of these graphs are proportional to the following correlators 
\bea
\Big(\langle \cO_1 \cO_2  \rangle \langle \cO_1 \cO_2 \cO_3 \cO_4  \rangle\langle \cO_3 \cO_4  \rangle \,\,{\rm + \,\, permutations\,\, of }\,\, \{1,2,3,4\} / {\rm \sim equivalence} \Big)\, ,
\eea
where the number of terms is given by the number of different ways of labelling each of these graphs with four letters taking into account that lines connecting the same pair of sheets are indistinguishable. Thus we have ${4!}/{(2!2!)}=6$ configurations. 
The coefficients associated to each graph can be grouped together into the following coefficient 
\bea\label{C4:3:2,2}
c_{4:3}^{(2,2)}\equiv \lim_{n\to 1}\frac{1}{n-1}\sum_{i<j<k}\[C^2_{ij}C^2_{ik}+C^2_{ij}C^2_{jk}+C^2_{ik}C^2_{jk}\]\,,
\eea
where each term corresponds with the contribution of one of the graphs in Fig.\,\ref{Figure-C(2,2:3)}, from left to right. Thus, we find the total contribution to $I_4$ of these terms is given by 
\bea\label{C43:22}
I_4\supset C_{4:3}^{(2,2)}\Big(\langle \cO_1 \cO_2  \rangle \langle \cO_1 \cO_2 \cO_3 \cO_4  \rangle\langle \cO_3 \cO_4  \rangle \,\,{\rm + \,\, permutations\,\, of }\,\, \{1,2,3,4\} / {\rm \sim equivalence} \Big) R^{8\Delta} \, .
\eea

For ${\cal N}=4$, there are two different configurations. The first one is of the form 
\begin{eqnarray}\label{conf3}
\text{${\cal N}=4$ } \quad 
\left(\begin{array}{cccccc}
1 &\cdots \cO^i_1 & \cdots \cO^j_1 & \cdots 1 &\cdots 1 &\cdots 1 \\
1 &\cdots 1 & \cdots \cO^j_2& \cdots  \cO^k_2 &\cdots 1 &\cdots 1\\
1 &\cdots 1& \cdots 1 & \cdots  \cO^k_3 &\cdots \cO^l_3 &\cdots 1\\
1 &\cdots \cO^i_4& \cdots 1 & \cdots 1 &\cdots \cO^l_4 &\cdots 1 \\
\end{array}
\right)\, ,
\end{eqnarray} 
\,\\ which corresponds to the left graph in Fig.\,\ref{Figure-C(1,1,1,1)}. However, as in the previous cases, for this given ordering of indices there would be a total of three independent graphs as illustrated in that figure. 
\begin{figure}[h!]
\centering
\begin{tikzpicture}
\draw [thick] (0,0) --(0,1.7);
 \draw [thick] (0,1.7)-- (2,1.7); 
 \draw [thick] (2,0) --(0,0);
  \draw [thick] (2,1.7) --(2,0);
\draw [fill] (0,0) circle[radius=0.06]; 
\draw [fill] (0,1.7) circle[radius=0.06];
\draw [fill] (2,1.7) circle[radius=0.06];
\draw [fill] (2,0) circle[radius=0.06]; 
\node [below, left] at (-0.05,-0.2) {$i$};
\node [above] at (-0.24,1.6) {$l$};
\node [below, right] at (2.05,1.9) {$k$};
\node [below, right] at (2.05,-0.2) {$j$};
\end{tikzpicture}
\begin{tikzpicture}
\draw [thick] (0,0) --(0,1.7);
 \draw [thick] (0,1.7)-- (2,0); 
 \draw [thick] (2,1.7) --(0,0);
  \draw [thick] (2,1.7) --(2,0);
\draw [fill] (0,0) circle[radius=0.06]; 
\draw [fill] (0,1.7) circle[radius=0.06];
\draw [fill] (2,1.7) circle[radius=0.06];
\draw [fill] (2,0) circle[radius=0.06]; 
\node [below, left] at (-0.05,-0.2) {$i$};
\node [above] at (-0.24,1.6) {$l$};
\node [below, right] at (2.05,1.9) {$k$};
\node [below, right] at (2.05,-0.2) {$j$};
\end{tikzpicture}\begin{tikzpicture}
 \draw [thick] (0,1.7)-- (2,1.7); 
 \draw [thick] (2,0) --(0,0);
   \draw [thick] (0,1.7)-- (2,0); 
 \draw [thick] (2,1.7) --(0,0);
\draw [fill] (0,0) circle[radius=0.06]; 
\draw [fill] (0,1.7) circle[radius=0.06];
\draw [fill] (2,1.7) circle[radius=0.06];
\draw [fill] (2,0) circle[radius=0.06]; 
\node [below, left] at (-0.05,-0.2) {$i$};
\node [above] at (-0.24,1.6) {$l$};
\node [below, right] at (2.05,1.9) {$k$};
\node [below, right] at (2.05,-0.2) {$j$};
\end{tikzpicture}
\caption{Graph representation of one class of contributions of the correlator of four twist operators to $I_4$ when the twist operators have non-trivial support on four sheets.}
\label{Figure-C(1,1,1,1)}
\end{figure}
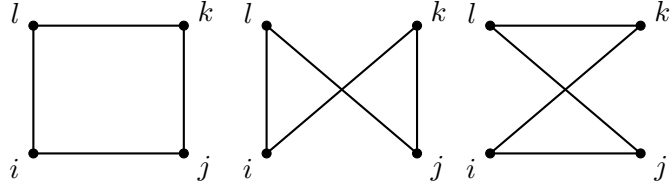
All these graphs give contributions proportional to the following correlators
\bea
\Big(\langle \cO_1 \cO_2 \rangle \langle \cO_2 \cO_3  \rangle\langle \cO_3 \cO_4 \rangle \langle \cO_1 \cO_4 \rangle\,\, +{\rm permutations\, of }\,\, \{1,2,3,4\}\Big)\, ,
\eea
where in this case, every labelling of the lines for each graph gives rise to a non-equivalent configuration and thus there is a total of $4!=24$ configurations. The associated coefficients can then be put together in the following single coefficient 
\bea\label{C4:4:1,1,1,1}
c_{4:4}^{(1,1,1,1)}=\lim_{n\to 1}\frac{1}{n-1}\sum_{i<j<k<l}\[C_{ij}C_{jk}C_{kl}C_{li}+C_{ij}C_{jl}C_{lk}C_{ki}+C_{il}C_{lj}C_{jk}C_{ki}\]\, ,
\eea
where each term in the right hand side comes from a graph in Fig.\,\ref{Figure-C(1,1,1,1)} from left to right. Thus the total contribution to $I_4$ from the configurations described by the above graphs is 
\bea\label{C4:4:1111}
I_4\supset \,C_{4:4}^{(1,1,1,1)}\Big(\langle \cO_1 \cO_2 \rangle \langle \cO_2 \cO_3  \rangle\langle \cO_3 \cO_4 \rangle \langle \cO_1 \cO_4 \rangle\,\, +{\rm permutations\, of }\,\, \{1,2,3,4\}  \Big) R^{8\Delta}\, .
\eea
The second set of configurations for ${\cal N}=4$ are described in the graphs of Fig.\,\ref{Figure-C(2,2:4)}.
\begin{figure}[h!]
\centering
\begin{tikzpicture}
 \draw [thick] (0,1.7) to [out=30, in=150] (1.7,1.7);
  \draw [thick] (0,1.7) to [out=-30, in=-150] (1.7,1.7);
 \draw [thick] (0,0) to [out=30, in=150] (1.7,0);
  \draw [thick] (0,0) to [out=-30, in=-150] (1.7,0);
\draw [fill] (0,0) circle[radius=0.06]; 
\draw [fill] (0,1.7) circle[radius=0.06];
\draw [fill] (1.7,1.7) circle[radius=0.06];
\draw [fill] (1.7,0) circle[radius=0.06]; 
\node [below, left] at (-0.05,-0.2) {$i$};
\node [above] at (-0.24,1.6) {$l$};
\node [below, right] at (1.85,1.9) {$k$};
\node [below, right] at (1.85,-0.2) {$j$};
\end{tikzpicture}
\begin{tikzpicture}
 \draw [thick] (0,0) to [out=120, in=-120] (0,1.7);
  \draw [thick] (0,0) to [out=60, in=-60] (0,1.7);
 \draw [thick] (1.7,0) to  [out=120, in=-120]  (1.7,1.7);
  \draw [thick] (1.7,0) to  [out=60, in=-60] (1.7,1.7);
\draw [fill] (0,0) circle[radius=0.06]; 
\draw [fill] (0,1.7) circle[radius=0.06];
\draw [fill] (1.7,1.7) circle[radius=0.06];
\draw [fill] (1.7,0) circle[radius=0.06]; 
\node [below, left] at (-0.05,-0.2) {$i$};
\node [above] at (-0.24,1.6) {$l$};
\node [below, right] at (1.85,1.9) {$k$};
\node [below, right] at (1.85,-0.2) {$j$};
\end{tikzpicture}
\begin{tikzpicture}
 \draw [thick] (0,0) to [out=75, in=-165] (1.7,1.7);
  \draw [thick] (0,0) to [out=15, in=-105] (1.7,1.7);
\draw [thick] (1.7,0) to [out=105, in=-15] (0,1.7);
\draw [thick] (1.7,0) to [out=175, in=-75] (0,1.7);
\draw [fill] (0,0) circle[radius=0.06]; 
\draw [fill] (0,1.7) circle[radius=0.06];
\draw [fill] (1.7,1.7) circle[radius=0.06];
\draw [fill] (1.7,0) circle[radius=0.06]; 
\node [below, left] at (-0.05,-0.2) {$i$};
\node [above] at (-0.24,1.6) {$l$};
\node [below, right] at (1.85,1.9) {$k$};
\node [below, right] at (1.85,-0.2) {$j$};
\end{tikzpicture}
\caption{Graph representation of another class of contributions of the correlator of four twist operators to $I_4$ when the twist operators have non-trivial support on four sheets.}
\label{Figure-C(2,2:4)}
\end{figure}
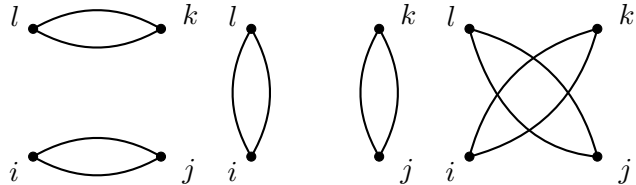

\noindent
Here each graph gives a contribution proportional to the following correlators 
\bea
\Big(\langle \cO_1 \cO_2  \rangle^2 \langle \cO_3 \cO_4  \rangle^2  \,\,{\rm + \,\, permutations\,\, of }\,\, \{1,2,3,4\} / {\rm \sim equivalence} \Big) \, ,
\eea
where the number of terms equal the number of non-equivalent ways of labelling the four lines of the associated graph. This is a total of ${4!}/{(2! 2!)}=6$ configurations.
The coefficients to which each of the above graphs contributes can be grouped together into 
\bea\label{C4:4:2,2}
c_{4:4}^{(2,2)}=\lim_{n\to 1}\frac{1}{n-1}\sum_{i<j<k<l}\[C^2_{il}C^2_{jk}+C^2_{ij}C^2_{kl}+C^2_{ik}C^2_{jl}\]\,.
\eea
This gives a total contribution to $I_4$ of the form 
\bea\label{I4-cont-4:4-2:2}
I_4\supset c_{4:4}^{(2,2)}\Big(\langle \cO_1 \cO_2  \rangle^2 \langle \cO_3 \cO_4  \rangle^2  \,\,{\rm + \,\, permutations\,\, of }\,\, \{1,2,3,4\} / {\rm \sim equivalence} \Big) R^{8\Delta}\, .
\eea

For ${\cal N}>4$ since there is a total of $8$ operators there would necessarily be a sheet with only one operator insertion and therefore it would vanish for all those configurations. Putting all the contributions from eqs. (\ref{C(4:4)}), (\ref{C:43:112}), (\ref{C43:22}), (\ref{C4:4:1111}) and (\ref{I4-cont-4:4-2:2}) together leads to the following formula for $I_4$ in the long-distance regime
\begin{align}\label{Final-I4}
\frac{I_4}{R^{8\Delta}} =& +c_{4:2}^{(4)}\,  \langle \cO_1 \cO_2\cO_3 \cO_4\rangle^2 \nonumber \\ 
&+ \,c_{4:3}^{(1,1,2)}\, \Big(\langle \cO_1 \cO_2 \cO_3 \rangle\langle \cO_3 \cO_4  \rangle\langle \cO_1 \cO_2 \cO_4 \rangle +{\rm permutations \,\, of }\,\, \{1,2,3,4\}/ {\rm \sim equivalence} \Big)\nonumber\\
& +\,c_{4:3}^{(2,2)} \Big(\langle \cO_1 \cO_2  \rangle \langle \cO_1 \cO_2 \cO_3 \cO_4  \rangle\langle \cO_3 \cO_4  \rangle \,\,{\rm + \,\, permutations\,\, of }\,\, \{1,2,3,4\} / {\rm \sim equivalence} \Big) \nonumber\\
& +\,c_{4:4}^{(1,1,1,1)}\Big(\langle \cO_1 \cO_2 \rangle \langle \cO_2 \cO_3  \rangle\langle \cO_3 \cO_4 \rangle \langle \cO_1 \cO_4 \rangle\,\, +{\rm permutations\, of }\,\, \{1,2,3,4\}  \Big) \nonumber\\
& +\,c_{4:4}^{(2,2)}\Big(\langle \cO_1 \cO_2  \rangle^2 \langle \cO_3 \cO_4  \rangle^2  \,\,{\rm + \,\, permutations\,\, of }\,\, \{1,2,3,4\} / {\rm \sim equivalence} \Big)\,.
\end{align}
After a close inspection one realizes that the terms appearing in the last line should be absent from a four-partite information, since those will not have the required scaling behaviour as one separates regions. In other words, they represent disconnected contributions. However, similar disconnected terms will also come from certain contributions to the four-point functions appearing on the other terms. We should be able to write down the final answer in terms of connected contributions only.\footnote{We thank Horacio Casini for pointing this out to us.} We do so next
\begin{align}
\frac{I_4}{R^{8\Delta}}=& +c_{4:2}^{(4)}\, \Big(\langle \cO_1 \cO_2\cO_3 \cO_4\rangle^2 -\langle \cO_1 \cO_2 \rangle^2 \langle \cO_3 \cO_4 \rangle^2-\langle \cO_1 \cO_3 \rangle^2 \langle \cO_2 \cO_4 \rangle^2-\langle \cO_1 \cO_4 \rangle^2 \langle \cO_2 \cO_3 \rangle^2 \Big)  \nonumber \\
&+2\, c_{4:3}^{(1,1,2)}\, \Big(\langle \cO_1 \cO_3 \cO_4 \rangle\langle \cO_1 \cO_2  \rangle \langle \cO_2 \cO_3 \cO_4 \rangle +\langle \cO_1 \cO_2 \cO_4 \rangle\langle \cO_1 \cO_3  \rangle\langle \cO_2 \cO_3 \cO_4 \rangle\nonumber\\
& \qquad \qquad + \langle \cO_1 \cO_2 \cO_3 \rangle\langle \cO_1 \cO_4  \rangle\langle \cO_2 \cO_3 \cO_4 \rangle +\langle \cO_1 \cO_2 \cO_4 \rangle\langle \cO_2 \cO_3  \rangle\langle \cO_1 \cO_3 \cO_4 \rangle \, \nonumber\\
& \qquad  \qquad + \langle \cO_1 \cO_2 \cO_3 \rangle\langle \cO_2 \cO_4  \rangle\langle \cO_1 \cO_3 \cO_4 \rangle +\langle \cO_1 \cO_2 \cO_3 \rangle\langle \cO_3 \cO_4  \rangle\langle \cO_1 \cO_2 \cO_4 \rangle \, \Big) \nonumber\\
& +\,2\,c_{4:3}^{(2,2)} \Big[ \Big(\langle \cO_1 \cO_2 \cO_3 \cO_4  \rangle-\langle \cO_1 \cO_2  \rangle \langle \cO_3 \cO_4  \rangle \Big)\langle \cO_1 \cO_2  \rangle \langle \cO_3 \cO_4  \rangle \, \nonumber\\
& \qquad \qquad  +\Big(\langle \cO_1 \cO_2 \cO_3 \cO_4  \rangle-\langle \cO_1 \cO_3 \rangle \langle \cO_2 \cO_4 \rangle \Big)\langle \cO_1 \cO_3 \rangle \langle \cO_2 \cO_4 \rangle \,\nonumber\\
& \qquad \qquad  + \Big(\langle \cO_1 \cO_2 \cO_3 \cO_4  \rangle-\langle \cO_1 \cO_4 \rangle \langle \cO_2 \cO_3 \rangle  \Big)\langle \cO_1 \cO_4 \rangle \langle \cO_2 \cO_3 \rangle  \Big]\, \nonumber\\
& +\,8\,c_{4:4}^{(1,1,1,1)}\Big(\langle \cO_1 \cO_2 \rangle \langle \cO_2 \cO_3  \rangle\langle \cO_3 \cO_4 \rangle \langle \cO_1 \cO_4 \rangle\,\, +\langle \cO_1 \cO_3 \rangle \langle \cO_2 \cO_3  \rangle\langle \cO_2 \cO_4 \rangle \langle \cO_1 \cO_4\rangle\, \nonumber \\
& \qquad \qquad  \quad + \langle \cO_1 \cO_2 \rangle \langle \cO_1 \cO_3  \rangle\langle \cO_2 \cO_4 \rangle \langle \cO_3 \cO_4 \rangle\, \Big) \nonumber\\
&+2\[ \,c_{4:4}^{(2,2)}+c_{4:3}^{(2,2)}+\frac{1}{2}c^{(4)}_{4:2}\,\,  \]\Big(\langle \cO_1 \cO_2 \rangle^2 \langle \cO_3 \cO_4 \rangle^2+\langle \cO_1 \cO_3 \rangle^2 \langle \cO_2 \cO_4 \rangle^2\nonumber \\ \label{pifoo}
& \qquad\qquad \qquad \qquad \qquad  \qquad  +\langle \cO_1 \cO_4 \rangle^2 \langle \cO_2 \cO_3 \rangle^2 \Big)\,,
\end{align}
where we have taken into account the number of terms associated to each coefficient and written them explicitly. Thus, consistency with the long-distance expected behavior (expectation from clustering principle) implies that the last combination of coefficients should exactly cancel. That is precisely what happens.

All the coefficients appearing in the above formula are computed in Appendix \ref{app-coeffs}, where we find  
\bea\label{coeffs-I4}
&&c_{4:2}^{(4)}=\frac{2^{8\Delta}\[\Gamma\(4\Delta+1\)\]^2}{2\,  \Gamma\(8\Delta+2\)}\,, \qquad c_{4:3}^{(1,1,2)}=-\frac{9\sqrt{\pi} \Gamma\(3\Delta\)^2}{4\Gamma\(\Delta\)^2 \Gamma\(4\Delta+\frac32\)}\,, \nonumber\\
&&c_{4:3}^{(2,2)}=-\frac{2^{8\Delta}\[\Gamma\(4\Delta+1\)\]^2}{2\,  \Gamma\(8\Delta+2\)}\,, \quad{\rm and}\quad
c_{4:4}^{(2,2)}=\frac{2^{8\Delta}\[\Gamma\(4\Delta+1\)\]^2}{4\,  \Gamma\(8\Delta+2\)}\,.
\eea
from eqs. (\ref{C42:4}), (\ref{C4:3(112)}), (\ref{C4:3(22)}), and (\ref{C4:4(22)}) respectively. Unfortunately, we were not able to find a completely explicit formula for $c_{4:4}^{(1,1,1,1)}$. The expression for this coefficient appears  in \req{c44:Splitting}, whose numerical evaluation for a few particular values of $\Delta$ (analytically for $\Delta=1/2$ and possibly also for $\Delta=1$) we present in table \ref{tablaC44}. As anticipated, we find that the last combination of coefficients in \req{pifoo} vanishes exactly, which represents a non-trivial consistency check of our calculations. 

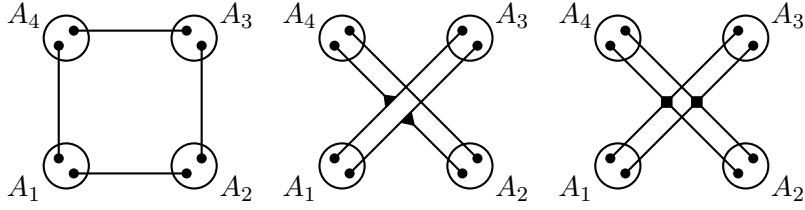
\begin{figure}[t]
\centering
\begin{tikzpicture}
 \draw [thick] (-0.1,0.1) to (-0.1,1.6);
  \draw [thick] (0.1,-0.1)  to (1.6,-0.1);
 \draw [thick] (1.8,0.1) to  (1.8,1.6) ;
  \draw [thick] (0.1,1.8) to  (1.6,1.8);
\draw [thick] (0,0) circle[radius=0.3]; 
\draw [fill] (-0.1,0.1) circle[radius=0.06]; 
\draw [fill] (0.1,-0.1) circle[radius=0.06]; 
\draw [thick] (0,1.7) circle[radius=0.3];
\draw [fill] (-0.1,1.6) circle[radius=0.06]; 
\draw [fill] (0.1,1.8) circle[radius=0.06]; 
\draw [thick] (1.7,1.7) circle[radius=0.3];
\draw [fill] (1.6,1.8) circle[radius=0.06]; 
\draw [fill] (1.8,1.6) circle[radius=0.06]; 
\draw [thick] (1.7,0) circle[radius=0.3]; 
\draw [fill] (1.6,-0.1) circle[radius=0.06]; 
\draw [fill] (1.8,0.1) circle[radius=0.06]; 

\node [below, left] at (-0.2,-0.3) {$A_1$};
\node [above,left] at (-0.2,2.0) {$A_4$};
\node [below, right] at (1.9,2.0) {$A_3$};
\node [below, right] at (1.9,-0.3) {$A_2$};
\end{tikzpicture}
\begin{tikzpicture}
  \draw [thick]  (-0.1,0.1) to  (1.6,1.8);
     \draw [thick]  (0.1,-0.1) to  (1.8,1.6);
  \draw [thick] (0.1,1.8) to (1.8,0.1);
    \draw [thick] (-0.1,1.6) to (0.65,0.85);
       \draw [thick] (0.85,0.65)  to (1.6,-0.1);
\draw [thick] (0,0) circle[radius=0.3]; 
\draw [fill] (-0.1,0.1) circle[radius=0.06]; 
\draw [fill] (0.1,-0.1) circle[radius=0.06]; 
\draw [thick] (0,1.7) circle[radius=0.3];
\draw [fill] (-0.1,1.6) circle[radius=0.06]; 
\draw [fill] (0.1,1.8) circle[radius=0.06]; 
\draw [thick] (1.7,1.7) circle[radius=0.3];
\draw [fill] (1.6,1.8) circle[radius=0.06]; 
\draw [fill] (1.8,1.6) circle[radius=0.06]; 
\draw [thick] (1.7,0) circle[radius=0.3]; 
\draw [fill] (1.6,-0.1) circle[radius=0.06]; 
\draw [fill] (1.8,0.1) circle[radius=0.06]; 
 
  \draw [fill] (0.58,0.78)--(0.72,0.92)--(0.56,0.94)--(0.58,0.78); 
   \draw [fill] (0.78,0.58)--(0.92,0.72)--(0.96,0.54)--(0.78,0.58); 

\node [below, left] at (-0.2,-0.3) {$A_1$};
\node [above,left] at (-0.2,2.0) {$A_4$};
\node [below, right] at (1.9,2.0) {$A_3$};
\node [below, right] at (1.9,-0.3) {$A_2$};
\end{tikzpicture}
\begin{tikzpicture}
  \draw [thick]  (-0.1,0.1) to  (1.6,1.8);
     \draw [thick]  (0.1,-0.1) to  (1.8,1.6);
  \draw [thick] (0.1,1.8) to (1.8,0.1);
   \draw [thick] (-0.1,1.6) to (1.6,-0.1);
\draw [thick] (0,0) circle[radius=0.3]; 
\draw [fill] (-0.1,0.1) circle[radius=0.06]; 
\draw [fill] (0.1,-0.1) circle[radius=0.06]; 
\draw [thick] (0,1.7) circle[radius=0.3];
\draw [fill] (-0.1,1.6) circle[radius=0.06]; 
\draw [fill] (0.1,1.8) circle[radius=0.06]; 
\draw [thick] (1.7,1.7) circle[radius=0.3];
\draw [fill] (1.6,1.8) circle[radius=0.06]; 
\draw [fill] (1.8,1.6) circle[radius=0.06]; 
\draw [thick] (1.7,0) circle[radius=0.3]; 
\draw [fill] (1.6,-0.1) circle[radius=0.06]; 
\draw [fill] (1.8,0.1) circle[radius=0.06]; 

 \draw [fill] (0.58,0.78) rectangle (0.72, 0.92); 
 \draw [fill] (0.98,0.78) rectangle (1.12,0.92); 

\node [below, left] at (-0.2,-0.3) {$A_1$};
\node [above,left] at (-0.2,2.0) {$A_4$};
\node [below, right] at (1.9,2.0) {$A_3$};
\node [below, right] at (1.9,-0.3) {$A_2$};
\end{tikzpicture}\caption{We provide a graph representation of the various terms contributing to $I_4$. Single lines represent two-point correlators, lines converging at a triangle vertex represent three-point correlators while lines converging at a square vertex represents connected four-point correlators. Observe that each diagram gives a completely connected contribution to $I_4$. \label{I4-space-rep}}
\end{figure}

Additionally, the relation between $c_{4:2}^{(4)}$ and $c_{4:3}^{(2,2)}$ from \req{coeffs-I4} suggests that the $I_4$ can be written in a more compact form in terms of the square of the connected four-point function, namely,
\begin{align}\label{I4-final}
\frac{I_4}{R^{8\Delta}}=& +\Big[\langle \cO_1 \cO_2\cO_3 \cO_4\rangle -\langle \cO_1 \cO_2 \rangle \langle \cO_3 \cO_4 \rangle-\langle \cO_1 \cO_3 \rangle \langle \cO_2 \cO_4 \rangle-\langle \cO_1 \cO_4 \rangle \langle \cO_2 \cO_3 \rangle \Big]^2 \left[\frac{2^{8\Delta}\[\Gamma\(4\Delta+1\)\]^2}{2\,  \Gamma\(8\Delta+2\)}\right] \ \nonumber \\
&- \Big[\langle \cO_1 \cO_3 \cO_4 \rangle\langle \cO_1 \cO_2  \rangle \langle \cO_2 \cO_3 \cO_4 \rangle +\langle \cO_1 \cO_2 \cO_4 \rangle\langle \cO_1 \cO_3  \rangle\langle \cO_2 \cO_3 \cO_4 \rangle\nonumber\\
& \qquad  \quad + \langle \cO_1 \cO_2 \cO_3 \rangle\langle \cO_1 \cO_4  \rangle\langle \cO_2 \cO_3 \cO_4 \rangle +\langle \cO_1 \cO_2 \cO_4 \rangle\langle \cO_2 \cO_3  \rangle\langle \cO_1 \cO_3 \cO_4 \rangle \, \nonumber\\
& \qquad  \quad + \langle \cO_1 \cO_2 \cO_3 \rangle\langle \cO_2 \cO_4  \rangle\langle \cO_1 \cO_3 \cO_4 \rangle +\langle \cO_1 \cO_2 \cO_3 \rangle\langle \cO_3 \cO_4  \rangle\langle \cO_1 \cO_2 \cO_4 \rangle \, \Big] \left[\frac{9\sqrt{\pi} \Gamma\(3\Delta\)^2}{2\Gamma\(\Delta\)^2 \Gamma\(4\Delta+\frac32\)}\right] \nonumber\\
&+\Big[\langle \cO_1 \cO_2 \rangle \langle \cO_2 \cO_3  \rangle\langle \cO_3 \cO_4 \rangle \langle \cO_1 \cO_4 \rangle +\langle \cO_1 \cO_3 \rangle \langle \cO_2 \cO_3  \rangle\langle \cO_2 \cO_4 \rangle \langle \cO_1 \cO_4\rangle\, \,\, \nonumber\\
& \qquad \quad +  \langle \cO_1 \cO_2 \rangle \langle \cO_1 \cO_3  \rangle\langle \cO_2 \cO_4 \rangle \langle \cO_3 \cO_4 \rangle\, \Big] \[\,8c_{4:4}^{(1,1,1,1)}-\frac{2^{8\Delta}\[\Gamma\(4\Delta+1\)\]^2}{  \Gamma\(8\Delta+2\)}\] \, .
\end{align}
 Using the explicit form of the two- and three-point functions, the final expression for the $I_4$ can be written as in \req{I4-final0}. Finally, observe that this final formula can be expressed via a space representation of the correlators contributing to it as shown in Fig.\,\ref{I4-space-rep}.

In view of \req{I4-final} and the results previously obtained for $N=2,3$, it is clear that the leading long-distance term of the $N$-partite information will be given by some linear combination of products of $2$-, $3$-, $\dots$, $(N-1)$-, $N$-point functions of the smallest-conformal-dimension primary of the corresponding theory. While such information should also be available (at least in principle) within the mutual information expansion, the results presented here support the idea of using the $N$-partite information expansion as a complementary approach. Given a CFT for which the input information are all the $N$-partite informations corresponding to ball regions in the long-distance regime, one could systematically identify the scaling dimension of the leading primary as well as all its $N$-point correlators as follows: first, $\Delta$ would be extracted from the mutual information expansion in \req{i2l}; then, \req{tripi} would be used to extract the three-point coefficient;  \req{I4-final0} would be used to extract the four-point function; analogous formulas would yield the higher $N$-point functions. Information about subleading primaries will start appearing at the next order in the long-distance expansions. Determining the precise way in which this will happen and how one should systematically proceed in order to extract the CFT data involving such operators requires further work. However, we know that an essential part in such process will require to upgrade the expansion in terms of local replica operators (\ref{OPE-primary-expansion}) to an expansion in terms of ``OPE blocks'' associated to each replica primary \cite{Long:2016vkg} ---remember that each ``OPE block'' includes the resumed contribution of a given primary as well as their associated descendants \cite{Czech:2016xec}. Doing so would reduce the type of operators included in the $N$-partite information expansion to the replica primaries. For a recent discussion on the  ``OPE block'' expansion applied to the tripartite information see \cite{Agon:2021lus}.


\section{Free scalar CFT\label{Free-CFT}}

In this section we obtain even more explicit formulas for the $N$-partite information (for $N=2,3,4$) in the particular case of a CFT which has a free scalar as its lowest-dimensional operator. We particularize our formulas to three dimensions, obtaining analytic results which in section \ref{sec:Lattice}  we compare with  lattice calculations. We close with some conjectures regarding the sign of $I_N$ for free-scalar CFTs  in general dimensions.

Let us consider then a CFT which has a free scalar as its lowest-dimensional primary operator. This includes, obviously, free scalar theories, but also theories which have a free sector.
Various simplifications occur in that case. First, the conformal dimension of a free scalar field is given by 
\bea
\Delta_{\rm free\,scalar}=\frac{d-2}2\, ,
\eea
and so it is a half-integer. Second, correlators including an odd number of operators vanish. Finally, correlators involving an even number of fields are completely determined by two-point correlators by virtue of Wick's theorem. For instance, the four-point function becomes
 \bea\label{4corr-wicks}
\langle \cO_1 \cO_2  \cO_3 \cO_4 \rangle=\langle \cO_1 \cO_2 \rangle \langle \cO_3 \cO_4 \rangle+\langle \cO_1 \cO_3 \rangle \langle \cO_2 \cO_4 \rangle+\langle \cO_1 \cO_4 \rangle \langle \cO_2 \cO_3 \rangle\,,
\eea
and similarly for higher-point correlators. As a consequence, the long-distance $N$-partite information of a free theory must be given by a single term of the form
\bea\label{IN-CN}
I^{\rm free\,scalar}_N(\{A_i\})=C_N \( \prod_{\alpha=1}^{N}\Big\langle\mathcal{O}_{A_{\alpha}}\mathcal{O}_{A_{\alpha+1}}\Big \rangle +\, {\rm permutations\,\, of\,}\, \{A_1, \cdots ,A_N\} \)R^{N(d-2)}\,,
\eea
where $A_{N+1}\equiv A_1$ and $R$ is the radii of the spheres. This is due to the fact that the only completely connected $2N$-point correlators on $N$ regions  that can be constructed with two-point correlators are the ones presented above. 

\subsection{Mutual information}
In the case of the mutual information, the general formula at long distances in the case of spherical entangling surfaces was presented in \req{I2-2}. For a free scalar such formula reduces to 
\bea
I^{\rm free\,scalar}_2=\frac{ 2^{2(d-2)}\Gamma\(d-1\)^2}{2\, \Gamma\(2(d-1)\)}\frac{R^{4\Delta}}{r^{4\Delta}}\, ,
\eea
where $R$ is the radius of the balls which we assumed to be the same, and $r$ is the center-to-center distance between the regions. 

For comparison with the lattice results we consider the case of $d=3$. Also, we rewrite the result using the parameters $R_{\rm latt}\equiv 2R$, $r_{\rm latt} \equiv r$. We have
\bea
\left.I^{\rm free\,scalar}_2\right|_{d=3}=\frac{1}{48} \frac{R_{\rm latt}^2}{r_{\rm latt}^2}\approx 0.08333x^{-2}\, ,
\eea
where $x\equiv r_{\rm latt}/R_{\rm latt}$. 

\subsection{Tripartite information}
The tripartite information at long distances is given by \req{I3-final}. For a free scalar this reduces to
\bea\label{I3-free}
I^{\rm free\,scalar}_3=\frac{2^{3(d-2)} \Gamma\(\tfrac{d-1}2\)^3}{2 \pi \Gamma\( \tfrac{3(d-1)}2\)}\langle \cO_1 \cO_3 \rangle \langle \cO_1 \cO_2  \rangle\langle \cO_2 \cO_3 \rangle \, R^{3(d-2)}\, .
\eea
Now, let us consider a geometric configuration where the balls are located at the vertices of an equilateral triangle. Then, we have
\bea
\langle \cO_1 \cO_2 \rangle=\langle \cO_2 \cO_3  \rangle=\langle \cO_1 \cO_3 \rangle=\frac{1}{r^{d-2}}\,,
\eea
where $r$ is the length of the side of triangle. In that case, \req{I3-free} reduces to 
\bea
I^{\rm free\,scalar}_3=\frac{2^{3(d-2)} \Gamma\(\frac{d-1}2\)^3}{2 \pi \Gamma\( \frac{3(d-1)}2\)} \frac{R^{3(d-2)}}{r^{3(d-2)}}\,,
\eea
where $R$ is the radius of the balls, which we assumed to be the same for the three.
 
 Again, for comparison with the lattice results, we consider the case of $d=3$ and use instead the parameters $R_{\rm latt}$ and $r_{\rm latt}$ which are related to the previous ones by $R_{\rm latt}=2R$ and $r_{\rm latt}=r/\sqrt{3}$. $R_{\rm latt}$ is just the diameter of each disk, while $r_{\rm latt}$ is the radius of the circumference  containing the vertices of the equilateral triangle.
 In terms of these parameters we have
\bea
\left.I^{\rm free\,scalar}_3\right|_{d=3}=\frac{1}{12\,\sqrt{3} \, \pi} \frac{R_{\rm latt}^3}{r_{\rm latt}^3}\approx 0.01531 x^{-3}\eea
where $x\equiv r_{\rm latt}/R_{\rm latt}$.

\subsection{Four partite information}
For the $I_4$, the long-distance expression for spherical entangling surfaces was given in \req{I4-final}. This expression simplifies now for two reasons. First, the three-point function of the free scalar exactly vanishes, which makes the whole second term vanish. Second, the four-point function reduces to products of two-point functions as in \req{4corr-wicks}, which makes the first term vanish as well. We are then left with 
\begin{align}\label{Final-I4-free-4}
I^{\rm free\,scalar}_4=&+\[\,8c_{4:4}^{(1,1,1,1)}\left(\frac{d-2}{2}\right)-\frac{2^{4(d-2)}\[\Gamma\(2d-3\)\]^2}{\,  \Gamma\(4d-6\)}\]\Big(\langle \cO_1 \cO_2 \rangle \langle \cO_2 \cO_3  \rangle\langle \cO_3 \cO_4 \rangle \langle \cO_1 \cO_4 \rangle+\,\, 
 \\ \notag
& +\, \langle \cO_1 \cO_3 \rangle \langle \cO_1 \cO_4  \rangle\langle \cO_2 \cO_3 \rangle \langle \cO_2 \cO_4 \rangle\,+ \langle \cO_1 \cO_2 \rangle \langle \cO_1 \cO_3  \rangle\langle \cO_2 \cO_4 \rangle \langle \cO_3 \cO_4 \rangle\, \Big) R^{4(d-2)}\,.
\end{align}
Now, let us consider a configuration in which the spherical regions are located at the corners of a square of side $r$. In that case the correlators become
\bea
\langle \cO_1 \cO_2 \rangle=\langle \cO_2 \cO_3  \rangle=\langle \cO_3 \cO_4  \rangle=\langle \cO_4 \cO_1 \rangle=\frac{1}{r}\,,
\eea 
and 
\bea
\langle \cO_1 \cO_3 \rangle=\langle \cO_2 \cO_4  \rangle=\frac{1}{\sqrt{2}r}\,.
\eea 
Plugging this into \req{Final-I4-free-4} results in 
\bea\label{Final-I4-square-1}
&&I^{\rm free\,scalar}_4=16\[\,c_{4:4}^{(1,1,1,1)}\left(\frac{d-2}{2}\right)-\frac{2^{4(d-2)}\[\Gamma\(2d-3\)\]^2}{8\,  \Gamma\(4d-6\)}\] \frac{R^{4(d-2)}}{r^{4(d-2)}}\, .
\eea
Like in the previous cases, we restrict to the case of $d=3$ and use the parameters $R_{\rm latt}=2 R $ and  $r_{\rm latt}=r/\sqrt{2}$ where $R_{\rm latt}$ is the diameter of the disks while $r_{\rm latt}$ is the radius of the circle that passes through the vertices of the square. 
Thus, in terms of these parameters we have 
\bea\label{Final-I4-square-2}
&&\left.I_4^{\rm free\,scalar}\right|_{d=3}= \left[\frac{1}{180}+\frac{1}{6\pi^2}\right] \frac{R_{\rm latt}^4}{r_{\rm latt}^4}\simeq 0.0224\frac{R_{\rm latt}^4}{r_{\rm latt}^4}\, ,
\eea
where we used the analytic result obtained for $c_{4:4}^{(1,1,1,1)}(\Delta=1/2)$ in \req{c44:FinalValueDelta12}.

Evaluating $c_{4:4}^{(1,1,1,1)}$ for other values of $\Delta$ we can determine the leading long-distance coefficient in the spheres $I_4$ for a free-scalar CFT in higher dimensions. In particular, using the results presented in table \ref{tablaC44}, we find
\begin{align}
\left.I_4^{\rm free\,scalar}\right|_{d=4}&=\frac{256}{105} \frac{R^{8}}{r^{8}}\, , \\
\left.I_4^{\rm free\,scalar}\right|_{d=5}&\simeq 5.70678  \frac{R^{12}}{r^{12}}\, , \\
\left.I_4^{\rm free\,scalar}\right|_{d=6}&\simeq 15.4904\frac{R^{16}}{r^{16}}\, .
\end{align}
As we can see, the coefficient is always positive and grows with $d$. In view of these results, it is tempting to conjecture that the $I_N$ for  $N=2,3,4$ will be positive for free-scalar CFTs in arbitrary dimensions and for general regions.
In fact, lattice results for $d=3$ and for $N=1,2,\dots , 6$ suggest the positivity of $I_N$ for the free scalar CFT for arbitrary $N$ and general regions. 
 Thus, we conjecture that the $N$-partite information for theories with a free scalar as their lowest-dimensional primary is always positive for arbitrary $N$ and $d$.  While such a claim in the $d\geq 3$ case relies exclusively on our result for the leading long-distance term in the case of spherical regions, it is natural to expect that  moving the regions closer should tend to monotonically increase the absolute value of the $N$-partite information on general grounds (and hence its overall sign would not change with respect to the long-distance situation). Similarly, it is reasonable to expect  that changing the shape of the regions will not modify the sign of the $I_N$ either. This is related to the question of whether or not the sign of $I_N$ is completely fixed for a given theory independently of the shape of the entangling regions and their relative separations. As far as we know, all available evidence supports a positive answer to this question, but we are not aware of a general proof.

\section{Lattice calculations in $d=3$}
\label{sec:Lattice}
In this section we compute the $N$-partite information (for $N=2,3,4,5,6$) for various entangling regions assembled forming regular $N$-gon configurations for a three-dimensional free scalar in the lattice. First we focus on the long-distance regime. We verify that the leading scaling  obtained for $I_N$ in this regime in Section \ref{ldb} is fulfilled in all cases and study the dependence on the shape of the entangling regions. In the case of disk regions, we obtain numerical approximations for the coefficient of the leading terms, which can be tested against analytic calculations using the results of Section \ref{ldb} for $N=2,3,4$, finding excellent agreement.  Our results suggest that $I_N$ for a three-dimensional free scalar is positive for all $N$ and for general regions and configurations.  In subsection \ref{i2arbi} we perform lattice calculations of the mutual information for pairs of regions both for a free scalar and a free fermion for general separations.  In all cases, we also present analytic results for two toy models which approximate the respective scalings of each of the free models at long separations. In the case of disk regions, we provide analytic approximations for the whole range of values both for the scalar and the fermion. We determine various scalings at small and large separations for the different regions and observe that, normalized by the respective disk entanglement entropy coefficients, the free scalar mutual information is always greater than the fermion one, a feature which we conjecture to be true for arbitrary regions.

\subsection{Lattice setup and models}\label{latticelong}
In the following subsection we perform lattice calculations for a free scalar CFT. In order to do that, we consider a set of scalar fields and  conjugate   momenta $\phi_i,   \pi_j$, $i,   j= 1, \dots, N$  labelled  by  their  positions at the square lattice ---each subindex $i$ makes reference  to a position $(x_i,y_i)$. The fields  satisfy   canonical  commutation  relations,  $[\phi_i ,\pi_j]=i\delta_{ij}$ and $[\phi_i,\phi_j]. =[\pi_i, \pi_j]=0$. Then, given some  Gaussian state   $\rho$, the  EE for some entangling region $A$ can be obtained  from the correlators
$
X_{ij}\equiv \tr (\rho \phi_i\phi_j)$  and    $P_{ij}\equiv \tr (\rho   \pi_i \pi_j)\, 
$
as follows \cite{2003JPhA...36L.205P,Casini:2009sr},
\begin{equation}\label{see}
S(A)=\tr \left[(C_A +1/2) \log (C_A    +1/2)- (C_A -1/2)   \log (C_A-1/2) \right]\, , 
\end{equation}
where $C_A \equiv  \sqrt{X_A    P_A}$ and   $(X_A)_{ij}\equiv X_{ij}$,   $(P_A)_{ij}=P_{ij}$  (with $i,j\in A$)  are the restrictions of the correlators   to the  sites inside    $A$. We compute the $N$-partite information using \req{see} along with \req{IN1}.
Setting the lattice spacing to one, the Hamiltonian is given by
\begin{equation}
H= \frac{1}{2}   \sum_{i,j=-\infty}^{\infty}   \left[\pi^2_{i,j}   + (\phi_{i+1,j}   -\phi_{i,j})^2  +(\phi_{i,j+1}-\phi_{i,j})^2 \right]\, ,
\end{equation}
and the correlators  corresponding to the vacuum state are given by \cite{Casini:2009sr} 
\begin{align}
X_{(0,0),(i,j)} & = \frac{1}{8\pi^2} \int_{-\pi} ^{\pi} d   x  \int_{-\pi}^{\pi} d y \frac{\cos (j y)  \cos (ix) }{\sqrt{2(1-\cos x)+2(1-\cos y)}}\, , \\
P_{(0,0),(i,j)} & = \frac{1}{8\pi^2}\int_{-\pi} ^{\pi} d   x \int _{-\pi}^{\pi} d y  \cos(j y)   \cos(ix) \sqrt{2(1-\cos x)+2(1-\cos y)}\, .
\end{align}
\begin{figure}[t!]
\center
\includegraphics[scale=.62]{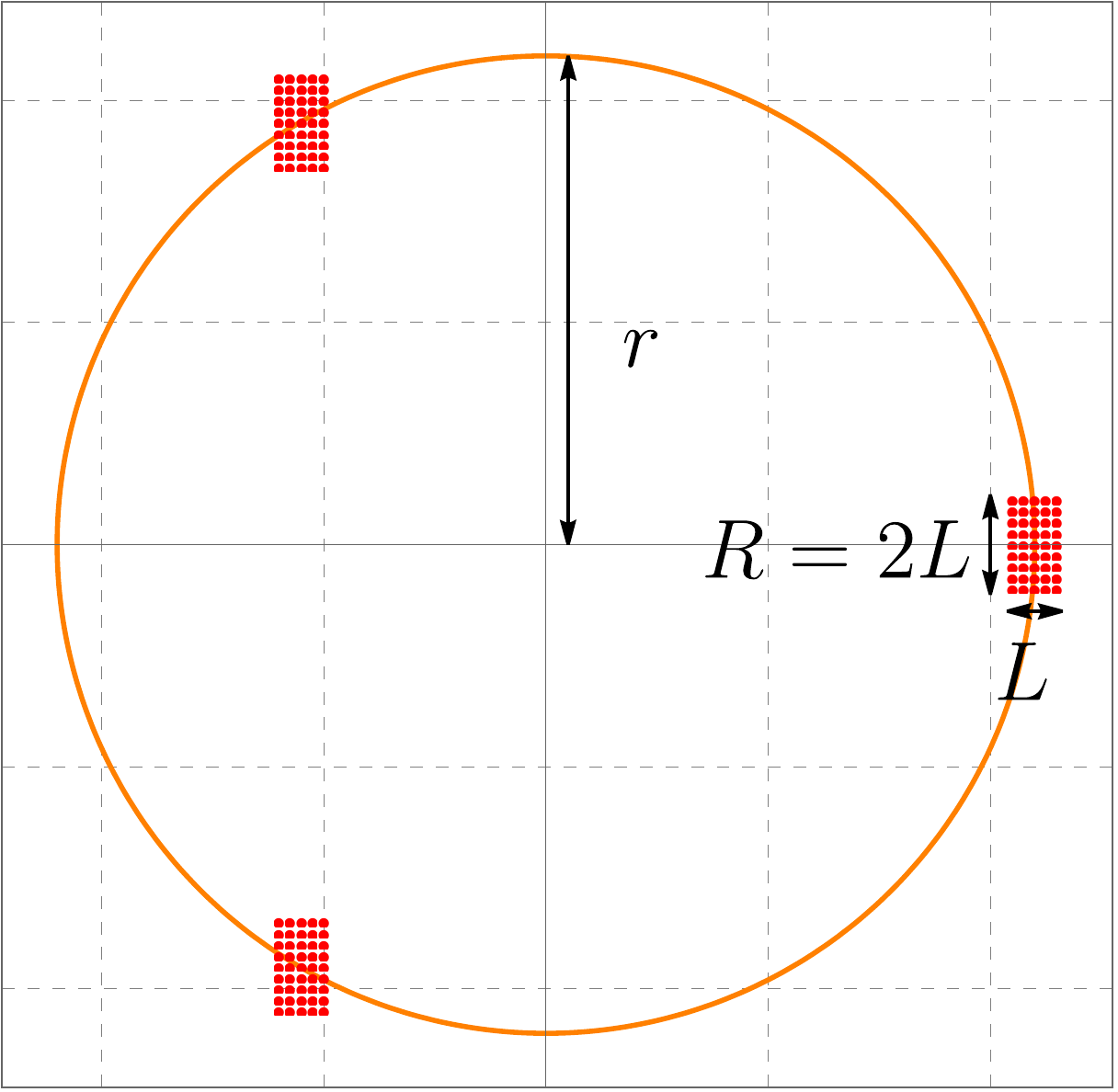} \hspace{0cm} 
\includegraphics[scale=.60]{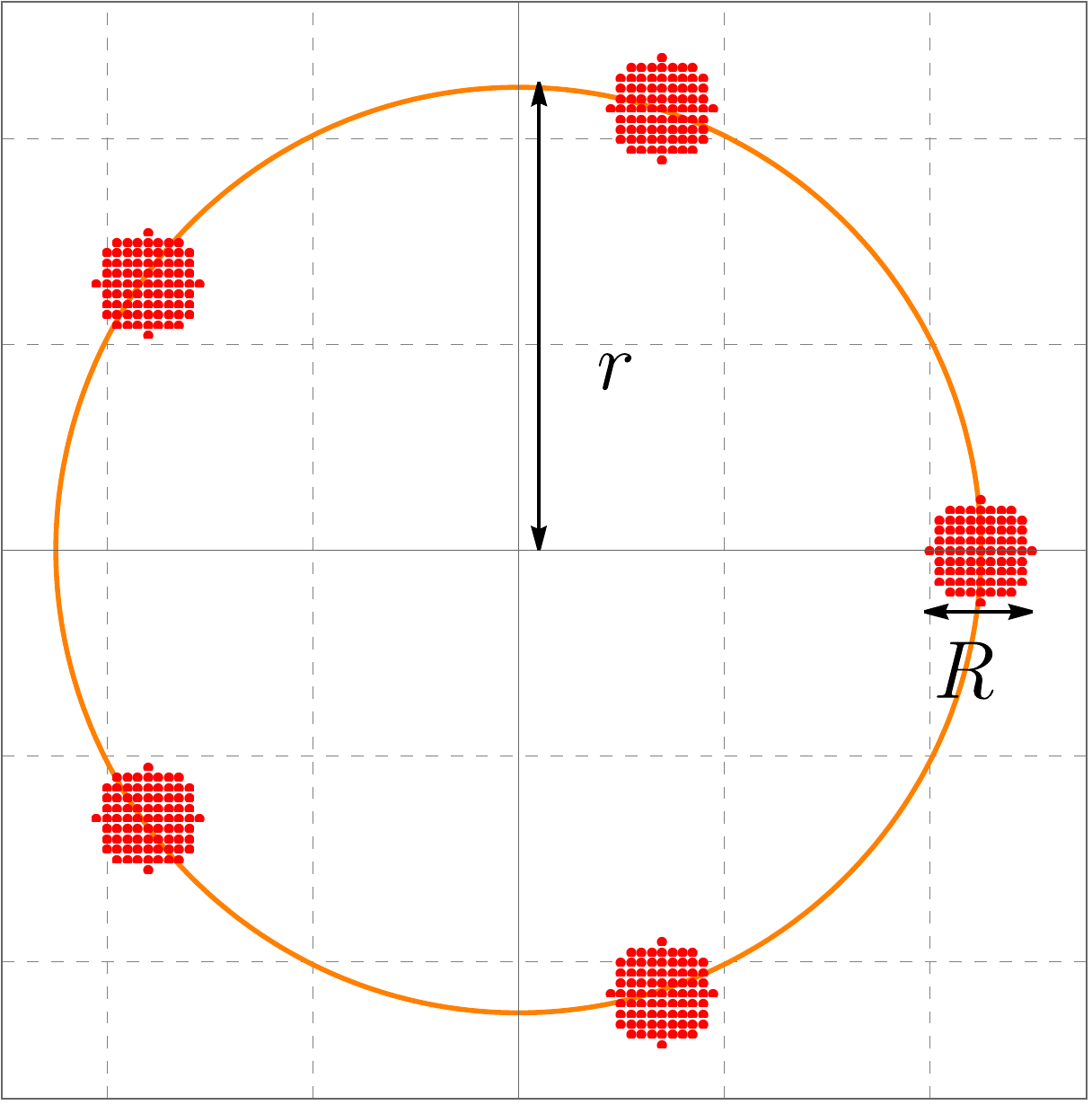} 
\caption{We show two examples of the $N$-gon arrangements of entangling regions considered in the lattice. The one in the left figure corresponds to $N=3$ for rectangles with dimensions $L\times 2L$. The one in the right figure corresponds to $N=5$ for disk regions. In each case, $r$ is defined as the radius of the circumference which shares center with the $N$-gon and passes through the center of the entangling regions.} 
\label{INscalaeeer}
\centering
\end{figure}

In the third subsection, where we compute mutual information of various regions for arbitrary separations, we also consider a free Dirac fermion CFT. For that, we take 
 fields  $\psi_i$,  $i=1,\dots, N$  defined at the  lattice  points, satisfying canonical   anticommutation   relations, $\{\psi_i,  \psi_j^{\dagger} \}=\delta_{ij}$. For a given Gaussian state  $\rho$, we define the correlators matrix  $D_{ij} \equiv \tr \small( \rho \psi_i \psi_j^{\dagger} \small)$. Then, similarly to the scalars case,  the EE for some entangling region $A$ can be obtained as \cite{Casini:2009sr}
 \begin{equation}
S (A)=- \tr   \left[  D_A   \log  D_A   + (1- D_A) \log (1- D_A)\right] \, ,
\end{equation}
where $D_A $ is the  restriction of $D_{ij}$ to the  lattice sites lying within  $A$. Our mutual information calculations are performed using this formula for the corresponding regions $A$, $B$ and $A\cup B$. The lattice Hamiltonian for the free fermion is given   by
\begin{equation}
H=-\frac{i}{2}  \sum_{n ,m} \left[  \left(\psi^{\dagger}_{m, n}   \gamma^0     \gamma^1 (\psi_{m+1,n}  -\psi_{m,n})+\psi^{\dagger}_{m,  n}   \gamma^0 \gamma^2   (\psi_{m,n+1}   -\psi_{ m,n}  ) \right)  - {\rm h.c.}\right] \, ,
\end{equation}
and the correlators in the vacuum state read \cite{Casini:2009sr} 
\begin{equation}
D_{(n,k), (j,l)} =   \frac{1}{2} \delta_{n,  j} \delta_{kl}     - \int_{-\pi}^{\pi} d x  \int_{-\pi}^{\pi } d y   \frac{\sin (x)   \gamma^0  \gamma^1+   \sin(y) \gamma^0  \gamma^2}{8\pi^2 \sqrt{   \sin^2   x +   \sin^2   y}}  e^{i(x (n-j)  +y(k-l))}\, .
\end{equation}

For our numerical calculations, we use a lattice of size 200, so we can consider points separated at most 200 units. In the second subsection, we analyze the behavior of the $N$-partite information at long distances for $N =2,3,4,5,6$, for several (simple) entangling regions and arrangements ---\ie relative positions of the blocks with respect to one another. For a given $N$, we consider two arrangements: one in which the blocks are located at the vertices of a regular $N$-gon, and one in which they are all aligned. For clarity reasons, we only present explicit results for the $N$-gon arrangement ---see Fig.\,\ref{INscalaeeer}--- but we also mention the analogous linear-arrangement results when something qualitatively different is obtained in that case. For each setup, we consider five different elementary blocks: disks, squares, rectangles with dimensions $L \times 2L$ (along the horizontal and vertical axes, respectively), rectangles with dimensions $L \times 4L$, and rectangles with dimensions $L \times 6L$. We study the $N$-partite information for sets of $N$ identical elementary blocks of those kinds. We  do this as a function of the ratio $x\equiv r_{\rm latt}/R_{\rm latt}$, where $R_{\rm latt}$ is the characteristic size of the blocks and $r_{\rm latt}$ is the characteristic separation among them. On the other hand, in the third subsection, we focus on the mutual information $I_2$, but we compute it for general separations, both for the scalar and the fermion. In that case, results are presented for: disks, squares, rectangles with dimensions $L \times 2L$, rectangles with dimensions $L \times 3L$, rectangles with dimensions $L \times 4L$ and rectangles with dimensions $L\times 5L$.  


In the continuum limit, the above lattice models yield results corresponding to a free scalar field and a free fermion, respectively. In order to obtain such values, we proceed as follows. Given a particular configuration and keeping $x$ fixed, we compute $I_N$ for increasingly greater values of $R_{\rm latt}$ and $r_{\rm latt}$. In every case, we find that such values clearly tend to certain constant asymptotic values as  $R_{\rm latt},r_{\rm latt}\rightarrow \infty$. In order to obtain those, we perform fits of the data points using functions $\{1,1/x,1/x^2,\dots\}$. Reliable values are obtained whenever the resulting constant values hardly depend on the order at which we stop adding fitting functions. This is the situation encountered for the present models in all cases considered and therefore the results presented can be directly attributed to the continuum theories. In the case of the free fermion, we  need to take into account the ``doubling'' in the number of fermionic degrees of freedom which occurs in the lattice. In the present three-dimensional case, the Dirac fermion result is obtained by dividing the final result by 4.

The quantities $R_{\rm latt}$ and $r_{\rm latt}$ (which for the remainder of the Section we denote simply by $R$ and $r$, respectively) are precisely defined as follows ---see Fig.\,\ref{INscalaeeer}. For square and rectangle blocks, $R$ is the number of lattice links of the largest side. For instance, this means that in a square of $5 \times 5$ lattice \emph{points}, $R = 4$; while in a $L \times 4 L$ rectangle of $4 \times 13$ lattice points, $R = 12$ ($L = 3$ in this example). For disk blocks, $R$ is the number of lattice links of the diameter. Thus, a circle with $R = 12$ has $13$ lattice points along its diameter. The definition of a circle in a lattice is not precise, so we define the circle of size $R$ to be the set of lattice points at a distance less than or equal to $R/2$ to the central point. Regarding the magnitude $r$, for $N=2$ we define it to be the distance between the rightmost boundary point of the left region and the leftmost boundary point of the right one, whereas for an $N\geq 3$-gon arrangement, it corresponds to the radius of the circle in which we circumscribe the polygon. This is not the precise distance between adjacent sites except in a particular case (the hexagon), but the difference is always an $\mathcal{O}(1)$ number which does not change the functional dependence we want to identify. On the other hand, for the linear arrangement and  $N\geq 3 $, $r$ is the separation between the centers of adjacent blocks, measured in number of lattice links.

When presenting the mutual information results in subsection \ref{i2arbi}, we will normalize the results by the universal coefficient appearing in the EE of a disk region. For general CFTs, the EE of an arbitrary smooth region $A$ is given by
\begin{equation}
S (A)= c_0\frac{ {\rm perimeter}(\partial A) }{\delta}-F(A) +\mathcal{O}(\delta)\, ,
\end{equation}
where $c_0$ is a non-universal constant, $\delta$ is a UV cutoff and $F(A)$ is a dimensionless and universal coefficient. In the case of a disk region, we denote this quantity by $F_0$, namely, $F(\partial A=\mathbb{S}^1) \equiv F_0$.\footnote{$F_0$ and its properties have been further studied in different contexts and for various theories  ---see \eg \cite{Ryu:2006ef,Liu:2012eea,Fonda:2014cca,Allais:2014ata,Casini:2015woa,Anastasiou:2020smm,Huerta:2022cqw} and references therein.} As it turns out, $F_0$ satisfies a number of interesting properties. On the one hand, it coincides with the Euclidean free energy of the corresponding CFT on a three-sphere \cite{Casini:2011kv}. In addition, an entropic monotonicity theorem for general three-dimensional theories has been proven in \cite{Casini:2012ei} and the c-function coincides with $F_0$ at the fixed points ---see also \cite{Myers:2010xs,Casini:2015woa,Casini:2017vbe}. Finally, as shown in 
 \cite{Bueno:2021fxb}, for arbitrary CFTs it happens that
\begin{equation}
F(A)/F_0 \geq 1 \quad \text{for all regions }  A \quad \text{and } \quad F(A)=F_0 \Leftrightarrow A=\text{disk}\, ,
\end{equation}
namely, disks globally minimize the EE universal term for general three-dimensional CFTs. In view of this, in order to compare the mutual information results of different theories, it is natural to normalise the results by $F_0$ in each case, and that is what we do. For the free scalar and the free Dirac fermion we have, respectively \cite{Klebanov:2011gs,Marino:2011nm}
\begin{equation}
F_0^{\rm s}=\frac{1}{16}\left[2\log 2-\frac{3}{\pi^2}\zeta(3) \right] \, , \quad F_0^{\rm f}=\frac{1}{8}\left[2\log 2+\frac{3}{\pi^2}\zeta(3) \right]  \, .
\end{equation}

%
%
%
%

\subsection{Long-distance $I_N$ for a free scalar}
Based on the results obtained in Section \ref{ldb}, we expect the $N$-partite information for a free scalar in general dimensions to scale as
\begin{equation}\label{scalarscaling}
I_N^{\rm free\, scalar} \sim (1/x)^{(d-2)N} \quad \text{for}\quad x\equiv r/R \gg 1\, .
\end{equation}
Here, we used the result for the free-scalar conformal dimension,  $\Delta_{\rm free\,scalar}=(d-2)/2$. Hence, in $d=3$ we expect $I_N^{\rm free\, scalar} \sim (1/x)^{N} $. In order to test this, for each set of entangling regions we produce points at five distinct (large) values of $x$, and we fit them to $(1/x)^{N-1}$, $(1/x)^N$, and $(1/x)^{N+1}$ curves. In Table \ref{tabla1}, which we present in Appendix \ref{subsec:ScalarNumerics}, we have compiled the resulting coefficients of determination $\mathcal R^2$ ---not to be confused with the size of the regions--- for each fit and highlight the one which corresponds to the best one. 

The scaling predicted by \req{scalarscaling} is strongly favoured by all the sets of regions and for all the values of $N$ considered. Considering even greater values of $N$ becomes increasingly challenging due to technical reasons associated to the lattice, but the pattern is clear.  Repeating the analysis using the linear arrangement instead, we find very similar results and the same evidence in favor of \req{scalarscaling}.\footnote{In that case, for $N=6$ the $1/x^5$ scaling is slightly favoured with respect to $1/x^6$ for a couple of sets of regions. However, this is most likely due to the fact that in the linear arrangement setup, studying the long-distance regime is more problematic in a finite lattice because, as opposed to the $N$-gon case, all separations between regions have to be large along the same direction.  }

\begin{figure}[t!]
\center
\includegraphics[scale=.589]{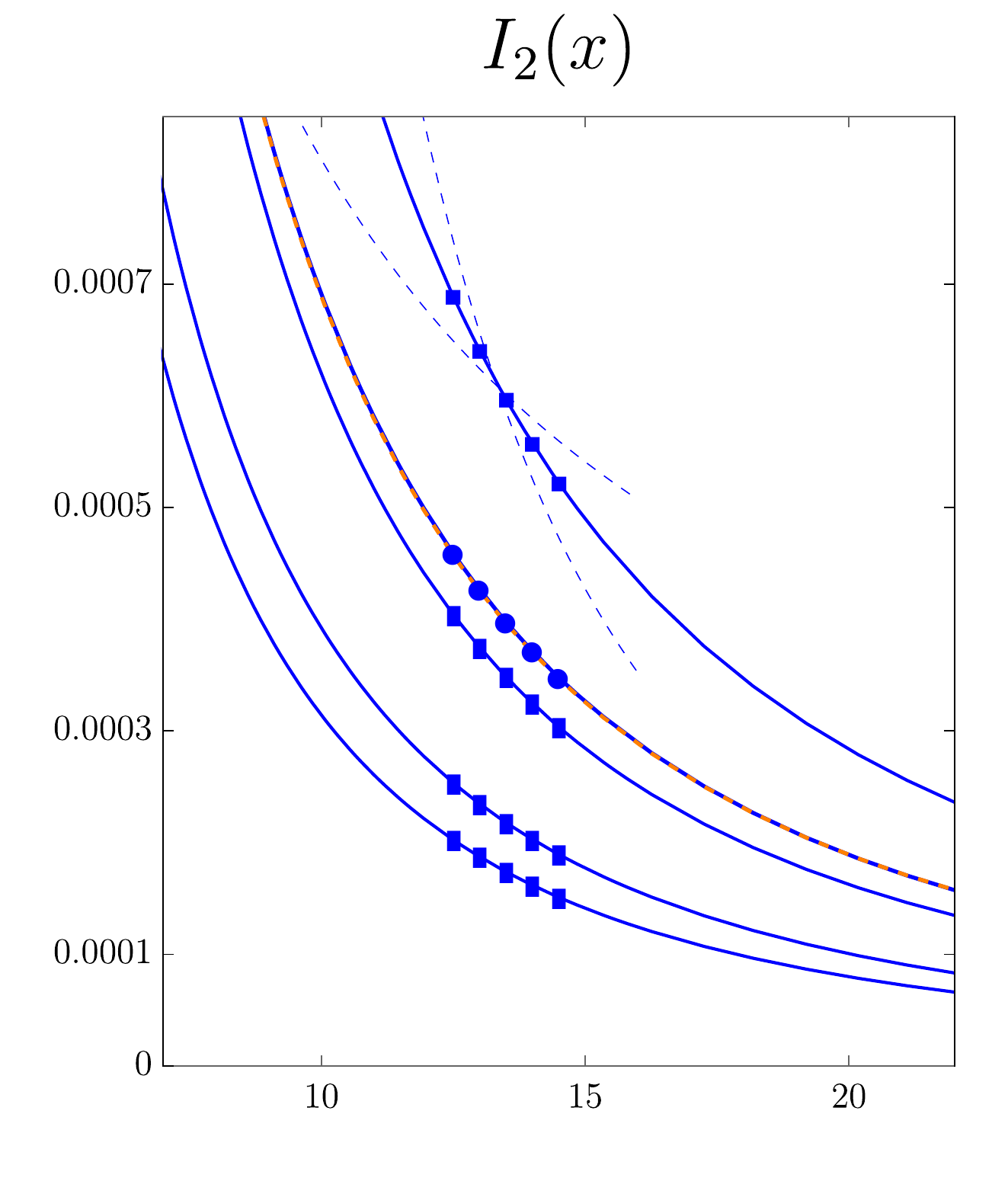} \hspace{0cm} 
\includegraphics[scale=.636]{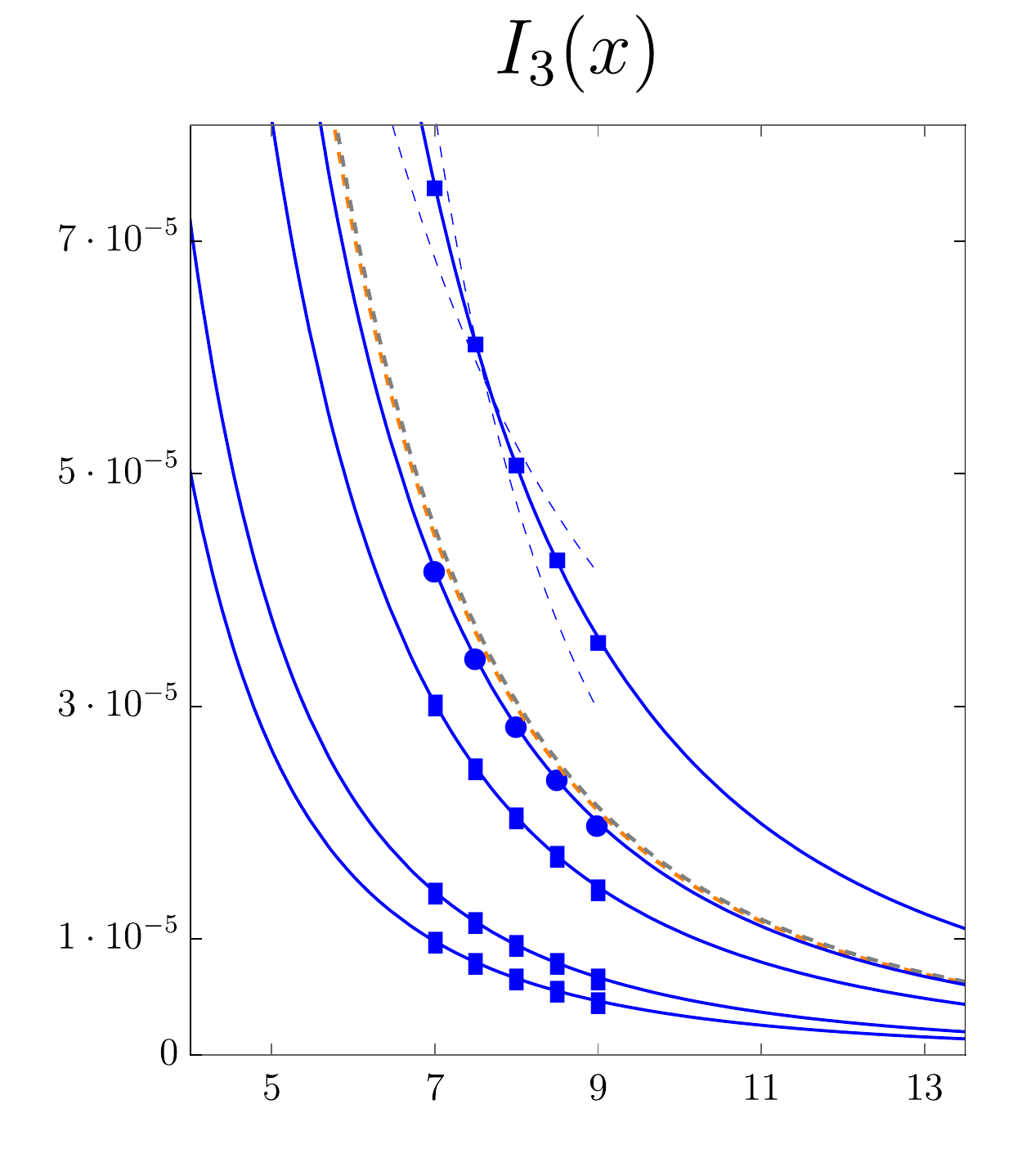} 
\vspace{-0.2cm}
\includegraphics[scale=.6]{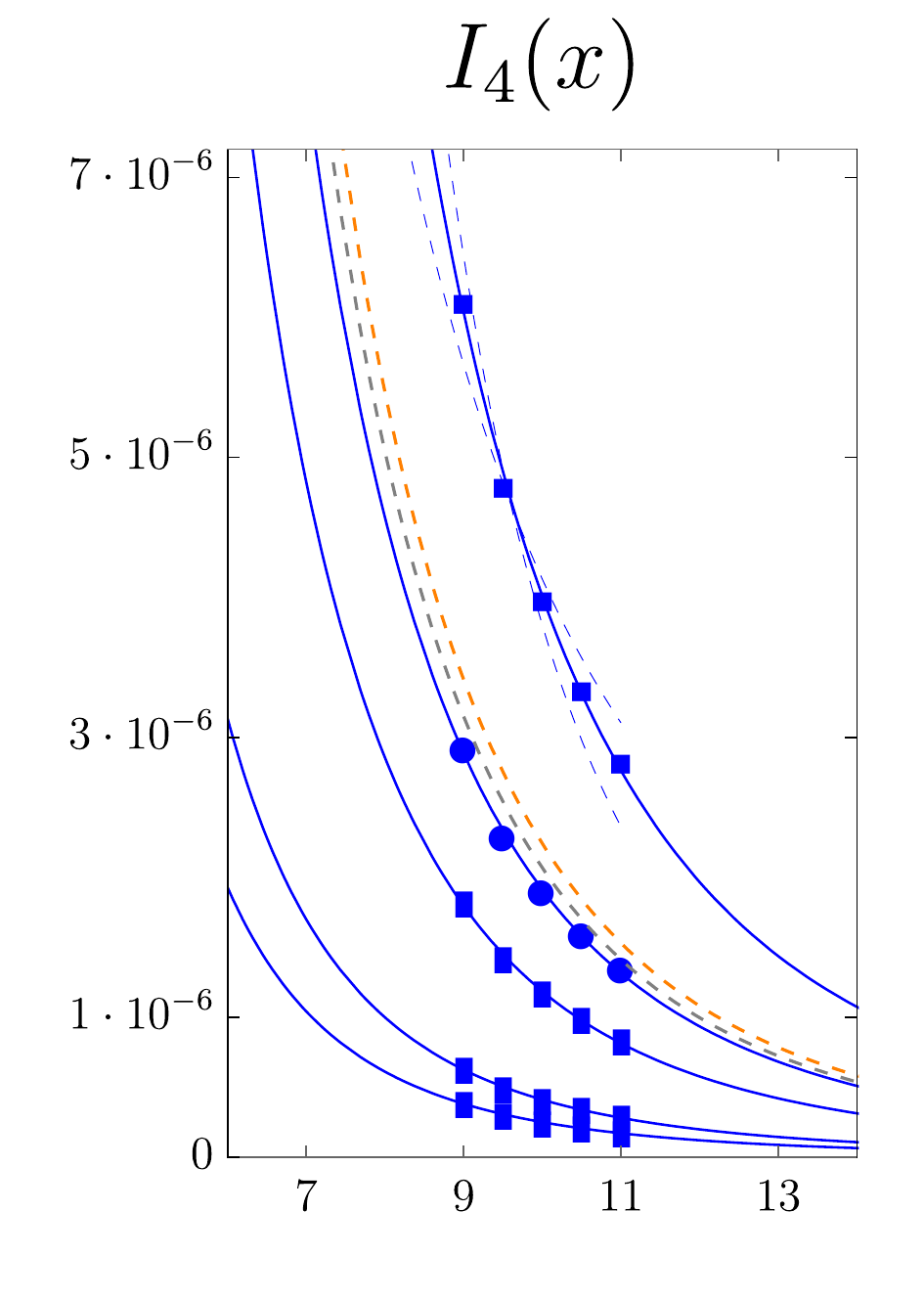} \hspace{-0.7cm} 
\includegraphics[scale=.65]{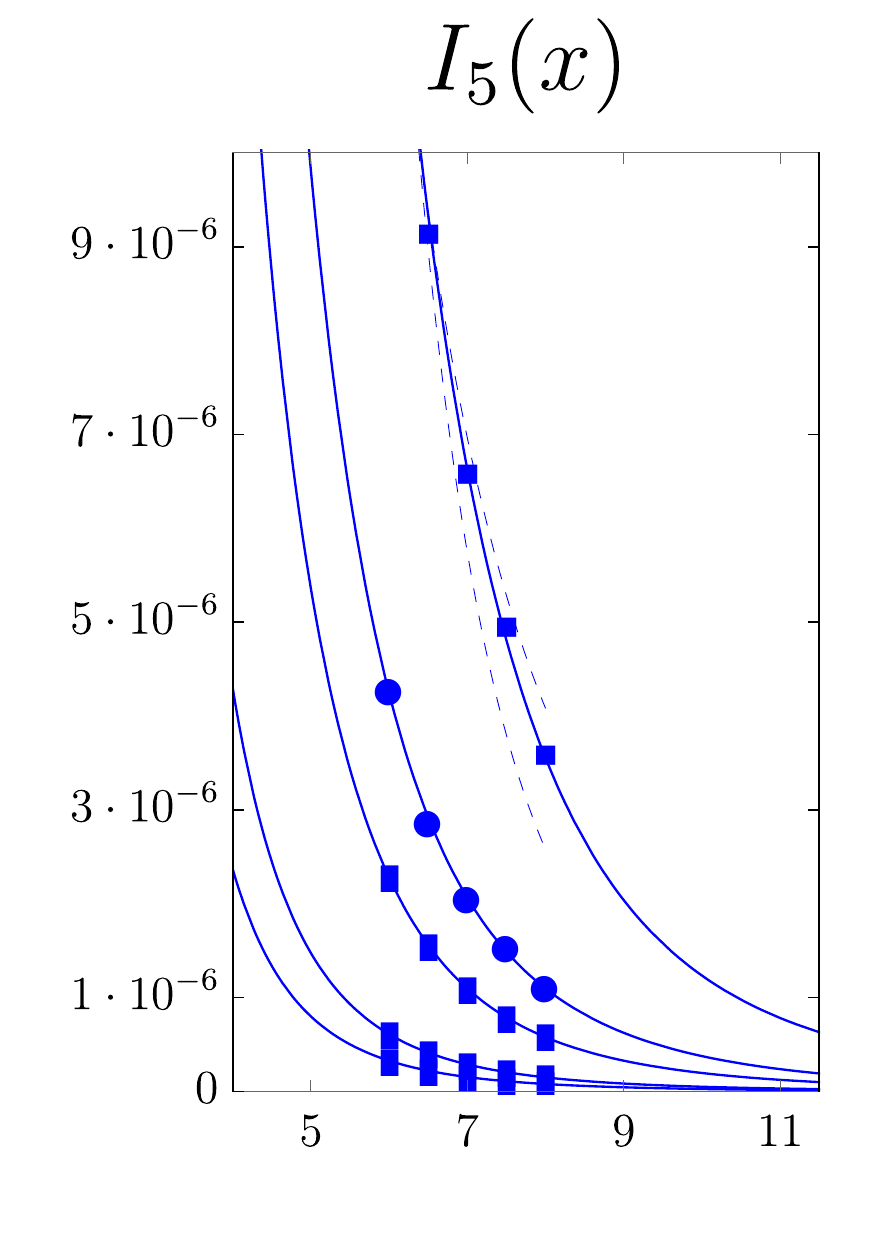} \hspace{-0.7cm} 
\includegraphics[scale=.605]{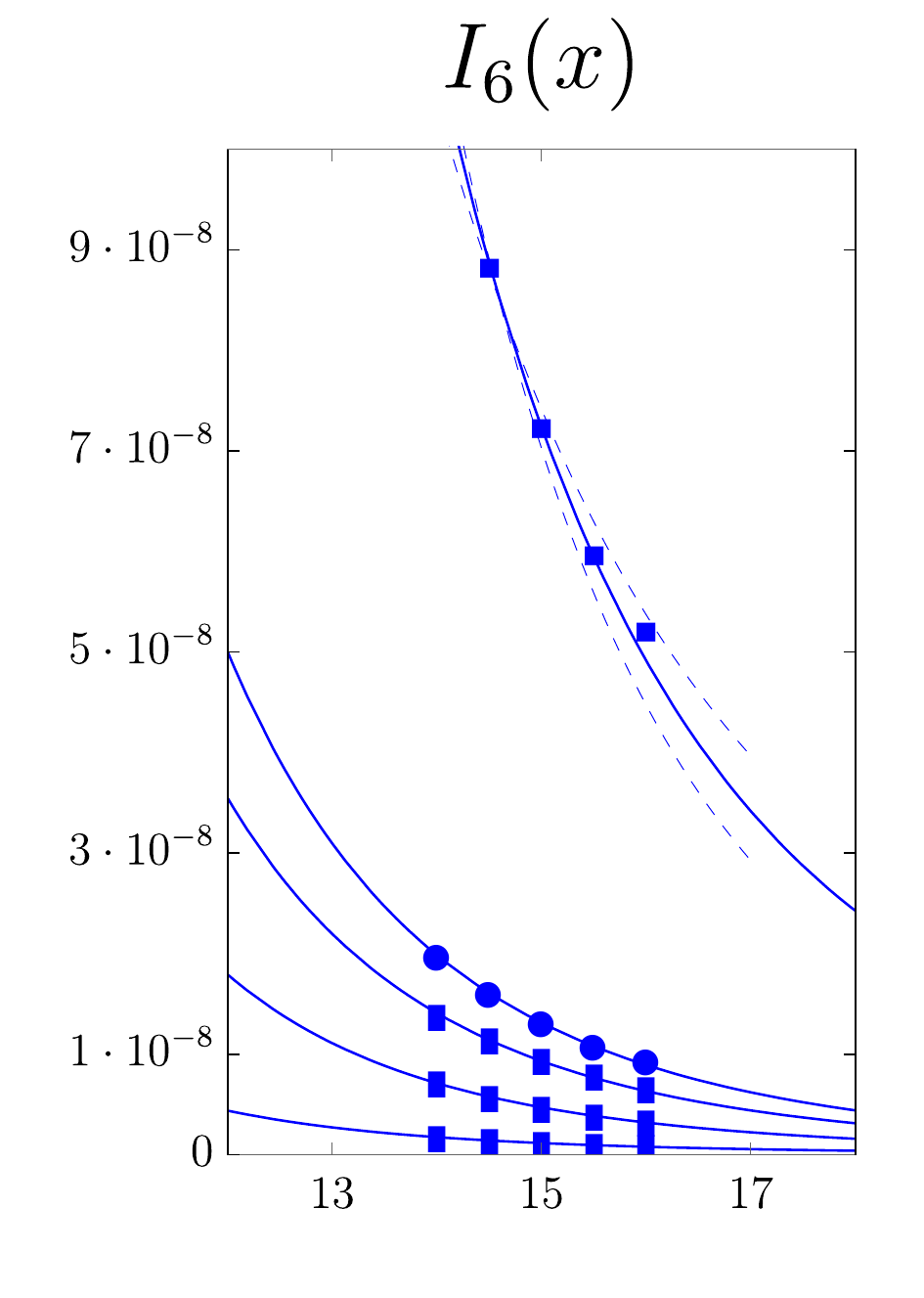} 
\caption{For a free scalar  in the lattice, we plot the $I_N$ (continuum values) for $N=2,3,4,5,6$ for sets of: squares, disks, $L\times 2L$ rectangles,  $L\times 4L$ rectangles and  $L\times 6L$ rectangles (in each plot moving from the uppermost curve to the bottom one) placed in a regular $N$-gon configuration. The solid curves correspond to fits of the form \req{INF}, including two terms. The orange dashed lines in the $I_2$, $I_3$ and $I_4$ plots correspond to curves $(1/12)/(x+1)^2$, $1/(12 \sqrt{3} \pi) \cdot (1/x^3)$ and $(1/180+1/(6\pi^2))/x^4$ respectively, which are the exact leading behaviours. In the last two cases, the subleading term (unknown) has a non-negligible effect, which explains the discrepancy with the points and the numerical fits (the dashed gray lines represent the leading terms in the numerical approximations which include two terms, namely: $0.0832/(x+1)^2$, $0.0155/x^3$ and $0.0207/x^4$). } 
\label{INscalar}
\centering
\end{figure}

Let us now present the explicit results found for the $N$-partite information of the free scalar corresponding to the configurations and regions explained above. The data points obtained appear plotted in Fig.\,\ref{INscalar}. 
As explained in the previous subsection, for every data set we find that a function of the form $I_N \propto (1/x)^N$ fits the points very well. Naturally, we expect subleading corrections to this behavior, namely,
\begin{equation}\label{INF}
I_N^{\rm free\, scalar} =\alpha_{(N)}(1/x)^N+\beta_{(N)}  (1/x)^{N+1}+\dots \quad (x\gg 1)
\end{equation}
Now, the values of $\alpha_{(2)}$ and $\alpha_{(3)}$ in the case of disk regions have been obtained analytically  in \cite{Agon:2015ftl} and \cite{Agon:2021lus}, respectively. They read
\begin{align}
\left.\alpha_{(2)}\right|_{\rm disks}&=\frac{1}{12}\simeq 0.083333\, , \\ \left.\alpha_{(3)}\right|_{\rm disks}&=\frac{1}{12 \sqrt{3}\pi}\simeq  0.0153147\, .
\end{align}
In these two cases, performing a fit to the data points which includes the subleading contribution ---\ie of the form (\ref{INF}) without the dots--- results in a numerical prediction for $\alpha_{(2)}$ and $\alpha_{(3)}$ which is slightly better than the one obtained when only the leading term is considered when compared to the analytic result. In particular, we find\footnote{Observe that in the case of the mutual information of the disk regions, defining $r$ as the distance between the leftmost point of the right disk and the rightmost point of the left disk, rather than the separation between centers, the leading scaling is $\sim \alpha_{(2)}|_{\rm disks}/(x+1)^2$. For $x\gg 1$, this is equivalent to $\alpha_{(2)}|_{\rm disks}\cdot [1/x^2-4/x^3+\mathcal{O}(x^{-4})]$, so we could have chosen to present the numerical results as an expansion on $1/x$ rather than $1/(x+1)$. However, the definition of $x$ we use here is more convenient for the analysis of the following subsection, and performing a meaningful comparison with the leading analytic behavior of our lattice points represented as a function of $x$ in Fig.\,\ref{INscalar}  requires using $1/(x+1)$.  }
\begin{align}
\left.I_2^{\rm free\, scalar} \right|_{\rm disks}& \simeq    {0.0832} \,/(x+1)^2+0.00629 \, /(x+1)^3\, , \quad (x\gg 1) \\
\left.I_3^{\rm free\, scalar} \right|_{\rm disks}& \simeq   { 0.0155} \, (1/x)^3- 0.00861 \, (1/x)^4\, ,  \quad (x\gg 1)
\end{align}
as the best fits. Hence, the lattice calculations for the disks mutual and tripartite informations reproduce the analytic results with remarkable accuracy (they differ by $\sim 0.16\%$ and $\sim 1.21\%$, respectively).   For comparison, fits which only include the leading term yield $0.0836$ and $0.0144$ respectively for the leading coefficients (discrepancies $\sim 0.32\%$ and $\sim 6.35\%$ respectively).
In view of this, we proceed similarly for the $N=4$ case, finding 
\begin{align}
\left.I_4^{\rm free\, scalar} \right|_{\rm disks} &\simeq    0.0207 \,(1/x)^4- 0.0155 \, (1/x)^5\, ,\quad (x\gg 1)
\end{align}
The leading coefficient can be compared with the exact formula result in  \req{Final-I4-square-2} obtained in the previous section. As we can see, both results differ by $\sim 8\%$, which is considerably more than for the $N=2,3$ cases, but still a very good agreement in view of the limitations of the lattice.\footnote{These limitations include the fact that disks are difficult to approximate in a square lattice, or the fact that increasing the number of regions makes it more difficult to reliably obtain the continuum limit in a finite lattice.} This is a highly non-trivial check of our general formula \req{I4-final0} for the long-distance four-partite information of ball regions.
On the other hand, for the $N=5,6$ cases, we find
\begin{align}
\left.I_5^{\rm free\, scalar} \right|_{\rm disks}&\simeq     0.0446 \, (1/x)^5-  0.0699\, (1/x)^6\, , \quad (x\gg 1)\\
\left.I_6^{\rm free\, scalar} \right|_{\rm disks}&\simeq    0.183\, (1/x)^6- 0.489  \, (1/x)^7\, , \quad (x\gg 1)
\end{align}
The leading-term coefficients could be eventually compared with formulas analogous to \req{I4-final0}.

Performing similar fits for the square and rectangular regions, we find for the respective mutual information coefficients\footnote{As mentioned above, in this case we define the $\beta_{(2)}$ so that $I_2^{\rm free\, scalar} \simeq  \alpha_{(2)} \,/(x+1)^2+ \beta_{(2)} \, /(x+1)^3+\dots$}
\begin{align}
\left.\alpha_{(2)}\right|_{L \times L}&\simeq 0.124\, , \quad \left.\alpha_{(2)}\right|_{L\times 2L}\simeq 0.0682 \, ,\quad  \left.\alpha_{(2)}\right|_{L\times 4L}\simeq 0.0412 \, , \quad  \left.\alpha_{(2)}\right|_{L\times 6L}\simeq 0.0325 \, , \\ \notag
\left.\beta_{(2)}\right|_{L \times L}&\simeq 0.0150 \, , \quad \left.\beta_{(2)}\right|_{L\times 2L}\simeq 0.00126 \, ,\quad  \left.\beta_{(2)}\right|_{L\times 4L} \simeq -0.000502 \, , \quad  \left.\beta_{(2)}\right|_{L\times 6L}\simeq -0.000636 \, .
\end{align}

\begin{figure}[t!]
\center
\includegraphics[scale=.75]{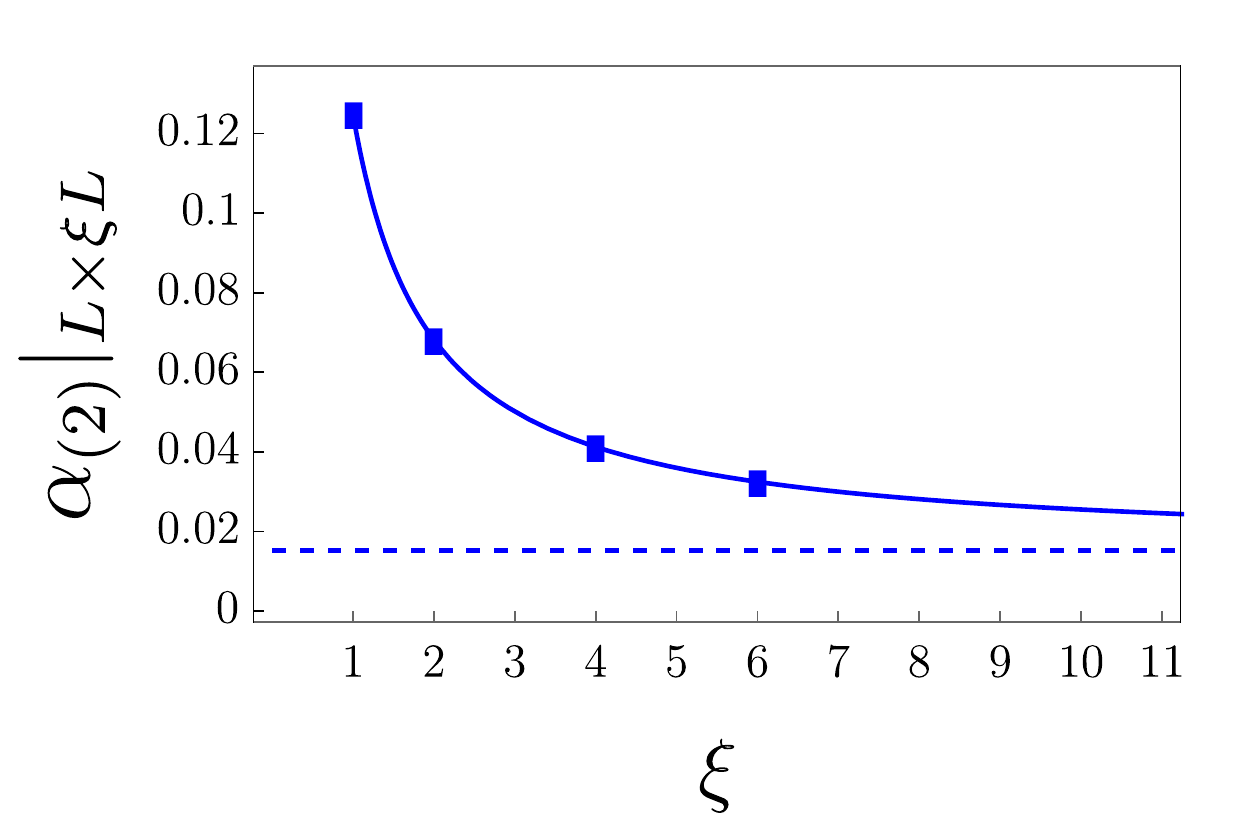}
\caption{We plot the values of the leading coefficient in the mutual information of a free scalar field in the long-distance regime for pairs of rectangle regions of dimensions $L\times \xi L$, as a function of $\xi$. The solid curve is the fit presented in the main text, and the dashed line, the asymptotic value corresponding to $\xi \rightarrow \infty$ (namely, very thin rectangles).} 
\label{alphaaa2}
\centering
\end{figure}

Observe that, as we make the rectangles thinner, the coefficients $\alpha_{(2)}$ tend to decrease. It is easy to check that they are very accurately approximated ---see Fig.\,\ref{alphaaa2}--- by a function of the form
\begin{equation}
\left.\alpha_{(2)}\right|_{L \times \xi L} \simeq 0.015215+
0.102340(1/\xi)+ 0.0073286(1/\xi)^2 \, .
\end{equation}
From this, we can extract the limiting value corresponding to arbitrarily thin rectangles, 
\begin{equation}\label{alpha2in}
\left.\alpha_{(2)}\right|_{L \times \xi L} \overset{\xi\gg 1}{\simeq} 0.015215\,.
\end{equation}
Note that in order for this to hold one needs to be careful with the order of limits (we are taking $x\gg 1$ and $\xi \gg 1$ at the same time). In particular, \req{alpha2in} is valid as long as we keep $r\gg \xi L$, \ie  the expression is valid for pairs of very thin rectangles which are separated a distance much greater than the length of their longest side. It would be interesting to compute this coefficient analytically.\footnote{An analogous calculation of the leading coefficient as a function of $\xi$ was performed in the case of a free fermion in \cite{Agon:2021zvp}. In that case, the leading scaling is $I_2^{\rm free\, fermion} \sim \mathcal{O}(x^{-4})$ instead.}

On the other hand, we find for the $I_3$,
\begin{align}
\left.\alpha_{(3)}\right|_{L \times L}&\simeq 0.0280 \, , \quad \left.\alpha_{(3)}\right|_{L\times 2L}\simeq 0.0112 \, ,\quad  \left.\alpha_{(3)}\right|_{L\times 4L}\simeq 0.00512  \, , \quad  \left.\alpha_{(3)}\right|_{L\times 6L}\simeq 0.00355 \, , \\ \notag
\left.\beta_{(3)}\right|_{L \times L}& \simeq -0.0168 \, , \quad \left.\beta_{(3)}\right|_{L\times 2L}\simeq -0.00551 \, ,\quad  \left.\beta_{(3)}\right|_{L\times 4L}\simeq -0.00211 \, , \quad  \left.\beta_{(3)}\right|_{L\times 6L}\simeq -0.00136 \, , 
\end{align}
for the $I_4$,
\begin{align}
\left.\alpha_{(4)}\right|_{L \times L}&\simeq 0.0442 \, , \quad \left.\alpha_{(4)}\right|_{L\times 2L}\simeq  0.0125\, ,\quad  \left.\alpha_{(4)}\right|_{L\times 4L}\simeq 0.00418 \, , \quad  \left.\alpha_{(4)}\right|_{L\times 6L}\simeq 0.00251 \, , \\ \notag
\left.\beta_{(4)}\right|_{L \times L}&\simeq -0.0409 \, , \quad \left.\beta_{(4)}\right|_{L\times 2L}\simeq -0.00626 \, ,\quad  \left.\beta_{(4)}\right|_{L\times 4L}\simeq -0.000682 \, , \quad  \left.\beta_{(4)}\right|_{L\times 6L}\simeq -0.0000901 \, ,  
\end{align}
for the $I_5$,
\begin{align}
\left.\alpha_{(5)}\right|_{L \times L}&\simeq 0.152 \, , \quad \left.\alpha_{(5)}\right|_{L\times 2L}\simeq 0.0226 \, ,\quad  \left.\alpha_{(5)}\right|_{L\times 4L}\simeq  0.00543\, , \quad  \left.\alpha_{(5)}\right|_{L\times 6L}\simeq 0.00279\, , \\ \notag
\left.\beta_{(5)}\right|_{L \times L}&\simeq -0.281 \, , \quad \left.\beta_{(5)}\right|_{L\times 2L}\simeq -0.0294 \, ,\quad  \left.\beta_{(5)}\right|_{L\times 4L}\simeq -0.00409 \, , \quad  \left.\beta_{(5)}\right|_{L\times 6L}\simeq -0.00144 \, ,  
\end{align}
and for the $I_6$,
\begin{align}
\left.\alpha_{(6)}\right|_{L \times L}&\simeq 1.10 \, , \quad \left.\alpha_{(6)}\right|_{L\times 2L}\simeq 0.158 \, ,\quad  \left.\alpha_{(6)}\right|_{L\times 4L}\simeq  0.0557\, , \quad  \left.\alpha_{(5)}\right|_{L\times 6L}\simeq 0.0171 \, , \\ \notag
\left.\beta_{(6)}\right|_{L \times L}&\simeq -4.02 \, , \quad \left.\beta_{(6)}\right|_{L\times 2L}\simeq -0.752 \, ,\quad  \left.\beta_{(6)}\right|_{L\times 4L}\simeq -0.0311 \, , \quad  \left.\beta_{(5)}\right|_{L\times 6L}\simeq -0.0591 \, .
\end{align}
The resulting curves for the various $I_N$ are shown in Fig.\,\ref{INscalar} and approximate the data points very well in all cases. Let us mention that, as a matter of fact, fits which only include the leading term in \req{INF} also do a very good job at fitting the data ---in most cases the curves are indistinguishable from the ones presented in the figure. However, as mentioned earlier, the numerical results for the coefficients $\alpha_{(2)}$ and $\alpha_{(3)}$ in the case of disk regions are more precisely obtained including also the subleading term, and so we choose to perform fits of that kind for all the rest of configurations. For comparison, in Fig.\,\ref{INscalar} we have included, in the case of square configurations, the best fits obtained from ansatze of the form $I_N \propto (1/x)^{N-1}$ and  $I_N \propto (1/x)^{N+1}$. It is obvious from the figure that those are clearly wrong.

Observe that for every $N$, the values of $I_N^{\rm free\, scalar}$ for all possible sets of regions are positive. It seems then reasonable to speculate that the free scalar $N$-partite information is positive for all possible regions and configurations. The results obtained replacing the $N$-gon arrangement by the aligned one also support this claim. In view of the exact results obtained in the previous section, it is in fact natural to conjecture that $I_N^{\rm free\, scalar} \geq 0$ for all possible regions and in general dimensions. It would be interesting to find additional support for this conjecture.

 \subsection{$I_2$ at arbitrary separations for free scalars and fermions}\label{i2arbi}
In this subsection we compute the mutual information of pairs of entangling regions of different shapes for three-dimensional free scalars and free fermions in the lattice for general separations. We also include results corresponding to two analytic toy models whose long-distance behaviours mimic the free fermion and free scalar ones, respectively. The toy models, which are introduced in detail in Appendix \ref{EMIMEMI}, correspond to the so-called ``Extensive Mutual Information'' (EMI) model  \cite{Casini:2008wt} and a modified version of it, which we call ``Modified EMI'' (MEMI) model. In the case of disk regions, we use the toy models to produce very precise analytic approximations to the free scalar and free fermion mutual informations for general separations. Based on the results, we conjecture that  the free scalar mutual information is greater than the fermion one for arbitrary regions (both conveniently normalized by their respective $F_0$). 


\begin{figure}
\center
\includegraphics[scale=.59]{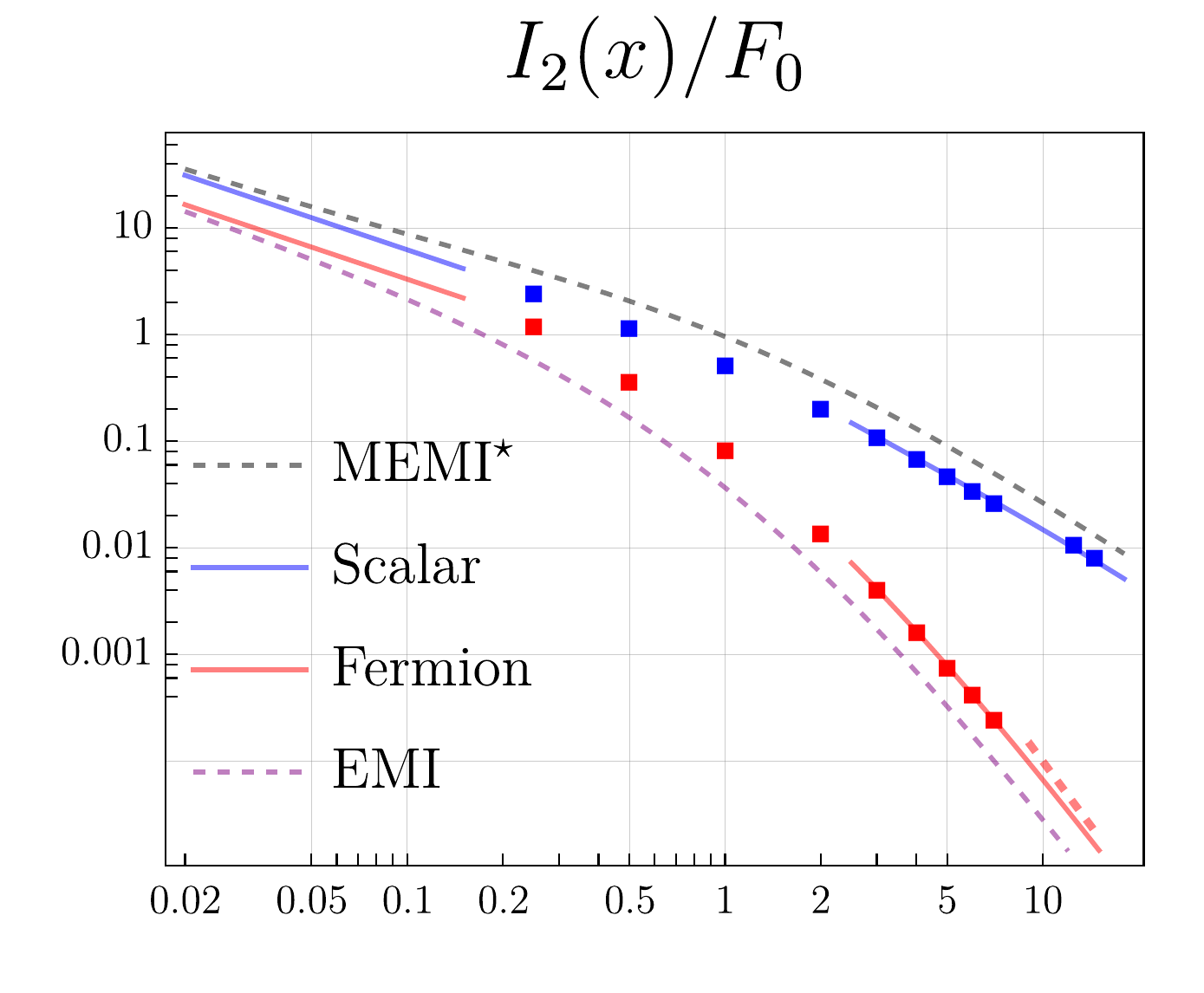} \hspace{-.8cm} 
\includegraphics[scale=.57]{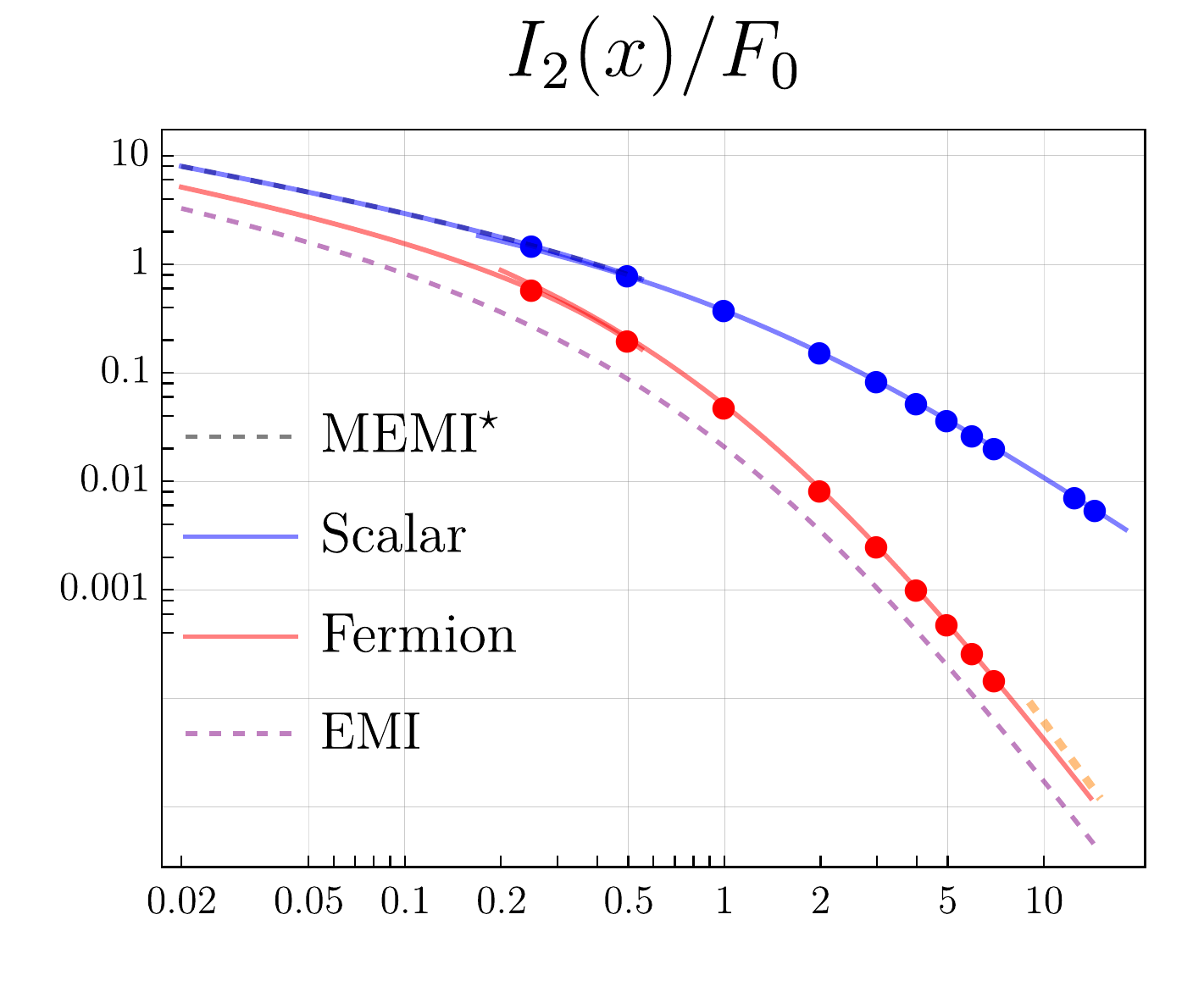}  \\ \vspace{-.1cm}
\includegraphics[scale=.625]{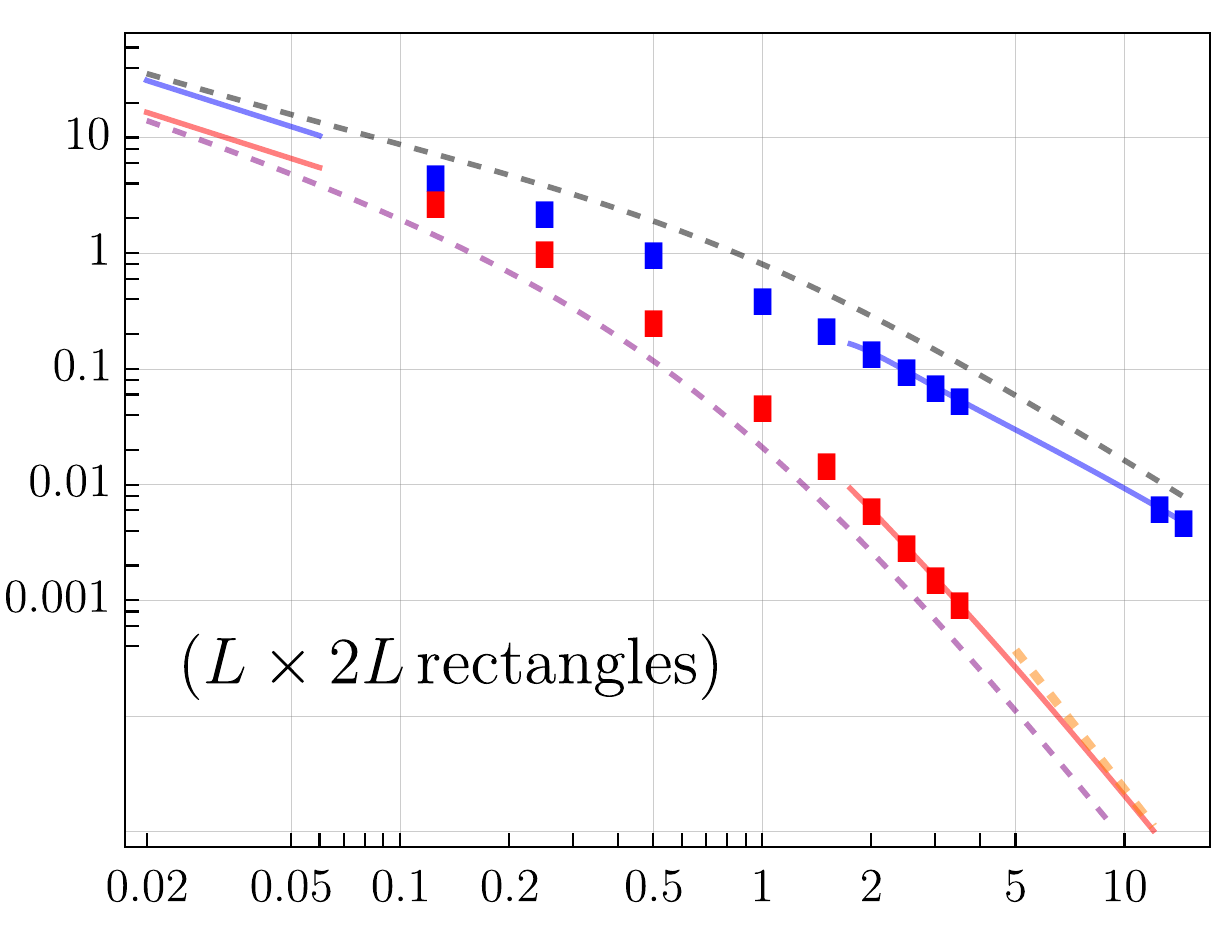}\hspace{0.1cm}\includegraphics[scale=.615]{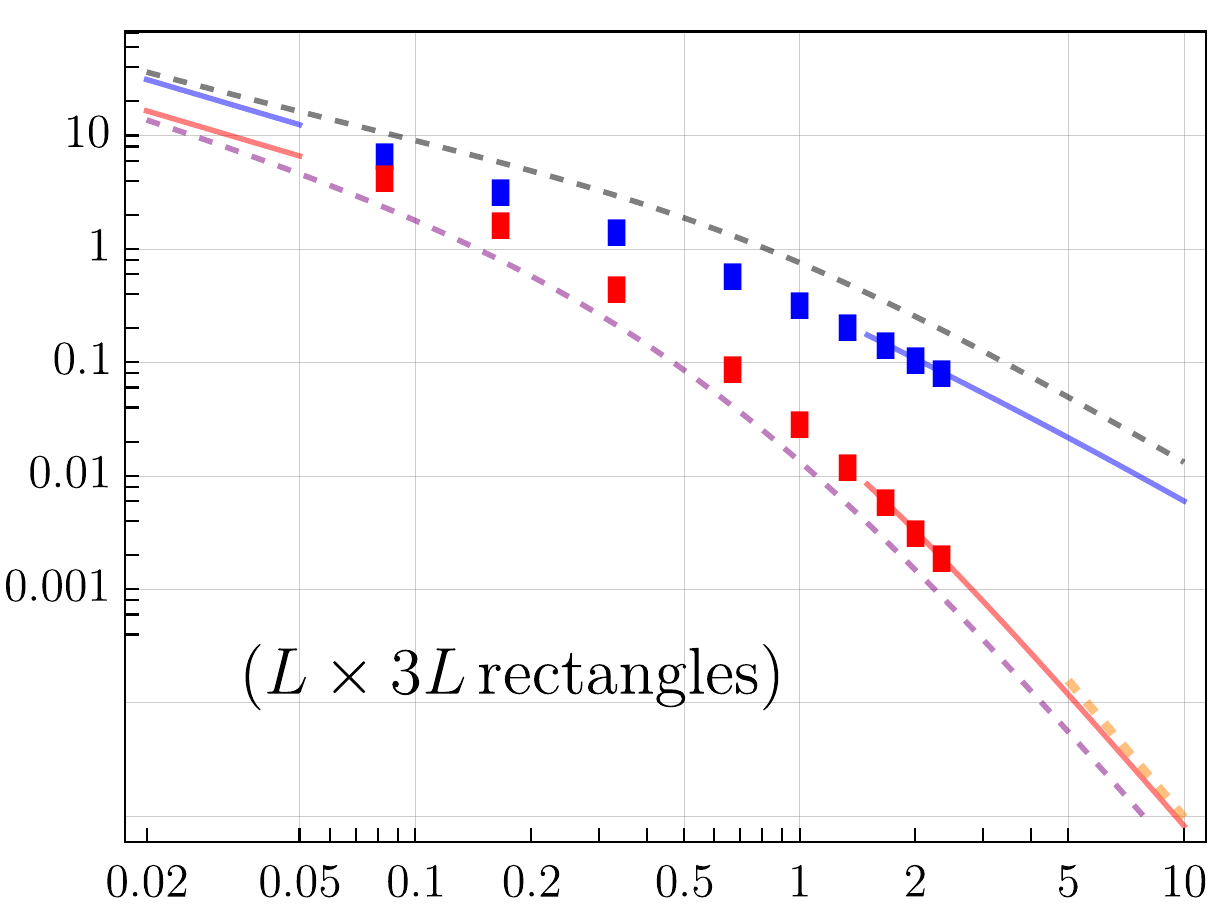}\\
\includegraphics[scale=.625]{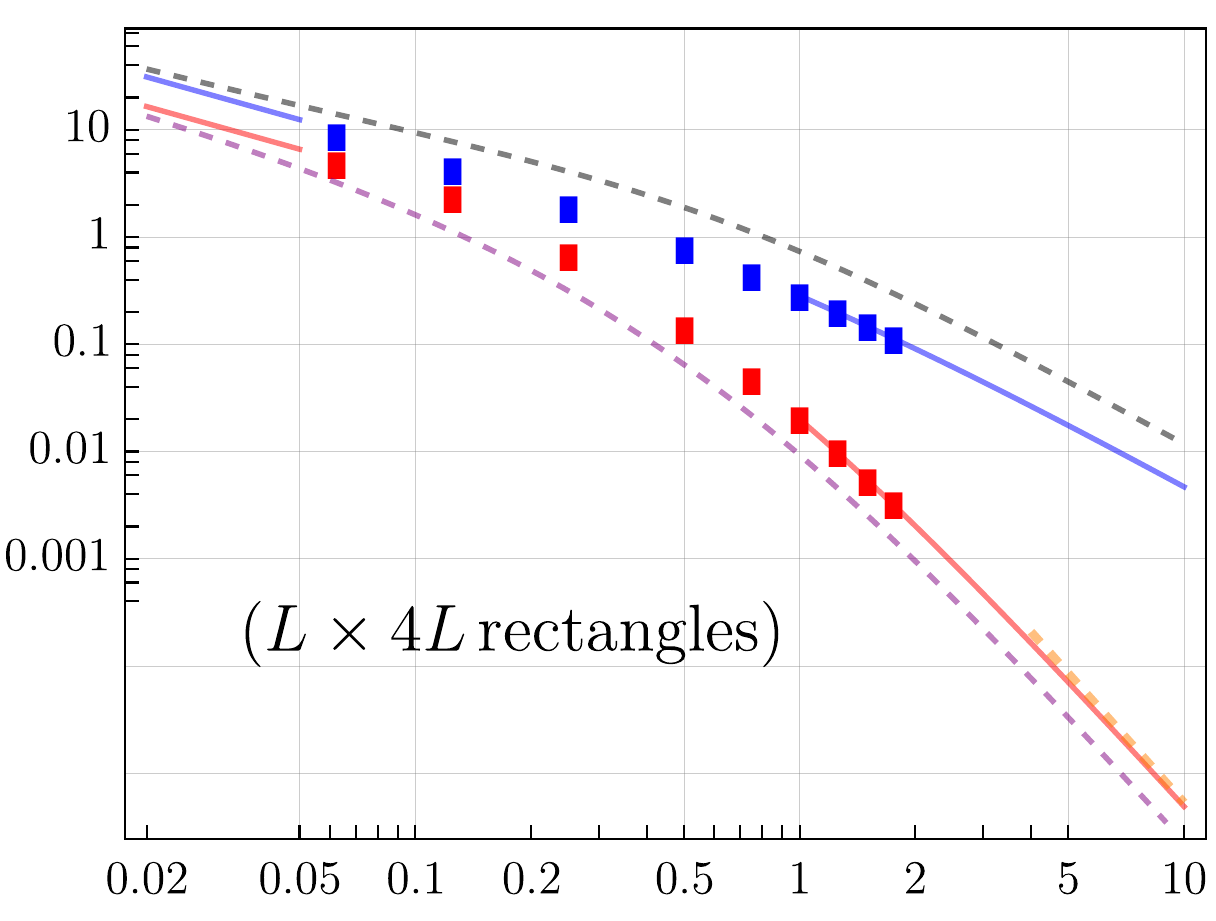}\hspace{0.1cm}\includegraphics[scale=.625]{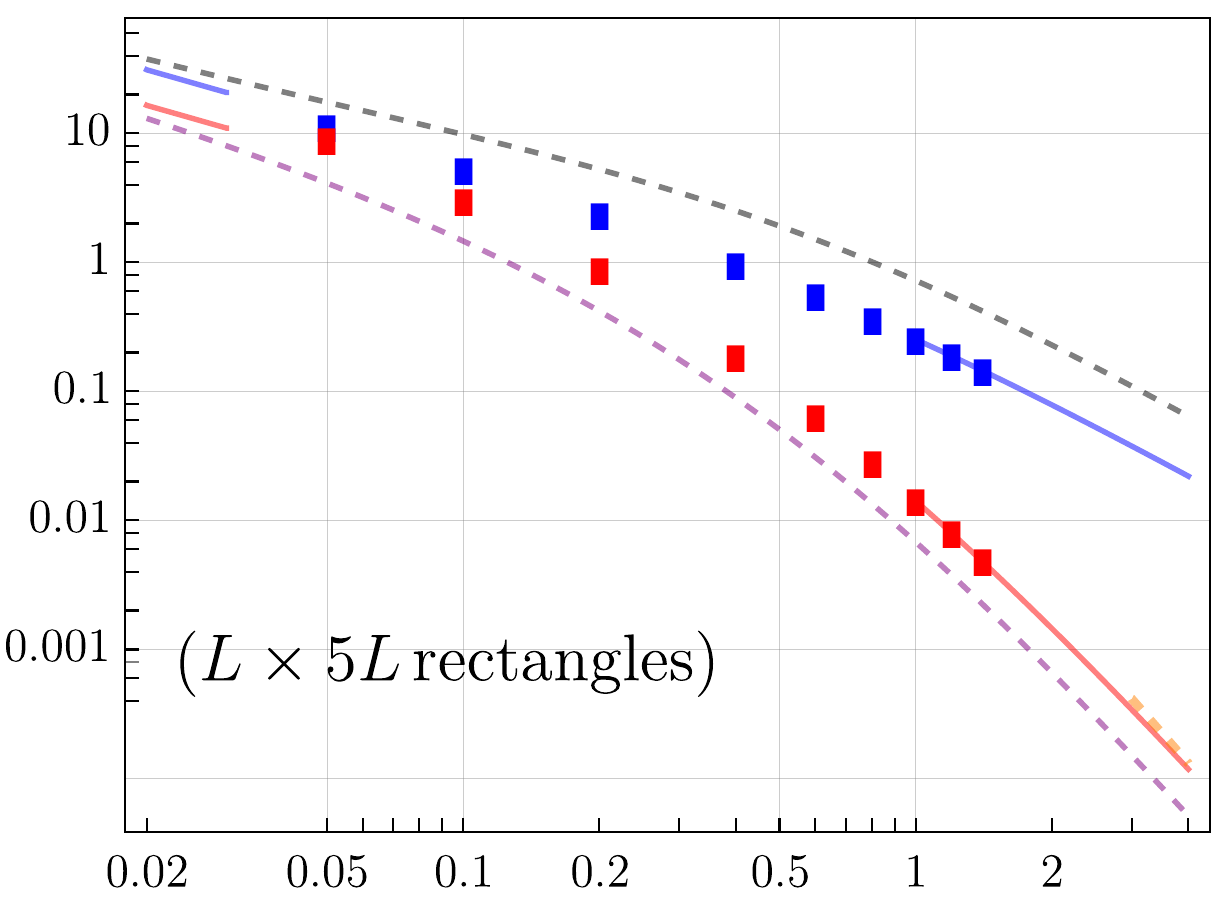}
\caption{We plot, in a $\log$-$\log$ scale, the mutual information (normalized by the disk EE coefficient $F_0$) for pairs of identical regions for free scalars (blue), free fermions (red) and the analytic toy models EMI and MEMI$^\star$ (gray and purple, respectively) as a function of the dimensionless separation $x\equiv r/R$, ($R\equiv \xi L$). The upper plots correspond to square (left) and disk (right) regions, respectively, whereas the middle and lower plots correspond to rectangle regions of sides $L\times (2,3,4,5)L$. The thick blue and red lines represent the analytic approximations explained in the main text. The orange dotted lines correspond to the leading long-distance approximations obtained for the free fermion in \cite{Agon:2021zvp}. } 
\label{I22p}
\centering
\end{figure} 

We present our results for the mutual information in a $\log$-$\log$ scale. This allows for a better appreciation of the short- and long-distance behaviours. On the one hand, observe that for rectangles of sides $L\times \xi  L$ separated a distance $r\equiv x \cdot \xi L$ along the $Y$ axis we have, for general CFTs,
\begin{equation}\label{shorti}
I_2(x)/F_0\overset{x\rightarrow 0}{=} \frac{ k/F_0}{x}+ \dots
\end{equation} 
where $k$ is the universal coefficient characterizing the EE of a thin strip region. Hence, in a doubly logarithmic scale, for any CFT we should have a linear behaviour with unit slope for short separations, since $\log (I_2/F_0)= \log ( k/F_0) - \log x + \dots$ in that case. Such straight lines should cut the $Y$ axis at a value $\log ( k/F_0)$. The values of such constants are known for free scalars and fermions. In particular, we have \cite{Casini:2009sr}
\begin{equation}
k^{\rm s}\simeq 0.0397\, , \quad k^{\rm f}\simeq 0.0722 \, .
\end{equation}
For two disk regions, the limit $r\rightarrow 0$ corresponds to the situation in which the disks touch at a point.\footnote{See \eg \cite{Bueno:2019mex} for examples of both finite and divergent mutual informations for pairs of regions with a single contact point.} In that case, the results for the analytic toy models we consider in Appendix \ref{EMIMEMI} suggest that the mutual information diverges, for general CFTs, as
\begin{equation}
I_2(x)/F_0\overset{x\rightarrow 0}{=} \frac{\gamma /F_0}{\sqrt{x}}+ \mathcal{O}(x^0)\,,
\end{equation}
where $\gamma$ is a new theory-dependent coefficient.
On the other hand, for long separations and general regions, we know that free scalars and fermions behave, respectively, as
\begin{equation}\label{longd}
\left.I_2(x)/F_0\right|_{{\rm free\, scalar}} \overset{x\rightarrow 0}{=} 
\alpha_2^{\rm s}\cdot (1/x)^2+\mathcal{O}(x^{-3}) \, , \quad  \left.I_2(x)/F_0\right|_{\rm free\, fermion} \overset{x\rightarrow 0}{=} 
\alpha_2^{\rm f}\cdot (1/x)^4+\mathcal{O}(x^{-5}) \, , 
\end{equation} 
and so, for long separations we expect the log-log curves to approach straight lines of slope $-2$ and $-4$, respectively.

We present the results in Fig.\,\ref{I22p}. The different plots correspond to the mutual information of pairs of identical regions separated along the $X$ axis corresponding to: squares, disks and increasingly thinner rectangles with lengths: $L\times 2L$, $L\times 3L$, $L\times 4L$ and $L\times 5L$, respectively. In all cases, we have included solid lines representing the short- and long-distance behaviors appearing in \req{shorti} and eq.\,(\ref{longd}). In the latter case, the curves represent fits of the form (\ref{longd}) ---including three subleading terms in each case--- to the data points corresponding to separations $x=3,4,5,6,7$. The dotted orange lines correspond to approximations of the form $\left.I_2(x)/F_0\right|_{\rm free\, fermion} \simeq 
\alpha_2^{\rm f}\cdot (1/x)^4$ with the coefficients $\alpha_2^{\rm f}$ obtained in \cite{Agon:2021zvp}. In all cases, the numerical fits tend to the corresponding curves.  In the case of the squares and the rectangles, the data points seem to tend, in all cases, to the short-distance approximations given by  \req{shorti}, which is a good check of the numerics. In the same plot we include the exact analytic results obtained in Appendix \ref{EMIMEMI} for each set of regions for the EMI and MEMI models ---see eqs.\,(\ref{diskEMI}), (\ref{rectEMI}), (\ref{diskMEMI}) and (\ref{rectMEMI}). For the MEMI model, the value of $F_0$ turns out to vanish identically, and so we normalize the corresponding curves in a way such that they tend to the free scalar results for $x\rightarrow 0$. As we can see, the EMI curves are quite similar to the fermion ones, and so are the MEMI ones to the free scalar ones. This is particularly so in the case of the disks. In that case, we can actually approximate the free scalar and free fermion results by the MEMI and EMI ones for rather general values of $x$. In particular, normalizing the MEMI and EMI formulas by the long-distance coefficients of the free scalar and free fermion, respectively, we find the approximations 
\begin{align}\label{fsss}
\left. I_2(x)/F_0\right|^{\rm disks}_{{\rm free\, scalar}}  &\simeq  \frac{1.29802}{(1+x) \sqrt{x (2 + x)} }\,, &(x \gtrsim 0.17)\\
\left. I_2(x)/F_0\right|^{\rm disks}_{{\rm free\, fermion}}  &\simeq  \frac{2.43574 \left[\sqrt{x(2+x)}-2x(2+x)(1+x-\sqrt{x(2+x)})\right]}{x(1+x)(2+x)} \,,  &(x \gtrsim 0.2)
\end{align}
These appear plotted as the thick blue and red lines respectively in the disks plot in Fig.\,\ref{I22p} and, as we can see, they approximate the data points very well for arbitrary values of $x$ greater than (roughly) $0.2$. This is no coincidence. Indeed, as shown in \cite{Agon:2021zvp}, in the case of spherical entangling surfaces in arbitrary dimensions, the EMI model result is exactly equal to the contribution to the free-fermion mutual information produced by the conformal block corresponding to the leading replica primary operator ---namely, a conserved current $\bar \psi_i\gamma_{\mu}\psi_j$, where $i,j$ are replica indices. The EMI model does not contain any additional terms which do contribute to the free fermion mutual information, but such leading primary  block contribution is large enough to reproduce most of the full free-fermion result for most values of $x$. As a matter of fact, something analogous happens with the free scalar and the MEMI model, namely, the conformal block making the largest contribution to the free-scalar mutual information in the case of spheres corresponds exactly to the MEMI result in general dimensions ---in this case, the leading conformal block corresponds to a replica operator of the form $\phi_i\phi_j$. Again, such contribution, given by \req{fsss} in the three-dimensional case, approximates the full free-scalar result very precisely for sufficiently large values of $x$.

Near the touching points, we find the following fits for the scalar and the fermion, respectively,
\begin{align}\label{i2x0}
\left. I_2(x)/F_0\right|_{{\rm free\, scalar}}^{\rm disks}  &\overset{x\rightarrow 0}{\simeq} + 1.3125/\sqrt{x} -1.3522+ 0.41962 \sqrt{x} \,, \\ \label{i2x0f}
\left. I_2(x)/F_0\right|_{{\rm free\, fermion}}^{\rm disks}  &\overset{x\rightarrow 0}{\simeq}+ 0.95599/\sqrt{x} -1.7427+ 0.83507 \sqrt{x} \,.
\end{align}
These appear represented as solid lines in the plot. Computing lattice data points near $x\rightarrow 0$ is increasingly challenging and so the values of  $\gamma/F_0|_{\rm free\, scalar}$ and $\gamma/F_0|_{\rm free\, fermion}$ reported here should be taken with a grain of salt. They should be improved by additional calculations near the touching region.  

We observe that the curves for both models are monotonically decreasing as a function of the separation for all sets of regions and that the scalar results lie above the fermion ones for all regions and separations, a trend which persists as we make the entangling regions thinner. It seems then reasonable to conjecture that
\begin{equation}
\left.I_2(A,B)/F_0\right|_{{\rm free\, scalar}}  > \left.I_2(A,B)/F_0\right|_{{\rm free\, fermion}}\, ,
\end{equation}
for arbitrary pairs of regions. Similar hierarchies ---with the scalar systematically producing greater results--- for  EE universal coefficients in three dimensions have been previously observed \eg in \cite{Bueno:2015rda,Bueno:2015ofa,Bueno:2021fxb}.

\section{$N$-partite information in holography}\label{holographysec}
In this section, we would like to reinterpret our analytic results of Section \ref{Npar} from the holographic perspective. 
According to the proposal presented in \cite{Faulkner:2013ana}, the entanglement entropy of an arbitrary region $A$ in a holographic theory dual to Einstein gravity can be obtained from the formula 
\bea{\label{FLM}}
S(A)=\underset{m\sim A}{\rm min}{\frac{{\rm area}(m(A))}{4G_N}}+S_b(A^b)\, ,
\eea
which includes the usual Ryu-Takayanagi term \cite{Ryu:2006bv} plus the leading quantum correction.
In this expression, $m(A)$ is the minimal area surface homologous to $A$, $A^b$ is the bulk homology region with boundary $\partial A^b=A\cup m(A)$ and $S_b(A^b)$ represents its associated bulk von Neumann entropy. 

For a boundary region made of disconnected subregions, $A=\cup_i^K A_i$, it is known that, for sufficiently large mutual separations,
 the corresponding minimal-area surface becomes disconnected and equal to the union of the individual extremal surfaces associated to the subregions 
\bea
m(\cup_{i=1}^K A_i)=\cup_{i=1}^Km(A_i)\,.
\eea
Thus, at long distances the minimal-area contribution to the holographic $N$-partite information cancels out from its definition and should be given entirely by the $N$-partite information of the bulk dual theory, namely, 
\bea
I_N(\{A_i\}_N)=I^b_{N}(\{A^b_i\}_N)\,.
\eea
In this section we will prove the above equality for largely separated boundary spherical regions and its associated bulk hemispheres. We will achieve this by employing the same techniques developed in Section \ref{Npar}  to the bulk theory in the AdS background. Our result generalizes the analogous one obtained in \cite{Agon:2015ftl} for the case of the holographic mutual information.

Our main observations are the following. First, following the arguments presented in \cite{Agon:2015ftl}, we conclude that, at large separations, the twist operator associated to a bulk region can be approximated as
\bea
\tilde{\Sigma}^{(n)}_{A^b}\approx \sum_{i<j}C^b_{ij} \phi^{i}(r^b_{A^b}) \phi^{j}(r^b_{A^b})\, ,
\eea
in the multicopy bulk theory, where the bulk OPE coefficients can be computed via 
\bea\label{Bulk-OPE-C-coeffs}
C^b_{ij}=\lim_{|x-x_A|\to \infty} G(r^b,r^b_{A^b})^{-2} \langle \tilde{\Sigma}^{(n)}_{A^b} \phi^{i}(r^b) \phi^{j}(r^b) \rangle \,,
\eea
where $r^b$ are bulk Poincar\'e AdS coordinates so that $r^b=(x,z)$, $x$ being the flat space boundary coordinates.  $G(r,r_{A^b})$ is the two-point function of the bulk field $\phi(r^b)$ dual to the operator ${ \cal O}(x)$. In particular, for large separations, namely, for $|x-x_A|\gg R_A$, we have 
\bea\label{2pf-bulk}
G(r^b,r^b_{A^b})\approx \alpha^2_\Delta \frac{z^\Delta  z_{A^b}^\Delta}{|x-x_A|^{2\Delta}}\, ,
\eea
where the coefficient $\alpha_\Delta$ comes from a normalization relation between bulk and boundary two-point functions \cite{Freedman:1998tz}. Indeed, we could arrive at the above formula from the bulk two-point function and the extrapolate dictionary \cite{Harlow:2018fse}, this is 
\bea \label{extra-dict-scalar}
\phi(x,z)\approx \alpha_\Delta z^\Delta {\cal O}(x),
\eea
which is the leading term in the near boundary expansion of the dual bulk operator. 

The above statement generalizes to arbitrary $N$-point correlation functions. Namely, the bulk $N$-point correlation function can be obtained using (\ref{extra-dict-scalar}) for each bulk operator  
\bea\label{bulk-N-correlator}
\langle \phi(x_1,z_1) \cdots \phi(x_N,z_N)\rangle \approx \alpha^N_\Delta z_1^\Delta \cdots z_N^\Delta \langle {\cal O}(x_1) \cdots {\cal O}(x_N)\rangle \, ,
\eea
provided all distances $|x_i-x_j|\gg |z_k|$ for all $\{i,j,k\}$. This is precisely the regime we are considering when computing correlation functions of bulk twist operators. The relation (\ref{bulk-N-correlator}) holds trivially for bulk replica operators as well with the obvious generalization.

In our derivation of the operator formula for $I_N(\{A_i\}_N)$ in terms of twist operators we did not make any assumption about the theory being conformal or being defined on flat space. Hence, the same formula should apply to $I^b_{N}(\{A^b_i\}_N)$, this is  
\bea\label{IN-bulk}
I^b_{N}(\{A^b_i\}_N)=\lim_{n\to 1}\frac{\(-\)^{N+1}}{1-n}\langle \tilde{\Sigma}_{A^b_1}^{(n)}\tilde{\Sigma}_{A^b_2}^{(n)} \cdots \tilde{\Sigma}_{A^b_N}^{(n)}\rangle\,.
\eea
The same happens with our derivation of the twist OPE coefficient $C_{ij}$ of Section \ref{Cij-coeffs}, more precisely, Eq. (\ref{twist-correlator-3}).  Thus, for the bulk twist operator we can similarly write 
\bea \label{twist-correlator-bulk}
\langle \Omega |\Sigma^{(n)}_{A^b} {\phi}^{l}(r^b) {\phi}^{k}(r^b)|\Omega\rangle\approx \Big\langle{\phi}_{A^b}\[r^b,i\frac{\tau_{kl}}{\pi}\]{ \phi}(r^b)  \Big\rangle \,,
\eea
where ${\phi}_{A^b}\[r^b,s\]$ is the bulk field operator evolved by the bulk modular flow associated to region $A^b$, this is
\bea
{\phi}_{A^b}\[r^b,s\]=\rho^{is}_{A^b} \phi(r^b)\rho^{-is}_{A_b}\,, \quad{\rm and}\quad \tau_{kl}= \frac{\pi (k-l)}{n}\,.
\eea
It turns out that the bulk modular flow for the region dual to a spherical region is also explicitly known.  The CFT modular flow for a spherical region is generated by the conformal Killing vector which preserves the associated causal cone, and it is given by (\ref{sphere-modular}).
Such Killing vector has a natural extension into the bulk where it becomes the AdS  
Killing vector field which preserves the associated dual bulk hemispherical causal cone. This flow is given by  \cite{Casini:2011kv}
\bea\label{sphere-modular-bulk}
r_b^{\pm}[s]=R\frac{\(R+r_b^{\pm}\)-e^{\mp 2\pi s} \(R-r^{\pm}\) }{\(R+r^{\pm}\)+e^{\mp 2\pi s} \(R-r_b^{\pm}\)}\, ,
\eea
where $r_b^\pm=r^b\pm t$ are standard null coordinates. Notice that this flow reduces to the boundary flow for $z=0$.  
The map generated by the bulk modular flow acts as a diffeomorphism of the AdS metric and thus the action on the bulk field is trivial, this is 
\bea
{\phi}_{A^b}\[r^b,i\frac{\tau_{kl}}{\pi}\]=\phi\(r^b\[i\frac{\tau_{kl}}{\pi}\]\)\, ,
\eea
since $\phi(r^b)$ transforms as a bulk scalar field. Using the above relation into (\ref{twist-correlator-bulk}), together with the extrapolate dictionary (\ref{extra-dict-scalar}) and the explicit modular flow (\ref{sphere-modular-bulk}), we find 
\bea
\langle \Omega |\Sigma^{(n)}_{A^b} {\phi}^{l}(r^b) {\phi}^{k}(r^b)|\Omega\rangle\approx \alpha_\Delta^2 \frac{\Omega^\Delta (x,i\frac{\tau_{kl}}\pi)z^{2\Delta}}{|x\[i\frac{\tau_{kl}}{\pi}\]-x|^{2\Delta}}\, ,
\eea
for large $|x-x_A|$. Plugging this formula back into (\ref{Bulk-OPE-C-coeffs}) and using (\ref{2pf-bulk}) we arrive at our final formula for the bulk twist OPE coefficients 
\bea
C^b_{lk}=\frac{1}{\alpha_\Delta^2 z_{A^b}^2} \lim_{|x-x_A|\to \infty}\frac{\Omega^\Delta (x,i\frac{\tau_{kl}}\pi)}{|x\[i\frac{\tau_{kl}}{\pi}\]-x|^{2\Delta}}=\frac{C_{lk}}{\alpha_\Delta^2 z_{A^b}^2}\, .
\eea
This is the main result of this section. It gives a simple relation between the OPE coefficients of the twist bulk operator and the ones of the boundary theory. Together with the extrapolate dictionary (\ref{extra-dict-scalar}) it allows us to conclude the equality between boundary and bulk twist operators
\bea
\tilde{\Sigma}^{(n)}_{A^b}\approx \tilde{\Sigma}^{(n)}_{A}\, ,
\eea 
which holds inside correlation functions with operator insertions far away from the locations of the twist operator. Therefore, at large separations we conclude that 
\bea
I^b_{N}(\{A^b_i\}_N)=I_N(\{A_i\}_N)\,.
\eea

\section{Final comments}
In this paper we have presented several new results involving the $N$-partite information of spacetime regions in conformal field theories. Our main results appear summarized in the introduction. Here, we close with a few comments regarding possible extensions.

As we have seen, our final formula for $I_4$ includes a coefficient which we left expressed in terms of a triple integral which we were not able to evaluate explicitly for general values of $\Delta$ ---see \req{c44:Splitting}. It would certainly be interesting to see it done.

On the other hand, it should be possible (although increasingly challenging) to obtain explicit formulas for the long-distance leading term of the $I_N$ for $N\geq 5$ in the case of spherical entangling regions. In each case, the formulas should involve a linear combination of products of $2$-,$3$-,$\dots$, $N$-point functions of the leading primary. In relation with the CFT reconstruction program mentioned above, obtaining information about subleading primaries requires going beyond leading order in the long-distance expansion. As a concrete question in this direction, it would be interesting to determine precisely how much information about a CFT one can reconstruct from the knowledge of the leading and first subleading terms in the mutual information long-distance expansion for spherical regions.

On a different front, it would be interesting to provide further evidence (or disprove) the two conjectures we have put forward in this paper involving free fields, namely: i) the fact that the free scalar $N$-partite information is positive-definite in general dimensions and for arbitrary regions; ii) the observation that, normalized by the disk EE universal coefficient, the mutual information of the three-dimensional free scalar is greater than the free fermion one for arbitrary regions. In relation with the latter, it would be interesting to explore the possible existence of similar hierarchies between theories beyond free fields and also beyond the mutual information case, \ie for $N\geq 3$.

In the context of holographic theories, it would be interesting to extend the arguments leading to the equality between boundary and bulk twist operators to general shapes and arbitrary distances. This could help us to understand further these mysterious objects, although, there is no guarantee that such equality holds in those cases. The former case might require further input on the bulk modular flow which might be possible in the bulk free field approximation. On the other hand, moving away from the long-distance regime for boundary spheres may be possible, given the universality of both, bulk and boundary modular flows. This second exploration is of special interest, since  understanding the behaviour of the $N-$partite information at moderate separation distances could elucidate the nature of the holographic phase transition of the Ryu-Takayanagi prescription.

\section*{Acknowledgements}

We thank Horacio Casini for useful discussions. The work of AVL is supported by the F.R.S.-FNRS Belgium through convention IISN 4.4514.08. The work of CA is supported by the Simons Foundation through \textit{It from Qubit: Simons Collaboration on Quantum Fields, Gravity, and Information.}


\appendix

\section{Coefficients for the $N$-partite information \label{app-coeffs}}
Throughout the paper, we are interested in computing sums of combinations of the following basic coefficient,
\bea
C_{ij}=\frac{1}{\sin^{2\Delta}\[\frac{\pi (i-j)}n\]} \, .
\eea
Following \cite{Chen:2017hbk, Chen:2016mya}, where similar sums over $C_{ij}$ were computed, we will make extensive use of the following representation for the sine function:
\bea\label{int-rep-sin-2}
\frac{1}{\sin^{\Delta}\(\frac{\theta}2\)}&=&\frac{2^{\Delta-2}}{\pi^2 \Gamma\(\Delta\)}\int_{-\infty}^\infty d q\,e^{-\frac{q}2}e^{\frac{q\, \theta}{2\pi}} \Gamma\(\frac{\Delta}{2}+i\frac{ q}{2\pi }\)\Gamma\(\frac{\Delta}{2}-i\frac{ q}{2\pi }\) \nonumber\\
&=&\frac{2^{\Delta-2}}{\pi^2 \Gamma\(\Delta\)}\int_{-\infty}^\infty d q\,e^{-\frac{q}2}e^{\frac{q\, \theta}{2\pi}} \, \Big|\Gamma\(\frac{\Delta}{2}+i\frac{ q}{2\pi }\)\Big|^2\,.
\eea
This formula can be written in a more compact form in terms of the beta function $B(x,y)$ due to its relation with the gamma function,
\bea\label{int-rep-Beta}
B(x,y)=\frac{\Gamma(x)\Gamma(y)}{\Gamma(x+y)}=\int_{0}^1  t^{x-1}(1-t)^{y-1} dt\,,
\eea
where in the last equality we present its integral definition. 
Thus, we can rewrite \req{int-rep-sin-2} as
\bea\label{int-rep-sin-3}
\frac{1}{\sin^{\Delta}\(\frac{\theta}2\)}=\frac{2^{\Delta-2}}{\pi^2} \int_{-\infty}^\infty d q\,e^{-\frac{q}2}e^{\frac{q\, \theta}{2\pi}} B\(\frac{\Delta}{2}+i\frac{ q}{2\pi }\,, \, \frac{\Delta}{2}-i\frac{ q}{2\pi }\)\,.
\eea
For the sake of simplicity, we introduce the following compact notation
\bea\label{compact-beta}
B_q(\Delta)\equiv B\(\Delta+i\frac{q}{2\pi},\Delta-i\frac{q}{2\pi}\)=\frac{\Big|\Gamma\(\Delta+i\frac{q}{2\pi}\)\Big|^2}{\Gamma\(2\Delta\)} \, ,
\eea
to represent the beta function of our interest. In terms of this function we have
\bea\label{int-rep-sin-3.5}
\frac{1}{\sin^{\Delta}\(\frac{\theta}2\)}=\frac{2^{\Delta-2}}{\pi^2} \int_{-\infty}^\infty d q\,e^{-\frac{q}2}e^{\frac{q\, \theta}{2\pi}} B_q\(\frac{\Delta}{2}\)\,,
\eea
where the integral representation of the beta function (\ref{compact-beta}) from  \req{int-rep-Beta} takes the simple form
\bea\label{compact-beta-2}
B_q(\Delta)=\int_{0}^1 dt\, t^{\Delta-1}(1-t)^{\Delta-1} e^{\frac{iq}{2\pi} \log\(\frac{t}{1-t}\)}\,.
\eea
We will make use of this notation in what follows.

\subsection{Computation of $I_2$ coefficients \label{App-mutual}}
We are interested in computing the sum 
\bea
c_{2:2}^{(2)}\equiv \lim_{n\to 1}\frac{1}{n-1}\sum_{i<j}C^2_{ij}=\lim_{n\to 1}\frac{n}{2(n-1)} \sum_{j=1}^{n-1}\frac{1}{\sin^{4\Delta}\(\frac{\pi j}n\)} \, .
\eea
For that purpose, we use the representation (\ref{int-rep-sin-3.5}) with $\theta\to 2\pi j/n$ in the above sum, which leads to
\bea
c_{2:2}^{(2)}=\lim_{n\to 1}\frac{1}{n-1} \sum_{j=1}^{n-1}
 \frac{2^{4\Delta-3}}{\pi^2}\int_{-\infty}^\infty d q\,e^{-\frac{q}2}e^{\frac{q j}{n}}  B_q\(2\Delta \) \, .
\eea
We can carry out the sum over $j$ explicitly using
\bea
 \sum_{j=1}^{n-1} e^{\frac{q j}n}=\frac{e^{q/n}-e^q}{1-e^{q/n}}\approx (n-1)\frac{q e^{q}}{e^q-1} \, ,
\eea
which cancels out the $(n-1)$ in the $n \to 1$ limit and leads to
\bea\label{C2-1}
c_{2:2}^{(2)}=\frac{2^{4\Delta-3}}{\pi^2}\int_{0}^1 dt\, t^{2\Delta-1}(1-t)^{2\Delta-1}\int_{-\infty}^\infty d q\,\frac{q e^{\frac{q}{2}}}{e^q-1} e^{\frac{iq}{2\pi} \log\(\frac{t}{1-t}\)}\,,
\eea
after plugging in the integral representation (\ref{compact-beta-2}). We can now perform the integral over $q$, which is just the Fourier transform presented in \req{Fourier-1} with $k\to [\log t-\log(1-t)]/2\pi$. This results in:
\bea\label{Fourier-1-1}
\int_{-\infty}^\infty d q\,\frac{q e^{\frac{q}{2}}}{e^q-1}e^{\frac{iq}{2\pi} \log\(\frac{t}{1-t}\)}=4\pi^2 t(1-t)\,.
\eea
Plugging this result back into (\ref{C2-1}) leads to the well-known expression for $c_{2:2}^{(2)}$,
\bea\label{C2:2(2)}
c_{2:2}^{(2)}=2^{4\Delta-1}\int_{0}^1 dt\, t^{2\Delta}(1-t)^{2\Delta}=2^{4\Delta-1}\frac{\Gamma\(2\Delta+1\)^2}{\Gamma\(4\Delta+2\)}\,.
\eea

\subsection{Computation of $I_3$ coefficients \label{App-tripartite}}

We are interested in computing the coefficients $c^{(3)}_{3:2}$ and $c^{(1,1,1)}_{3:3}$ which appear in the expression for the tripartite information at long distances. We start with
\bea
c^{(3)}_{3:2}&\equiv& \lim_{n\to 1}\frac{1}{n-1}\sum_{i<j}C^3_{ij} \, .
\eea
This computation is identical to the one of the coefficient $c^{(2)}_{2:2}$ which determines $I_2$ at long distances with the simple replacement $\Delta\to 3\Delta/2$, therefore, we conclude that
\bea \label{C3:2(3)}
c^{(3)}_{3:2}= \frac{2^{6\Delta - 1}\Gamma\(3\Delta+1\)^2}{\Gamma\(6\Delta+2\)}\,.
\eea

Our second coefficient of interest is
\bea
\label{intC3:SymmetryManipulation}
c^{(1,1,1)}_{3:3}&\equiv& \lim_{n\to 1}\frac{1}{n-1}\sum_{i<j<k}C_{ij}C_{jk}C_{ki}= \lim_{n\to 1}\frac{1}{3!(n-1)}\sum_{i\neq j\neq k}C_{ij}C_{jk}C_{ki}\nonumber\\
&=& \lim_{n\to 1}\frac{n}{3 (n-1)}\sum_{0< j< k}C_{0j}C_{jk}C_{k0}=\lim_{n\to 1}\frac{n}{3(n-1)} \sum_{k=2}^{n-1}\sum_{j=1}^{k-1}C_{0j}C_{jk}C_{k0} \, .
\eea
In the first line, we rewrite the ordered sum in a disordered form by including the permutation factor that accounts for the multiple counting of each configuration. In the second line, we fix one index to the particular value of zero and multiply by $n$ the result (making use of the replica symmetry). Furthermore, we reorder the resulting sum, which requires an extra permutation factor of $2!$. Plugging in the explicit expression for the coefficients, we get
\bea
c^{(1,1,1)}_{3:3}=\lim_{n\to 1}\frac{n}{3(n-1)} \sum_{k=2}^{n-1}\frac{1}{\sin^{2\Delta}\(\frac{\pi k}n\)}\sum_{j=1}^{k-1}\frac{1}{\sin^{2\Delta}\(\frac{\pi (k-j)}n\)}\frac{1}{\sin^{2\Delta}\(\frac{\pi j}n\)}\,.
\eea
Next, we use the integral representation (\ref{int-rep-sin-3.5}) for each of the sine functions above, leading to
\begin{align}
\label{intC3:BeforeSum}
c^{(1,1,1)}_{3:3} = & \lim_{n\to 1}\frac{n}{3(n-1)} \[\frac{2^{2\Delta-2}}{\pi^2}\]^3 \nonumber \\
& \times \sum_{k=2}^{n-1}\, \sum_{j=1}^{k-1} \int_{-\infty}^\infty dp \int_{-\infty}^\infty dq \int_{-\infty}^\infty  ds\, e^{-\frac{p}2}\,e^{-\frac{q}2}e^{-\frac{s}2}e^{\frac{p k}{n}} e^{\frac{q j}{n}}e^{\frac{s (k-j)}{n}} B_p\(\Delta\)  B_q\(\Delta\) B_s\(\Delta\)\,.
\end{align}
In this form, we can explicitly carry out the sums over $j$ and $k$ and take the $n \to 1$ limit of the resulting expression. This gives
\bea
 e^{-\frac{p}2}e^{-\frac{q}2}e^{-\frac{s}2}\sum_{k=2}^{n-1}\sum_{j=1}^{k-1}e^{\frac{p k}{n}}e^{\frac{q j}{n}}e^{\frac{s (k-j)}{n}} 
 \approx (n-1) f(p,q,s) \,,
 \eea
where
\bea\label{f-2}
f(p,q,s)\equiv -e^{\frac{p+q+s}2}\[\frac{p}{(e^{-p}-e^{q})(1-e^{p+s})}-\frac{q}{(e^{s}-e^{q})(1-e^{p+q})}+\frac{s}{(e^{s}-e^{q})(1-e^{p+s})}\] \,.
\eea
This function is to be introduced inside \req{intC3:BeforeSum}. We can therefore manipulate it to get a more symmetric form by changing $p\to -p$, obtaining
\bea
\tilde{f}(p,q,s)=e^{\frac{p+q+s}2}\[\frac{p}{(e^p-e^s)(e^p-e^q)}+\frac{q}{(e^q-e^s)(e^q-e^p)}+ \frac{s}{(e^s-e^q)(e^s-e^p)}\]\,.
\eea
The resulting function $\tilde{f}(p,q,s)$ is symmetric under permutations among $\{p,q,s\}$. We could deduce from here that inside the integral each term will contribute the same and thus it should be enough to focus on evaluating any of the three terms and multiplying the answer by three. However, this is not true because each term individually is not well defined due to spurious divergences at coincident points of the $\{p, q,s\}$ variables. Such poles would cancel in the full expression, the following form of the function makes it manifest
\bea\label{tilde-f-2}
\tilde{f}(p,q,s)= \frac{e^{\frac{p+q+s}2}}{(e^p-e^q)}\[\frac{p-s}{e^p-e^s}-\frac{q-s}{e^q-e^s}\] \, .
\eea

Thus, the integral of interest is
\bea\label{C3-4}
c^{(1,1,1)}_{3:3}&=&\frac{2^{6(\Delta-1)}}{3\pi^6 }\int_{-\infty}^\infty  \, dp \, e^{\frac p2}B_p\(\Delta\)\int_{-\infty}^\infty  \,dq \, e^{\frac q2}B_q\(\Delta\) \frac{1}{e^p-e^q}
\nonumber \\
&&\times  \int_{-\infty}^\infty \, ds \, e^{\frac s2}\,B_s\(\Delta\) \(\frac{p-s}{e^p-e^s}-\frac{q-s}{e^q-e^s}\) \, .
\eea
We start by doing the integral over $s$ of the $p$ dependent piece; the one that depends on $q$ is identical. We have
\bea\label{Int-s-1}
\int_{-\infty}^\infty  ds \, e^{\frac s2}\,B_s\(\Delta\)\frac{p-s}{e^p-e^s}=e^{-\frac{p}2}\,\int_{-\infty}^\infty ds \,\frac{ (s-p) \, e^{\frac{s-p}2}}{e^{s-p}-1} B_s\(\Delta\)\,=4\pi^2 e^{-\frac{p}2} B_p\(\Delta+1\)\, ,
\eea
where in the last equality we used \req{Int-linear-gamma2}. All in all, we obtain
\bea
c^{(1,1,1)}_{3:3}=\frac{2^{6\Delta}}{3\(2\pi\)^4}\int_{-\infty}^\infty  \, dp \int_{-\infty}^\infty  \,dq \,  B_p\(\Delta\) B_q\(\Delta\) \(\frac{e^{\frac{p}2} B_q\(\Delta+1\)-e^{\frac{q}2} B_p\(\Delta+1\)}{e^{q}-e^p}\) \, .
\eea
Once again, the resulting integral has a simple spurious pole if we split the terms which does not appear in the full expression. Nevertheless, for computational purposes we need to treat each term independently. We can do so with \req{int-simple-gamma2}, obtaining
\bea\label{C3-int}
c^{(1,1,1)}_{3:3}=\frac{2^{6\Delta}}{3\(2\pi\)^4 \Delta }\int_{-\infty}^\infty  dq\, q\, e^{-\frac q2} \, B^2_q\(\Delta\) B_q\(\Delta+1\) \, .
\eea
This final integral can also be carried out using \req{Int-Gamma-C43-appendix}, which leads to our final formula
\bea\label{C3:3(1,1,1)}
c^{(1,1,1)}_{3:3}=-\frac{2^{6 \Delta} \Gamma\(\Delta+\frac12\)^3}{12 \pi \Gamma\( 3\Delta+\frac32\)}\,.
\eea
We have been very explicit in the manipulations leading to the final form of this coefficient. We will be more sketchy from now on; the general strategy is always the same: replacing the inverse powers of the sine by the integral representation \eqref{int-rep-sin-3.5}, summing and extracting the piece that contributes in the $n \to 1$ limit, and then manipulating the resulting expression to be able to integrate it.
 
\subsection{Computation of $I_4$ coefficients} 
The first coefficient appearing in \req{Final-I4} is straightforward to obtain, as it is given by the single sum
\bea
c_{4:2}^{(4)}=\lim_{n\to 1}\frac{1}{n-1}\sum_{j<k}C^4_{ij}\,.
\eea
This can be written as a single sum using the symmetry under $i \leftrightarrow j$ and the ciclicity of the replica geometry,
\bea
c_{4:2}^{(4)}=\lim_{n\to 1}\frac{n}{2(n-1)}\sum_{j=1}^{n-1}C^4_{0j}\,.
\eea
The computation is now analogous to the one of $c_{2:2}^{(2)}$ which produced \req{C2:2(2)}, the only difference being the power of $C_{0j}$ in the sum. This is accounted for by changing $\Delta \to 2\Delta $, producing
\bea\label{C42:4}
c_{4:2}^{(4)}= 2^{8\Delta-1} \frac{\Gamma\(4\Delta+1\)^2}{\Gamma\(8\Delta+2\)}\,.
\eea

The remaining coefficients in \req{Final-I4} require more work, so before getting into the details of the computations, let us write them in a convenient form. Using arguments analogous to \eqref{intC3:SymmetryManipulation} based on replica symmetry and the symmetries of the expression being summed, we can write them as
\begin{align}
\label{C4:3:1,1,2-2}
& c_{4:3}^{(1,1,2)}=\lim_{n\to 1}\frac{n}{3(n-1)}\sum_{k=2}^{n-1}\sum_{j=1}^{k-1}\[C^2_{0j}C_{jk}C_{k0}+C_{0j}C^2_{jk}C_{k0}+C_{0j}C_{jk}C^2_{k0}\] \, , \\
\label{C4:3:2,2-2}
& c_{4:3}^{(2,2)}=\lim_{n\to 1}\frac{n}{3(n-1)}\sum_{k=2}^{n-1}\sum_{j=1}^{k-1}\[C^2_{0j}C^2_{0k}+C^2_{0j}C^2_{jk}+C^2_{0k}C^2_{jk}\] \,, \\
\label{C4:4:1,1,1,1-2}
& c_{4:4}^{(1,1,1,1)}=\lim_{n\to 1}\frac{n}{4(n-1)}\sum_{l=3}^{n-1}\sum_{k=2}^{l-1}\sum_{j=1}^{k-1}\[C_{0j}C_{jk}C_{kl}C_{l0}+C_{0j}C_{jl}C_{lk}C_{k0}+C_{0l}C_{lj}C_{jk}C_{k0}\] \, , \\
\label{C4:4:2,2-2}
& c_{4:4}^{(2,2)}=\lim_{n\to 1}\frac{n}{4(n-1)}\sum_{l=3}^{n-1}\sum_{k=2}^{l-1}\sum_{j=1}^{k-1}\[C^2_{0l}C^2_{jk}+C^2_{0j}C^2_{kl}+C^2_{0k}C^2_{jl}\] \,.
\end{align}

\subsubsection{Computation of $c_{4:3}^{(1,1,2)}$}

Replace in \req{C4:3:1,1,2-2} the integral representation for each coefficien. Summing and extracting the relevant piece in the $n \to 1$ limit, we get
\bea
e^{-\frac p2} e^{-\frac q2} e^{-\frac s2} \sum_{k=2}^{n-1}\sum_{j=1}^{k-1}
\(e^{\frac{p j}{n}} e^{\frac{q(k-j)}{n}} e^{\frac{s k}{n}}+e^{\frac{s j}{n}} e^{\frac{p(k-j)}{n}}e^{\frac{q k}{n}} +e^{\frac{s(k-j)}{n}} e^{\frac{q j}{n}} e^{\frac{p k}{n}} \) \approx  (n-1)f^{(1,1,2)}_{4:3}(p,q) \, ,
\eea
where 
\begin{align}
\label{f43:112}
f^{(1,1,2)}_{4:3}(p,q,s) = &\frac{e^{\frac{p+q+s}2}}{(e^q-e^p)}\[\frac{q+s}{e^{q+s}-1}-\frac{p+s}{e^{p+s}-1}\]+ \frac{e^{\frac{p+q+s}2}}{(e^s-e^p)}\[\frac{s+q}{e^{s+q}-1}-\frac{p+q}{e^{p+q}-1}\] \nonumber \\ &+\frac{e^{\frac{p+q+s}2}}{(e^s-e^q)}\[\frac{s+p}{e^{s+p}-1}-\frac{q+p}{e^{q+p}-1}\] \, .
\end{align}
Therefore, we are left with the following triple integral to evaluate
\bea
c_{4:3}^{(1,1,2)}=\frac{2^{8\Delta-6}}{3\, \pi^6}\int_{-\infty}^\infty d p\,\int_{-\infty}^\infty d q\,\int_{-\infty}^\infty d s\, f^{(1,1,2)}_{4:3}(p,q,s)\, B_p(2\Delta)B_q(\Delta)B_s(\Delta) \, .
\eea 
The form of $f^{(1,1,2)}_{4:3}(p,q,s)$ allows to integrate each piece by means of the same tricks than before. Integrals \eqref{Int-linear-gamma2} and \eqref{int-simple-gamma2} reduce the computation to the following single integral
\begin{align}\label{C43:112}
c_{4:3}^{(1,1,2)}= & \frac{2^{8\Delta-4}}{2\pi^4 \Delta\,} \int_{-\infty}^\infty d q\,q\, e^{-\frac q2}\, B_q(2\Delta)B_q(\Delta)B_q(\Delta+1) \nonumber \\
& +\frac{2^{8\Delta-4}}{3\pi^4 \Delta\,} \int_{-\infty}^\infty d q\,q\, e^{-\frac q2}\,  B^2_q(\Delta)B_q(2\Delta+1)\,.
\end{align}
We can now shift the argument of one of the beta functions of the first term by means of basic gamma function identities to obtain:
\bea
c_{4:3}^{(1,1,2)}&=&\frac{2^{8\Delta-3}(7\Delta+2)}{3\pi^4 \Delta\, \(2\Delta+1\)}  \int_{-\infty}^\infty d q\,q\, e^{-\frac q2} B^2_q\(\Delta\)B_q\(2\Delta+1\)\nonumber\\
&& -\frac{3\times 2^{8\Delta-6}}{\pi^4 \(2\Delta+1\)}  \int_{-\infty}^\infty d q\,q\, e^{-\frac q2} B^2_q\(\Delta\) B_q\(2\Delta\)\, .
\eea
Using the result \eqref{Int-Gamma-C43-appendix} from Appendix \ref{App-integrals}, we get
\bea\label{C4:3(112)}
c_{4:3}^{(1,1,2)}=-\frac{9\sqrt{\pi} \Gamma\(3\Delta\)^2}{4\Gamma\(\Delta\)^2 \Gamma\(4\Delta+\frac32\)}=-\frac{9\times 2^{8\Delta}\,\Gamma\(3\Delta\)^2
 \Gamma \(4\Delta+1\)}{2\,\Gamma\(\Delta\)^2\Gamma\(8\Delta+2\)} \, .
\eea

\subsubsection{Computation of $c_{4:3}^{(2,2)}$}

We start now from \req{C4:3:2,2-2} and use the integral representation of the coefficients. We carry out the sum over the indices, obtaining
\bea
e^{-\frac p2} e^{-\frac q2} \sum_{k=2}^{n-1}\sum_{j=1}^{k-1}
\(e^{\frac{p l}{n}} e^{\frac{qk}{n}}+e^{\frac{p j}{n}} e^{\frac{q(k-j)}{n}}+e^{\frac{p k}{n}} e^{\frac{q(k-j)}{n}}\) \approx  (n-1)f^{(2,2)}_{4:3}(p,q) \, ,
\eea
where 
\bea
\label{f43:22}
f^{(2,2)}_{4:3}(p,q)&= &\frac{e^{\frac{p+q}2}}{(e^p-1)}\[\frac{q+p}{e^{q+p}-1}-\frac{q}{e^{q}-1}\]+ \,\frac{e^{\frac{p+q}2}}{(e^q-e^p)}\[\frac{q}{e^{q}-1}-\frac{p}{e^{p}-1}\] \nonumber \\ &&+\,\frac{e^{\frac{p+q}2}}{(e^{p+q}-1)}\[\frac{q}{e^{q}-1}-\frac{p\, e^p}{e^{p}-1}\]\,.
\eea
We have an integral expression for our coefficient of the form
\bea
c_{4:3}^{(2,2)}=\frac{2^{8\Delta-4}}{3\pi^4}\int_{-\infty}^\infty d p\,\int_{-\infty}^\infty d q\, f^{(2,2)}_{4:3}(p,q)\,B_p(2\Delta)B_q(2\Delta) \, .
\eea
Symmetrizing the integrands in $q$ we can obtain
\bea
c_{4:3}^{(2,2)}=-\frac{2^{8\Delta-4}}{6\pi^4 \Delta  }\int_{-\infty}^\infty d q\, q^2 \, B^2_q(2\Delta)-\frac{2^{8\Delta-2}}{6 \pi^2}\int_{-\infty}^\infty d q\,\,B_q(2\Delta)B_q(2\Delta+1) \, ,
\eea
and from here, by shifting the argument of the last beta function, we can use \req{Int-Gamma-C43-appendix} to conclude:
\bea\label{C4:3(22)}
c_{4:3}^{(2,2)}=-\frac{2^{8\Delta}\[\Gamma\(4\Delta+1\)\]^2}{2\,  \Gamma\(8\Delta+2\)}\,.
\eea

\subsubsection{Computation of $c_{4:4}^{(1,1,1,1)}$}

Doing the sum like in the previous cases, we obtain:
\bea
\label{c44:original}
c_{4:4}^{(1,1,1,1)}=\frac{2^{8\Delta-8}}{4\, \pi^8 } \int_{-\infty}^\infty d p \int_{-\infty}^\infty d q \int_{-\infty}^\infty d r \int_{-\infty}^\infty d s \, f^{(1,1,1,1)}_{4:4}(p,q,r,s)B_p(\Delta)B_q(\Delta)B_r(\Delta)B_s(\Delta) \, \nonumber ,\\ 
\eea
where
\bea
f_{4:4}^{(1,1,1,1)}(p,q,r,s)&=&e^{\frac{p+q+r+s}2}\Bigg[ \frac{(p+s)}{(e^{p+s}-1)(e^p-e^r)(e^p-e^q)} +\frac{(q+s)}{(e^{q+s}-1)(e^q-e^r)(e^q-e^p)} \nonumber\\
&& \qquad \qquad \qquad  +\frac{(r+s)}{(e^{r+s}-1)(e^r-e^q)(e^r-e^p)} \Bigg] \nonumber \\
&+&e^{\frac{p+q+r+s}2}\Bigg[ \frac{e^{q}(p+r-q-s)}{(e^{p+r}-e^{q+s})(e^{p+r}-1)(e^p-e^q)} +\frac{e^{q+r}(q+s)}{(e^{q+s}-1)(e^{p+r}-1)(e^{q+r}-1)} \nonumber\\
&& \qquad \qquad \qquad  -\frac{(r-s)}{(e^{r}-e^s)(e^{q+r}-1)(e^p-e^q)} \Bigg] \nonumber\\
&+&e^{\frac{p+q+r+s}2}\Bigg[ \frac{e^{q}(p+s)}{(e^{p+s}-1)(e^{q+r}-1)(e^q-e^p)} +\frac{e^{p}(q+s)}{(e^{q+s}-1)(e^{p+r}-1)(e^{p}-e^{q})} \nonumber\\
&& \qquad \qquad \qquad  +\frac{(r+s+p+q)}{(e^{r+s+p+q}-1)(e^{q+r}-1)(e^{p+r}-1)} \Bigg] \, .
\eea
It is immediate to integrate over $s$ by means of \req{Int-linear-gamma2}. After some relabelling of the variables we arrive at the following expression
\bea
c_{4:4}^{(1,1,1,1)}\!\!\! &=&\!\!\! \frac{2^{8\Delta-8}}{ \pi^6 }\int_{-\infty}^\infty d p\,\int_{-\infty}^\infty d q\,\int_{-\infty}^\infty d r\, B_p(\Delta)B_q(\Delta)B_r(\Delta)\nonumber \\
&&\!\!\!\! 
\times  \Bigg[ \,  \frac{3\,e^{\frac{q+r}2}B_p(\Delta+1) }{(e^p-e^r)(e^p-e^q)}+ \frac{\( e^{q+r}+1\)\,e^{\frac{p+r}2} B_q(\Delta+1)}{(e^{q+r}-1)(e^{p+r}-1)} +\frac{e^q B_{q-p-r}(\Delta+1) }{(e^{p+r}-1)(e^p-e^q)} \nonumber \\ 
&&\!\!\!\!\qquad +\frac{B_{q+p+r}(\Delta+1)}{(e^{p+r}-1)(e^{q+r}-1)}+\frac{2\,e^{p+q}\,e^{\frac{p+r}2} B_q(\Delta+1)}{(e^{p+q}-1)(e^{p+r}-1)} \Bigg] \, .
\eea
We can write the $B_{q-p-r}(\Delta+1)$ as $B_{p+q+r}(\Delta +1)$ by sending $q \to -q$ and using the symmetry of the beta function $B_{-s}(\Delta) = B_s(\Delta)$. That term then combines with the one in which $B_{p+q+r}(\Delta +1)$ is already present, after some trivial relabelling. Similarly, we can replace
\bea
\frac{\( e^{q+r}+1\)\,e^{\frac{p+r}2} B_q(\Delta+1)}{(e^{q+r}-1)(e^{p+r}-1)}  +\frac{2\,e^{p+q}\,e^{\frac{p+r}2} B_q(\Delta+1)}{(e^{p+q}-1)(e^{p+r}-1)}\to \frac{3\,e^{\frac{p+r}2}B_q(\Delta+1)}{(e^{p+r}-1)}+\frac{4\,e^{\frac{p+r}2} B_q(\Delta+1)}{(e^{p+r}-1)(e^{q+r}-1)}\,. \nonumber
\eea
The first term is odd under $(p,r) \to (-p, -r)$, while the prefactor in the whole integral is even due to the symmetry of the beta functions. Therefore, it will not contribute and only the second term is relevant. Let us recap and split the integrals in a suitable way for further manipulation,
\begin{equation}
\label{c44:Splitting}
c_{4:4}^{(1,1,1,1)} = J_1 +J_2 + J_3 \, ,
\end{equation}
where, explicitly indicating the $\Delta$ dependence
\begin{align}
\label{c44:SplittingTerms}
J_1(\Delta) = & \frac{2^{8\Delta-8}}{ \pi^6 }\int_{-\infty}^\infty d p\,\int_{-\infty}^\infty d q\,\int_{-\infty}^\infty d r\,B_p(\Delta)B_q(\Delta)B_r(\Delta) \frac{2 B_{p+q+r}\(\Delta+1\)}{(e^{p+r}-1)(e^{p+q}-1)} \, , \\
J_2(\Delta) = &  \frac{2^{8\Delta-8}}{ \pi^6 }\int_{-\infty}^\infty d p\,\int_{-\infty}^\infty d q\,\int_{-\infty}^\infty d r\,  B_p(\Delta)B_q(\Delta)B_r(\Delta) \frac{3\,e^{\frac{q+r}2}B_p(\Delta+1) }{(e^p-e^r)(e^p-e^q)} \, , \\
J_3(\Delta) = & \frac{2^{8\Delta-8}}{ \pi^6 }\int_{-\infty}^\infty d p\,\int_{-\infty}^\infty d q\,\int_{-\infty}^\infty d r\,  B_p(\Delta)B_q(\Delta)B_r(\Delta) \frac{4\,e^{\frac{p+r}2} B_q(\Delta+1)}{(e^{p+r}-1)(e^{q+r}-1)} \, .
\end{align}
Notice that the expression we have obtained for $c_{4:4}^{(1,1,1,1)}$, though simpler than \req{c44:original}, is ill-defined due to the appearance of spurious poles. These are not present in the original expression, and we must remove them in order to obtain a sensible answer. We do it by symmetrizing under the variables of integration for each of the three pieces independently.

For $J_1$, after sending $p \to -p$, replace inside the integral,
\bea
\frac{2 e^{2p} B_{q+r-p}\(\Delta+1\)}{(e^{p}-e^r)(e^{p}-e^q)} \to \frac{2}{3}\Bigg[  \frac{e^{2p}\,B_{q+r-p}\(\Delta+1\) }{(e^{p}-e^r)(e^{p}-e^q)} +\frac{e^{2r}\,B_{q+p-r}\(\Delta+1\) }{(e^{r}-e^p)(e^{r}-e^q)}+\frac{e^{2q}\,B_{p+r-q}\(\Delta+1\) }{(e^{q}-e^p)(e^{q}-e^r)}\Bigg]\,. \nonumber
\eea
If we now use in the last term the following identity
\bea\label{id-exps}
\frac1{(e^p-e^r)(e^p-e^q)}+\frac1{(e^q-e^p)(e^q-e^r)}+\frac1{(e^r-e^q)(e^r-e^p)}=0\,,
\eea
we get:
\begin{align}
\label{c44:I1Intermediate}
J_1(\Delta) = & \frac{2}{3}\frac{2^{8\Delta-8}}{ \pi^6 } \int_{-\infty}^\infty dp \int_{-\infty}^\infty dq \int_{-\infty}^\infty dr \, \frac{B_p(\Delta)B_q(\Delta)B_r(\Delta)}{e^p-e^r}  \times \\
& \times \left[ \frac{e^{2p}\,B_{q+r-p}\(\Delta+1\)- e^{2q}\,B_{p+r-q}\(\Delta+1\) }{e^{p}-e^q} - \frac{e^{2r}\,B_{q+p-r}\(\Delta+1\) -e^{2q}\,B_{p+r-q}\(\Delta+1\) }{e^{r}-e^q} \right] \, .  \nonumber 
\end{align}
This formula is pole free and consequently it is well defined. Unfortunately, we were not able to integrate it in a closed form analytically; not even by means of the manipulations we will describe for $J_2$ and $J_3$. The most we managed to do is obtaining numerical results for particular (integer or semi-integer) values of $\Delta$. We present them after the results for $J_2$ and $J_3$.

The same manipulations in $J_2$ ---symmetrizing and using \eqref{id-exps}--- transform the integrand into the following pole-free form
\begin{align}
\label{c44:I2Intermediate}
J_2(\Delta) = & \frac{2^{8\Delta-8}}{ \pi^6 }\int_{-\infty}^\infty dp \int_{-\infty}^\infty dq \int_{-\infty}^\infty dr \, B_p(\Delta)B_q(\Delta)B_r(\Delta) \frac{e^{\frac{p+q+r}2}}{e^p-e^q} \times \\
& \times \left[ \frac{e^{-\frac{p}2}B_p(\Delta+1)-e^{-\frac{r}2}B_r(\Delta+1)  }{e^p-e^r} - \frac{e^{-\frac{q}2}B_q(\Delta+1) -e^{-\frac{r}2}B_r(\Delta+1) }{e^q-e^r} \right] \nonumber \, ,
\end{align}
which, after integrating over $q$ in the first term and $p$ in the second one, we get
\begin{equation}
\label{c44:I2Intermediate2}
J_2(\Delta) = \frac{2^{8\Delta-8}}{ \pi^6 \Delta } \int_{-\infty}^\infty dp \int_{-\infty}^\infty dr \,  B^2_p(\Delta) B_r(\Delta) \, p \, e^{\frac r2}\(\frac{e^{-\frac{p}2}B_p(\Delta+1)-e^{-\frac{r}2}B_r(\Delta+1)  }{e^p-e^r} \) \, .
\end{equation}
This integral can be done for integer or semi-integer $\Delta$ through the following procedure. We will present the details for $\Delta = (2k+1)/2$ with $k \in \mathbb{N}$; analogous manipulations provide the results for $\Delta = k$. First of all, write the difference in the denominator as $\sinh ((p-r)/2)$ and use the symmetries of the integrand to obtain
\begin{align}
\label{c44:I2HalfIntegerIntermediate}
J_2\left(\frac{2k+1}{2}\right) = \frac{2^{8k-4}}{(2k+1) \pi^6} \int_{-\infty}^\infty dp & \int_{-\infty}^\infty dr \, \frac{p \, B_p\left(\frac{2k+1}{2}\right)^2 B_r \left(\frac{2k+1}{2}\right)}{\sinh\left(\frac{p-r}{2}\right)}  \\
& \times \left[ \cosh(p) B_p \left(\frac{2k+3}{2}\right) - \cosh \left( \frac{p+r}{2} \right) B_r \left(\frac{2k+3}{2}\right) \right] \, . \nonumber
\end{align}
Now, use the shift formula for the beta function \eqref{Int:ShiftBeta} to write all arguments as $(2k+1)/2$. The final step is the most involved one; we will illustrate it for the first term within brackets. The basic idea is that we can decouple the $p$ and $r$ integrals by introducing an extra integration variable $u$ with a $\delta$-function at $u = (p-r)/2$; so that we can write the denominator as $\sinh(u/2)$. Writing then the $\delta$-function by means of its integral Fourier representation, we get, dropping the global prefactor in the previous integral
\begin{align}
\int_{-\infty}^{\infty} \frac{d \xi}{2\pi} & \left[ \int_{-\infty}^{\infty} dp \, p \cosh(p) \frac{p^2 + \pi^2 (2 k + 1)^2}{4 \pi^2 (2k+1)(2k+2)} B_p \left(\frac{2k+1}{2}\right)^3 e^{- i \xi p}  \right] \nonumber \\
 & \times \left[ \int_{-\infty}^{\infty} dr \, B_r \left(\frac{2k+1}{2}\right) e^{ i \xi r} \right] \left[ \int_{-\infty}^{\infty} du \, \frac{e^{ i \xi u}}{\sinh(u/2)} \right] \, .
\end{align}
The integral over $r$ can be done directly by using the hyperbolic representation of the beta function in \eqref{Int:RepresentationsBeta}, while the one over $u$ is just \req{Fourier-2}. The only really challenging one is the integral over $p$. We can use the expansion \eqref{Int:PowerBetaHalfInteger} to write the integrand as some polynomial in $p$ times $\cosh(p)/\cosh(p/2)$. The powers of $p$ can be traded for derivatives $i \partial_{\xi}$ and taken outside of the integral, while using $\cosh(p) = 2 \cosh^2(p/2) - 1$ we can write the remaining integral as a combination of \req{Int:InvCosh} and \req{Int:InvCosh2and3}. All in all, the previous expression becomes
\begin{equation}
- \frac{\pi^{2(2-3k)}}{2^{8k+1} (2k)!^3 (2k+1)(2k+2)} \int_{-\infty}^{\infty} d\xi \, F_2^{(1)} (\xi, k) \frac{\tanh(\pi \xi)}{\cosh^{2k+1}(\pi \xi)} \, ,
\end{equation}
where $F_2^{(1)} (\xi, k)$ is the result of extracting powers of $p$ as we described
\begin{equation}
\label{c44:I2FunctionFirstTerm}
F_2^{(1)} (\xi, k) = \left[ \prod_{l=0}^{k-1} \D_{2l+1}^3 \right] \D_{2k+1} \partial_{\xi} \left( \frac{3 - 4\xi^2}{\cosh(\pi \xi)} \right) \, ,
\end{equation}
and $\D_n$ is the following differential operator:
\begin{equation}
\label{c44:DiffOperator}
\D_n \equiv - \partial_{\xi}^2 + n^2 \pi^2 \, .
\end{equation}
This expression can be evaluated for any non-negative integer $k$ algorithmically and, in particular, it is not difficult to introduce it in a standard symbolic algebra software. The remaining integral over $\xi$ can be done by means of the general formulas presented in Appendix \ref{App-integrals}, because it is just an inverse power of $\cosh(\pi \xi)$ times some complicated function of $\xi$ and hyperbolic functions $\cosh (\alpha \xi)$ or $\sinh(\alpha \xi)$.\footnote{For this one might need to express powers of hyperbolic functions in the numerator as hyperbolic functions with different argument, \emph{e.g.}, $4 \cosh^3(x) =  3 \cosh(x) + \cosh(3x)$.} Alternatively, for low enough $k$, Mathematica is able to do the resulting integral over $\xi$ directly. One can show, as an example, that for $k = 0$ ($\Delta = 1/2$) the previous integral evaluates to $4 \pi^4 +16 \pi^6/15$.

The second term in \eqref{c44:I2HalfIntegerIntermediate} can be treated analogously (after separating the $p$ and $r$ parts of the $\cosh$ by means of the identity for the $\cosh$ of a sum), although in this case both the $p$ and the $r$ integrals have to be done by means of the expansion \eqref{Int:PowerBetaHalfInteger}. The remaining manipulations are identical, so we will just write the final result,
\begin{align}
\label{c44:I2HalfIntegerFinalValue}
J_2\left(\frac{2k+1}{2}\right) = - \frac{1}{16 \pi^{6k+2} (2k)!^3 (2k+1)^2(2k+2)} & \int_{-\infty}^{\infty} d\xi \, \tanh(\pi \xi) \nonumber \\
& \times \left[ \frac{F_2^{(1)} (\xi, k)}{2 \cosh^{2k+1}(\pi \xi)} - \frac{F_2^{(2)} (\xi, k)}{(2k)! \pi^{2k}} \right] \, ,
\end{align}
where $F_2^{(1)}$ was presented in \eqref{c44:I2FunctionFirstTerm} and $F_2^{(2)}$ is
\begin{align}
\label{c44:I2FunctionSecondTerm}
F_2^{(2)} (\xi, k) = & \left[ \left( \prod_{l=0}^{k-1} \D_{2l+1}^2 \right) \partial_{\xi} \left( \frac{1}{\cosh(\pi \xi)} \right) \right] \left[ \left( \prod_{l=0}^{k-1} \D_{2l+1}^2 \right) \D_{2k+1} \left( \frac{1}{\cosh(\pi \xi)} \right) \right] \nonumber \\
 & + 4 \left[ \left( \prod_{l=0}^{k-1} \D_{2l+1}^2 \right) \partial_{\xi} \left( \frac{\xi}{\cosh(\pi \xi)} \right) \right] \left[ \left( \prod_{l=0}^{k-1} \D_{2l+1}^2 \right) \D_{2k+1} \left( \frac{\xi}{\cosh(\pi \xi)} \right) \right] \, .
\end{align}
The case $\Delta = k$ is completely analogous, obtaining as a final result,
\begin{align}
\label{c44:I2ntegerFinalValue}
J_2\left(k\right) = \frac{1}{16 \pi^{6k-1} (2k-1)!^3 (2k)^2(2k+1)} & \int_{-\infty}^{\infty} d\xi \, \tanh(\pi \xi) \nonumber \\
& \times \left[ \frac{G_2^{(1)} (\xi, k)}{2 \cosh^{2k}(\pi \xi)} + \frac{G_2^{(2)} (\xi, k)}{(2k-1)! \pi^{2k-1}} \right] \, ,
\end{align}
where
\begin{align}
G_2^{(1)} (\xi, k) = & \left[ \prod_{l=1}^{k-1} \D_{2l}^3 \right] \D_{2k} \partial^4_{\xi} \left((3 - 4\xi^2)\tanh(\pi \xi) \right) \, , \label{c44:I2GFunctionFirstTerm} \\
G_2^{(2)} (\xi, k) = & \left[ \left( \prod_{l=1}^{k-1} \D_{2l}^2 \right) \partial^3_{\xi} \left( \tanh(\pi \xi) \right) \right] \left[ \left( \prod_{l=1}^{k-1} \D_{2l}^2 \right) \D_{2k} \partial_{\xi}^2 \left( \tanh(\pi \xi) \right) \right] \nonumber \\
 & + 4 \left[ \left( \prod_{l=1}^{k-1} \D_{2l}^2 \right) \partial^3_{\xi} \left( \xi \tanh(\pi \xi) \right) \right] \left[ \left( \prod_{l=1}^{k-1} \D_{2l}^2 \right) \D_{2k} \partial_{\xi}^2 \left( \xi \tanh(\pi \xi) \right) \right] \, . \label{c44:I2GFunctionSecondTerm}
\end{align}

We come finally to the last piece, $J_3$. We first manipulate a bit the integrand: changing $(p, q) \to (-p, -q)$ and symmetrizing afterwards over $p$ and $r$, the final part of the integral becomes
\begin{equation}
\frac{4\,e^{\frac{p+r}2} B_q(\Delta+1)}{(e^{p+r}-1)(e^{q+r}-1)}  \to 2\,e^{\frac{p+q+r}2} e^{q/2} B_q(\Delta+1) \left[ \frac{1}{(e^{r}-e^p)(e^{r}-e^q)} + \frac{1}{(e^{p}-e^r)(e^{p}-e^q)} \right] \, .
\end{equation}
Applying now \eqref{id-exps}, symmetrizing the result over the three variables and then using \eqref{id-exps} again to write the integrand in a pole free form like in the previous $J_1$ and $J_2$, we get
\begin{align}
\label{c44:I3Intermediate}
J_3(\Delta) = & - \frac{2}{3} \frac{2^{8\Delta-8}}{ \pi^6 }\int_{-\infty}^\infty dp \int_{-\infty}^\infty dq \int_{-\infty}^\infty dr \, B_p(\Delta)B_q(\Delta)B_r(\Delta) \frac{e^{\frac{p+q+r}2}}{e^p-e^q} \times \\
& \times \left[ \frac{e^{\frac{p}2}B_p(\Delta+1)-e^{\frac{r}2}B_r(\Delta+1)  }{e^p-e^r} - \frac{e^{\frac{q}2}B_q(\Delta+1) -e^{\frac{r}2}B_r(\Delta+1) }{e^q-e^r} \right] \nonumber \, .
\end{align}
Integration over $q$ in the first term and over $p$ in the second one produces
\begin{equation}
\label{c44:I3Intermediate2}
J_3(\Delta) = - \frac{2}{3} \frac{2^{8\Delta-8}}{ \pi^6 \Delta } \int_{-\infty}^\infty dp \int_{-\infty}^\infty dr \,  B^2_p(\Delta) B_r(\Delta) \, p \, e^{\frac r2}\(\frac{e^{\frac{p}2}B_p(\Delta+1)-e^{\frac{r}2}B_r(\Delta+1)  }{e^p-e^r} \) \, .
\end{equation}
Already from the form of this integral we can see that it is very similar to $J_2(\Delta)$. The technique we use is exactly the same one, so we will not give details about the intermediate steps, but instead we will present the final result directly. We need to separate semi-integer and integer $\Delta$. In the first case, we have
\begin{align}
\label{c44:I3HalfIntegerFinalValue}
J_3\left(\frac{2k+1}{2}\right) = & \frac{1}{6 \pi^{6k+2} (2k)!^3 (2k+1)^2 (2k+2)} \nonumber \\
& \times \int_{-\infty}^{\infty} d\xi \, \left[ \frac{F_3^{(1)} (\xi, k)}{8} \frac{\tanh(\pi \xi)}{\cosh^{2k+1}(\pi \xi)} - \frac{F_3^{(2)} (\xi, k)}{(2k)! \pi^{2k}} \coth(\pi \xi) \right] \, ,
\end{align}
where
\begin{align}
F_3^{(1)} (\xi, k) = & \left[ \prod_{l=0}^{k-1} \D_{2l+1}^3 \right] \D_{2k+1} \partial_{\xi} \left( \frac{1 + 4\xi^2}{\cosh(\pi \xi)} \right) \, , \label{c44:I3FunctionFirstTerm} \\
F_3^{(2)} (\xi, k) = & \left[ \left( \prod_{l=0}^{k-1} \D_{2l+1}^2 \right) \partial_{\xi} \left( \frac{\xi}{\sinh(\pi \xi)} \right) \right] \left[ \left( \prod_{l=0}^{k-1} \D_{2l+1}^2 \right) \D_{2k+1} \left( \frac{\xi}{\sinh(\pi \xi)} \right) \right] \, . \label{c44:I3FunctionSecondTerm}
\end{align}
While for integer $\Delta = k$ we get
\begin{align}
\label{c44:I3ntegerFinalValue}
J_3\left(k\right) = & \frac{1}{6 \pi^{6k-1} (2k-1)!^3 (2k)^2 (2k+1)} \nonumber \\
& \times \int_{-\infty}^{\infty} d\xi \, \left[ \frac{G_3^{(1)} (\xi, k)}{8} \frac{\tanh(\pi \xi)}{\cosh^{2k}(\pi \xi)} - \frac{G_3^{(2)} (\xi, k)}{(2k-1)! \pi^{2k-1}} \coth(\pi \xi) \right] \, ,
\end{align}
where
\begin{align}
G_3^{(1)} (\xi, k) = & \left[ \prod_{l=1}^{k-1} \D_{2l}^3 \right] \D_{2k} \partial^4_{\xi} \left((1 + 4\xi^2)\tanh(\pi \xi) \right) \, , \label{c44:I3GFunctionFirstTerm} \\
G_3^{(2)} (\xi, k) = & \left[ \left( \prod_{l=1}^{k-1} \D_{2l}^2 \right) \partial^3_{\xi} \left( \xi \coth(\pi \xi) \right) \right] \left[ \left( \prod_{l=1}^{k-1} \D_{2l}^2 \right) \D_{2k} \partial_{\xi}^2 \left( \xi \coth(\pi \xi) \right) \right] \, . \label{c44:I3GFunctionSecondTerm}
\end{align}

We conclude with Table \ref{tablaC44}, where we show numerical results for $J_1$ ---obtained from \req{c44:I1Intermediate}--- as well as exact results for $J_2$ and $J_3$, for the smallest semi-integer and integer values of $\Delta$. We also compute the total value of the coefficient $c_{4:4}^{(1,1,1,1)}$ for those cases.
\begin{table}[t!]
 \centering
\begin{tabular}{|c|c|c|c|c|}  \hline
 $\Delta$ & $J_1$ & $J_2$ & $J_3$ & $c_{4:4}^{(1,1,1,1)}$ \\ \hline 
 $1/2$ & $\frac{8}{135}+\frac{1}{9\pi^2} 
 $ 
 & $\frac{2}{45} + \frac{1}{3 \pi^2}  
 $ & $-\frac{2}{135}+\frac{2}{9 \pi^2}  
 $ & $\frac{4}{45}+\frac{2}{3\pi^2}\simeq 0.156436$ \\[.3em]
 $1$ & $\frac{64}{945}-\frac{4}{27\pi^2}$ & $\frac{32}{315} + \frac{4}{9 \pi^2} 
 $ & $\frac{32}{945}-\frac{8}{27 \pi^2} 
 $ & $\frac{64}{315}\simeq 0.203175$ \\[.3em]
 $3/2$ & $0.04382$ & $\frac{10496}{45045} + \frac{32}{27 \pi^2} 
 $ & $-\frac{10496}{135135}+\frac{64}{81 \pi^2} 
 $ & $0.399298$ \\[.3em]
 $2$ & $0.03827$ & $\frac{1482752}{2297295} + \frac{256}{81 \pi^2} 
 $ & $\frac{1482752}{6891885}-\frac{512}{243 \pi^2} 
 $ & $1.005593$ \\[.3em]
 $5/2$ & $0.03440$ & $\frac{1507328}{793611} + \frac{26624}{2835 \pi^2} 
 $ & $-\frac{1507328}{2380833}+\frac{53248}{8505 \pi^2} 
 $ & $2.886498$ \\[.3em]
 $3$ & $0.03151$ & $\frac{1097859072}{185910725} + \frac{32768}{1125 \pi^2} 
 $ & $\frac{365953024}{185910725}-\frac{65536}{3375 \pi^2} 
 $ & $8.888975$ \\ \hline
\end{tabular}
\caption{Values of the different terms contributing to $c_{4:4}^{(1,1,1,1)}$. $J_1$ is obtained by numerical integration of \eqref{c44:I1Intermediate}, while $J_2$ and $J_3$ are obtained from the expressions \eqref{c44:I2HalfIntegerFinalValue}, \eqref{c44:I2ntegerFinalValue}, \eqref{c44:I3HalfIntegerFinalValue}, and \eqref{c44:I3ntegerFinalValue}.}\label{tablaC44}
\end{table}
For $\Delta = 1/2$, there is a fortunate numerical coincidence, easy to discover, which allows us to write an analytic value of the coefficient. The total value of $c_{4:4}^{(1,1,1,1)}(\Delta=1/2)$ is twice that of $J_2(1/2)$ to an extremely good level of precision, and this last one we know analytically. Therefore, we conclude\footnote{This coefficient can alternatively be read from \cite{Agon:2015twa} ---more precisely, from the last two terms on the right hand side of their equation (4.19). Similarly, one can also find that expression in eqs (79) and (80) of \cite{Chen:2017hbk}. In order to identify the exact result, it is useful to notice the following relations $C_{ij}=2\langle \phi_i(1)\phi_j(1)\rangle$ and $C_{ij}=1/s_{ij}$ between our coefficients and the correlators in \cite{Agon:2015twa} and \cite{Chen:2017hbk} respectively. In their context, the computation appears in the subleading contribution to the mutual information of two disjoint disks for a free scalar in three space-time dimensions.}
\begin{equation}
\label{c44:FinalValueDelta12}
c_{4:4}^{(1,1,1,1)} (\Delta = 1/2) = \frac{4}{45} + \frac{2}{3 \pi^2} ~ .
\end{equation}
Unfortunately, it is not obvious that a similar relation holds for other values of $\Delta$. Still, for $\Delta=1$ we are reasonably confident that the numerical value obtained corresponds to
\begin{equation}
\label{c44:FinalValueDelta12333}
J_1 (\Delta = 1) \overset{?}{=} \frac{64}{945}-\frac{4}{27\pi^2}\quad  \Rightarrow  \quad c_{4:4}^{(1,1,1,1)} (\Delta = 1) \overset{?}{=} \frac{64}{315}  \,  .
\end{equation}
More generally, we do believe $J_1 (\Delta)$ might have a reasonably simple closed form, much like the others. This is a technically challenging problem which we leave open for further investigation.\footnote{It is also conceivable that all the integrals, or at least the coefficient as a whole, can be done for general $\Delta$ in a closed form. This is what happens for the other coefficients $c_{4:2}^{(4)}$, $c_{4:3}^{(1,1,2)}$, etc.} 

\subsubsection{Computation of $c_{4:4}^{(2,2)}$}

Finally, we come to the coefficient \eqref{C4:4:2,2-2}. With the integral representation for the coefficients, we must perform the following sum
\bea
e^{-\frac p2} e^{-\frac q2} \sum_{l=3}^{n-1}\sum_{k=2}^{l-1}\sum_{j=1}^{k-1}
\(e^{\frac{p l}{n}} e^{\frac{q( k-j)}{n}}+e^{\frac{p j}{n}} e^{\frac{q(l- k)}{n}}+e^{\frac{p k}{n}} e^{\frac{q( l-j)}{n}}\) \approx  (n-1)f^{(2,2)}_{4:4}(p,q) \, ,
\eea
where 
\bea
f^{(2,2)}_{4:4}(p,q)=\frac{e^{\frac{p+q}2}}{e^p-1}\[\frac{e^{p}\(p+q\)}{e^{p+q}-1}-\frac{p-q}{e^p-e^q}\]\,.
\eea
The full integral form of the coefficient is
\bea
c_{4:4}^{(2,2)}=\frac{1}{4}\[\frac{2^{8\Delta-4}}{\pi^4 }\]\int_{-\infty}^\infty d p\,\int_{-\infty}^\infty d q\, f^{(2,2)}_{4:4}(p,q) \, B_p(2\Delta)B_q(2\Delta) \, ,
\eea
and the integral over $q$ is trivial to do by means of the same tricks as in the previous cases. The final integral to be done is
\bea
c_{4:4}^{(2,2)}=\frac{2^{8\Delta-4}}{\pi^2 } \int_{-\infty}^\infty d p\, B_p(2\Delta)B_p(2\Delta+1) \, ,
\eea
but this is of the form (\ref{Int-gammas-complex}), so
\bea\label{C4:4(22)}
c_{4:4}^{(2,2)}=\frac{2^{8\Delta}\[\Gamma\(4\Delta+1\)\]^2}{4\,  \Gamma\(8\Delta+2\)}\,.
\eea

\section{Some useful integrals\label{App-integrals}}

We present here some very general Fourier transforms that are needed at different points of our computations. Defining the function
\begin{equation}
f_n(z) = \frac{e^{(a + i b)z}}{\cosh^n(z)} \, ,
\end{equation}
we can compute the integral over the real line for $|a| < n$ by means of a rectangular contour, with basis on the real line and height $i \pi$. The only pole enclosed is that at $z = i \pi / 2$, and the periodicity properties of the $\cosh$ allow to relate the horizontal integrals to obtain
\begin{equation}
\int_{-\infty}^{\infty} dx \, f_n(x) = \begin{cases} \displaystyle
2 \pi i R_n (a, b) \frac{\sinh \left( (b + i a) \pi/2 \right)}{\cosh(b \pi) - \cos (a \pi)} \, , \quad \text{if } n \text{ is even}. \vspace{0.2cm} \\
\displaystyle 2 \pi i R_n (a, b) \frac{\cosh \left( (b + i a) \pi/2 \right)}{\cosh(b \pi) + \cos (a \pi)} \, , \quad \text{if } n \text{ is odd}.
\end{cases} \,
\end{equation}
Here $R_n(a, b)$ is defined as
\begin{equation}
\label{Int:ResidueCosh}
R_n(a,b) \equiv e^{-i(a + i b) \pi /2} \, {\rm Res} \left[ f_n(z) , z= i \pi / 2 \right] ~ .
\end{equation}
This is a polynomial in $(a, b)$ which can be easily computed for any $n$ with the aid of a symbolic algebra software. Separating real and imaginary parts, and rescaling the variables of integration and the parameters we arrive at the general Fourier transforms for $k = 1, 2, \dots$,
\begin{align}
& \int_{-\infty}^{\infty} dx \, \frac{\cosh(\alpha x)}{\cosh^{2k-1}(\gamma x)} e^{i \xi x} = \frac{2 \pi / \gamma}{\cosh(\frac{\pi \xi}{\gamma}) + \cos (\frac{\pi \alpha}{\gamma})} {\rm Re} \left[ i R_{2k-1} \left( \frac{\alpha}{\gamma}, \frac{\xi}{\gamma} \right)\cosh \left( (\xi + i \alpha) \frac{\pi}{2 \gamma} \right) \right] \, , \\
& \int_{-\infty}^{\infty} dx \, \frac{\cosh(\alpha x)}{\cosh^{2k}(\gamma x)} e^{i \xi x} = \frac{2 \pi / \gamma}{\cosh(\frac{\pi \xi}{\gamma}) - \cos (\frac{\pi \alpha}{\gamma})} {\rm Re} \left[ i R_{2k} \left( \frac{\alpha}{\gamma}, \frac{\xi}{\gamma} \right)\sinh \left( (\xi + i \alpha) \frac{\pi}{2 \gamma} \right) \right] \, , \\
& \int_{-\infty}^{\infty} dx \, \frac{\sinh(\alpha x)}{\cosh^{2k-1}(\gamma x)} e^{i \xi x} = \frac{2 \pi i / \gamma}{\cosh(\frac{\pi \xi}{\gamma}) + \cos (\frac{\pi \alpha}{\gamma})} {\rm Im} \left[ i R_{2k-1} \left( \frac{\alpha}{\gamma}, \frac{\xi}{\gamma} \right)\cosh \left( (\xi + i \alpha) \frac{\pi}{2 \gamma} \right) \right] \, , \\
& \int_{-\infty}^{\infty} dx \, \frac{\sinh(\alpha x)}{\cosh^{2k}(\gamma x)} e^{i \xi x} = \frac{2 \pi / \gamma}{\cosh(\frac{\pi \xi}{\gamma}) - \cos (\frac{\pi \alpha}{\gamma})} {\rm Im} \left[ i R_{2k} \left( \frac{\alpha}{\gamma}, \frac{\xi}{\gamma} \right)\sinh \left( (\xi + i \alpha) \frac{\pi}{2 \gamma} \right) \right] \, .
\end{align}
We can play the same game with
\begin{equation}
\tilde{f}_n(z) = \frac{e^{(a + i b)z}}{\sinh^n(z)} \,.
\end{equation}
In this case, there are poles lying on the contour at $z = 0, i \pi$. We circumvent them in the usual way, thus computing the principal value of the integrals, and obtain
\begin{align}
& \int_{-\infty}^{\infty} dx \, \frac{\cosh(\alpha x)}{\sinh^{2k-1}(\gamma x)} e^{i \xi x} = \frac{i \pi}{\gamma} {\rm Im} \left[ i \tilde{R}_{2k-1} \left( \frac{\alpha}{\gamma}, \frac{\xi}{\gamma} \right) \tanh \left( (\xi - i \alpha) \frac{\pi}{2 \gamma} \right) \right] \, , \\
& \int_{-\infty}^{\infty} dx \, \frac{\cosh(\alpha x)}{\sinh^{2k}(\gamma x)} e^{i \xi x} = \frac{\pi}{\gamma} {\rm Re} \left[ i \tilde{R}_{2k} \left( \frac{\alpha}{\gamma}, \frac{\xi}{\gamma} \right) \coth \left( (\xi - i \alpha) \frac{\pi}{2 \gamma} \right) \right] \, , \\
& \int_{-\infty}^{\infty} dx \, \frac{\sinh(\alpha x)}{\sinh^{2k-1}(\gamma x)} e^{i \xi x} = \frac{\pi}{\gamma} {\rm Re} \left[ i \tilde{R}_{2k-1} \left( \frac{\alpha}{\gamma}, \frac{\xi}{\gamma} \right) \tanh \left( (\xi - i \alpha) \frac{\pi}{2 \gamma} \right) \right] \, , \\
& \int_{-\infty}^{\infty} dx \, \frac{\sinh(\alpha x)}{\sinh^{2k}(\gamma x)} e^{i \xi x} = \frac{i \pi}{\gamma} {\rm Im} \left[ i \tilde{R}_{2k} \left( \frac{\alpha}{\gamma}, \frac{\xi}{\gamma} \right) \coth \left( (\xi - i \alpha) \frac{\pi}{2 \gamma} \right) \right] \, ,
\end{align}
where, in this case, 
\begin{equation}
\label{Int:ResidueSinh}
\tilde{R}_n(a,b) \equiv {\rm Res} \left[ \tilde{f}_n(z) , z= 0 \right] ~ .
\end{equation}

We will be interested in particular cases of the previous Fourier transforms, trivially obtained from the previous general expressions. Let us list them here. The simplest ones are
\begin{align}
 & \int_{-\infty}^{\infty} dx \, \frac{e^{i \xi x}}{\cosh(x/2)} = 2 \int_{-\infty}^{\infty} dx \, \frac{e^{x/2}}{e^x + 1} e^{i \xi x} = \frac{2 \pi}{\cosh(\pi \xi)} ~ , \label{Int:InvCosh} \\
 & \int_{-\infty}^{\infty} dx \, \frac{e^{i \xi x}}{\sinh(x/2)} = 2 \int_{-\infty}^{\infty} dx \, \frac{e^{x/2}}{e^x - 1} e^{i \xi x} = 2 \pi i \tanh(\pi \xi) ~ . \label{Fourier-2}
\end{align}
Derivatives of these functions allow to include polynomials in $x$. As an example, we will need the integral
\begin{equation}
\label{Fourier-1}
\int_{-\infty}^\infty dx \,\frac{x e^{\frac{x}{2}}}{e^x - 1}e^{i \xi x} = \frac{\pi^2}{\cosh^2 (\pi \xi)} ~ .
\end{equation}
We list some other relevant integrals for the computations in Appendix \ref{app-coeffs},
\begin{align}
\label{Int:InvCosh2and3} & \int_{-\infty}^{\infty} dx \, \frac{e^{i \xi x}}{\cosh^2(x/2)} = \frac{4 \pi \xi}{\sinh(\pi \xi)} ~ , && \int_{-\infty}^{\infty} dx \, \frac{e^{i \xi x}}{\cosh^3(x/2)} = (1 + 4 \xi^2) \frac{\pi}{\cosh(\pi \xi)} \, , \\
\label{Int:InvSinh2and3} & \int_{-\infty}^{\infty} dx \, \frac{e^{i \xi x}}{\sinh^2(x/2)} = - \frac{4 \pi \xi}{\tanh(\pi \xi)} ~ , && \int_{-\infty}^{\infty} dx \, \frac{e^{i \xi x}}{\sinh^3(x/2)} = - i \pi (1 + 4 \xi^2) \tanh(\pi \xi) \, , \\
\label{Int:CoshOvSinh} & \int_{-\infty}^{\infty} dx \, \frac{\cosh(x/2)}{\sinh(x/2)} e^{i \xi x} = \frac{2 \pi i}{\tanh(\pi \xi)} ~ , && \int_{-\infty}^{\infty} dx \, \frac{\cosh(x/2)}{\sinh^2(x/2)} e^{i \xi x} = - 4 \pi \xi \tanh(\pi \xi) \, , \\
\label{Int:SinhOvCosh} & \int_{-\infty}^{\infty} dx \, \frac{\sinh(x/2)}{\cosh^2(x/2)} e^{i \xi x} = \frac{4 \pi i \xi}{\cosh(\pi \xi)} \, .
\end{align}

In Appendix \ref{app-coeffs}, we follow essentially two strategies when dealing with the integrals. Whenever we find it computationally feasible, we work with the beta functions for general argument. When doing this, it is convenient to remember the integral representations of the beta function defined in \req{compact-beta},
\begin{equation}
\label{Int:RepresentationsBeta}
B_q (\Delta) = \int_{0}^1 dt\, t^{\Delta-1}(1-t)^{\Delta-1} e^{\frac{iq}{2\pi} \log\(\frac{t}{1-t}\)} = \frac{1}{4^{\Delta}} \int_{-\infty}^{\infty} dx \frac{e^{i \frac{q}{2 \pi} x}}{\cosh^{2 \Delta}(x/2)} \,.
\end{equation}
We will refer to the first one as the \emph{logarithmic} representation, while the second one will be the \emph{hyperbolic} one. From properties of the gamma function it is trivial to show that
\begin{equation}
\label{Int:ShiftBeta}
B_q(\Delta + 1) = \frac{1}{4 \pi^2 (2 \Delta) (2 \Delta +1)} \left(q^2 + 4 \pi^2 \Delta^2 \right) B_q(\Delta) \, .
\end{equation}
From the logarithmic representation (or from the hyperbolic one, for what matters), and using the previous Fourier transforms, it is possible to prove a number of useful identities. The first one is
\begin{equation}
\label{Int-linear-gamma2}
\int_{-\infty}^\infty  dp \,\frac{ e^{\frac{p\pm q }2}\(p\pm q\)}{e^{p\pm q}-1}\, B_p(\Delta)=4\pi^2 B_q(\Delta+1) \, ,
\end{equation}
which follows directly from \req{Fourier-1} after introducing the beta function integral representation. From \req{Fourier-2} we similarly obtain
\begin{equation}
\label{int-simple-gamma2}
\int_{-\infty}^\infty  dp \,\frac{ e^{\frac{p\pm q }2}}{e^{p\pm q}-1}\, B_p(\Delta)=\pm \frac{q}{2\Delta}B_q(\Delta) \, .
\end{equation}
It is also straightforward to obtain the identity
\begin{equation}
\label{Int-gammas-complex}
\int_{-\infty}^\infty d q\, B_{q+Q}(\Delta)B_{q+\bar{Q}}(\bar{\Delta})=4\pi^2 B_{Q-\bar{Q}}(\Delta+\bar{\Delta})\,,
\end{equation}
Finally, a bit more involved is the following result
\begin{equation}
\label{Int-Gamma-C43-appendix}
\int_{-\infty}^\infty d q\,q\, e^{-\frac q2}B^2_q(\Delta_1) B_q(\Delta_2)=-\frac{2^{4-2\Delta_1-2\Delta_2} \pi^4 \Gamma\(\Delta_1+\frac12\) \(\Gamma\(\Delta_1+\Delta_2\)\)^2}{\Gamma\(\Delta_1\)\Gamma\(\Delta_2\)\Gamma\(2\Delta_1+\Delta_2+\frac12\)}\, .
\end{equation}
It can be proved to be valid by considering the integral
\begin{equation}
I(\alpha) = \int_{-\infty}^\infty dq \, e^{i \alpha q}B^2_q(\Delta_1) B_q(\Delta_2) \, ,
\end{equation}
which can be integrated over $q$ and one of the variables appearing in the representation of the beta function straightforwardly for ${\rm Im}(\alpha) \in (-1/2, 1/2)$. Using the trick presented in Appendix B of \cite{Sarosi:2017rsq}, taking $-i\partial_{\alpha}$ to recover the $q$ term in the integrand of \req{Int-Gamma-C43-appendix}, and approaching $\alpha \to i/2$ from below, it is possible to reproduce the value of the integral in terms of gamma functions.

This last computation shows very clearly that things become arbitrarily involved when one starts adding beta functions and/or more complicated prefactors. At some point in the computation of the $c_{4:4}^{(1,1,1,1)}$ coefficient in Appendix \ref{app-coeffs}, we were not able to carry out the integrals in full generality, for any value of $\Delta$. In those cases, we will need to restore to another approach, still able to generate results for integer or semi-integer $\Delta$, but in an algorithmic way. This approach is therefore well-suited for implementation in a standard symbolic algebra software, and it is based on the following observation. For $a = 0$, the $R_n(a,b)$ notably simplify,
\begin{align}
& R_{2k+1} (0, p/\pi) = - \frac{i}{(2k)! \, \pi^{2k}} \prod_{l=0}^{k-1} \left( p^2 + (2l+1)^2 \pi^2 \right) \, , \\
& R_{2k} (0, p/\pi) = - \frac{i p}{(2k-1)! \, \pi^{2k-1}} \prod_{l=1}^{k-1} \left( p^2 + 4 l^2 \pi^2 \right) \, .
\end{align}
From the hyperbolic representation of the beta function, it is then immediate to obtain
\begin{align}
& B_p \left( \frac{2 k + 1}{2} \right)^n = \frac{\pi^{(1-2k)n}}{2^{2kn} (2 k)!^n} \left[ \prod_{l=0}^{k-1} \left( p^2 + (2l+1)^2 \pi^2 \right) \right] \frac{1}{\cosh^n(p/2)} \, , \label{Int:PowerBetaHalfInteger} \\
& B_p \left( k \right)^n = \frac{\pi^{2 (1-k)n}}{2^{(2k-1)n} (2 k-1)!^n} \left[ \prod_{l=1}^{k-1} \left( p^2 + 4 l^2 \pi^2 \right) \right] \frac{p^n}{\sinh^n(p/2)} \, . \label{Int:PowerBetaInteger}
\end{align}
It is also possible to obtain these results by repeated application of \eqref{Int:ShiftBeta}, knowing the seed values
\begin{equation}
\label{Int:BetaValues}
B_p (1/2) = \frac{\pi}{\cosh(p/2)} \, , \qquad B_p(1) = \frac{1}{2} \frac{p}{\sinh(p/2)} ~ .
\end{equation}

\section{Identifying the leading scaling in the lattice}
\label{subsec:ScalarNumerics}
In this appendix we present a table with the coefficient of determination $\mathcal R^2$ obtained from different linear fits to the continuum lattice results for $I_N$ in the case of the free scalar in the long-distance regime. We do this for the scaling predicted from our general discussion as well as for the immediately greater and the immediately smaller power.  In all cases, we observe that the scaling predicted by \req{scalarscaling} is strongly favoured.\\
\noindent
\begin{table}[h!]
\centering
\begin{tabular}{|c|c| c|c|c || c| c| c| c|} 
 \hline
$N=2$ & Blocks & $\mathcal R^2[1/x]$ & $\mathcal R^2[1/x^2]$ & $\mathcal R^2[1/x^3]$ & $N=3$& $\mathcal R^2[1/x^2]$ & $\mathcal  R^2[1/x^3]$ & $\mathcal R^2[1/x^4]$ \\  \hline
& $L \times L$ & $0.99759$ & $\mathbf 1.$ & $0.99772$&  & $0.99387$ & $\mathbf{0.99994}$ & $0.99229$   \\ 
 &   Circles & $0.99761$ & $\mathbf 1.$ & $0.99770$ &  &  $0.99378$ & $ \mathbf{0.99995}$ & $0.99239$ \\ 
 &   $L \times 2L$ & $0.99763$ & $ \mathbf 1.$ & $0.99769$ &  & $0.99366$ & $ \mathbf{0.99996}$ & $0.99252$  \\ 
 &   $L \times 4L$ & $0.99764$ & $ \mathbf 1.$ & $0.99768$ &   & $0.99350$ & $\mathbf{0.99997}$ & $0.99269$ \\ 
 &   $L \times 6L$ & $0.99764$ & $\mathbf 1.$ & $0.99767$ &  &  $0.99345$ & $\mathbf{0.99997}$ & $0.99275$ \\ \hline
\end{tabular}\vspace{0.1cm}
\begin{tabular}{|c|c| c|c|c || c| c| c| c|} 
 \hline
$N=4$ & Blocks & $\mathcal R^2[1/x^3]$ & $\mathcal R^2[1/x^4]$ & $\mathcal R^2[1/x^5]$ & $N=5$& $\mathcal R^2[1/x^4]$ & $\mathcal R^2[1/x^5]$ & $\mathcal R^2[1/x^6]$ \\  \hline
& $L \times L$ &$0.99630$ & $\mathbf{0.99974}$ & $0.99479$ &  &$0.99726$ & $\mathbf{0.99884}$ & $0.98852$   \\ 
 &   Circles & $0.99613$ & $\mathbf{0.99975}$ & $0.99498$ &  & $0.99655$ & $\mathbf{0.99920}$ & $0.98984$  \\ 
 &   $L \times 2L$ &  $0.99589$ & $\mathbf{0.99977}$ & $0.99524$ &  & $0.99599$ & $\mathbf{0.99960}$ & $0.99103$ \\ 
 &   $L \times 4L$ &$0.99558$ & $\mathbf{0.99978}$ & $0.99555$ &   & $0.99474$ & $\mathbf{0.99987}$ & $0.99268$ \\ 
 &   $L \times 6L$ &  $0.99546$ & $\mathbf{0.99978}$ & $0.99567$  &  &  $0.99422$ & $\mathbf{0.99993}$ & $0.99326$ \\ \hline
\end{tabular}\vspace{0.1cm}
\begin{tabular}{|c|c|c|c|c|} 
 \hline
$N=6$ & Blocks & $\mathcal R^2[1/x^5]$ & $\mathcal R^2[1/x^6]$ & $\mathcal R^2[1/x^7]$ \\  \hline
& $L \times L$ & $0.99903$ & $\mathbf{0.99968}$ & $0.99665$ \\ 
 &  Circles & $0.99873$ & $\mathbf{0.99979}$ & $0.99716$ \\ 
 &   $L \times 2L$ & $0.99944$ & $\mathbf{0.99955}$ & $0.99598$ \\ 
 &   $L \times 4L$ & $0.99812$ & $\mathbf{0.99987}$ & $0.99791$ \\ 
 &   $L \times 6L$ & $0.99903$ & $\mathbf{0.99981}$ & $0.99686$ \\ \hline
\end{tabular}
\caption{We show the values of the coefficient of determination, $\mathcal R^2$, of different fits to the long-distance results obtained for $I_N$, with $N=2,3,4,5,6$, for a free scalar field in the lattice. In each case, we consider sets of $N$ identical entangling regions (squares, disks and increasingly thinner rectangles) placed on a regular $N$-gon arrangement.  $\mathcal R^2$ is computed for fits of the form $I_N \sim (1/x)^\alpha$ with $\alpha=N-1,N,N+1$. }\label{tabla1}
\end{table}

\section{Analytic mutual-information toy models}\label{EMIMEMI}
In this appendix we present two toy models which resemble the mutual information results corresponding to free scalars and fermions, respectively, while allowing for analytic calculations. For both models, we present analytic formulas for the mutual information corresponding to the different shapes considered in Section \ref{i2arbi}.

\subsection{EMI model}
The mutual information is a positive semi-definite quantity for general regions and theories, $I_2\geq 0$. On the other hand, as we mentioned in the introduction, the tripartite information
does not have a definite sign and, in fact, there exist theories for which $I_3(A_1, A_2,A_3)\geq 0$ and others with $I_3(A_1, A_2, A_3)\leq 0$, corresponding to ``subextensive'' and ``superextensive'' mutual informations, respectively. On the other hand, a theory with a vanishing tripartite information for arbitrary regions, $I_3(A_1, A_2,A_3)= 0$ for all $A_1,A_2,A_3$, would have a mutual information which is extensive in its arguments, since in that case $I_2(A_1,A_2 \cup A_3)=I_2(A_1,A_2)+ I_2(A_1,A_3)$.



Interestingly, imposing $I_3(A_1, A_2,A_3)= 0$ for general regions along with the set of known axioms satisfied by mutual information in a relativistic QFT, strongly restricts the form of the mutual information in general dimensions. The result defines the so-called ``Extensive Mutual Information'' (EMI) model.
Further imposing conformal invariance, the expression for its mutual information is fully determined up to a global multiplicative constant and reads  \cite{Casini:2008wt}\footnote{One can also arrive at the EMI formula by assuming the R\'enyi twist operators to be exponentials of free fields  \cite{Swingle:2010jz}. The EMI model formula can also be interpreted naturally as a ``partial entanglement entropy'' \cite{Han:2019scu}.} 
\begin{equation}
I_{\rm \ssc EMI}(A_1,A_2)=2\kappa_{(d)}\,\int_{\partial A_1} d \sigma_1 \, \int_{\partial A_2} d \sigma_2 \, \frac{(n_1\cdot n_2 )(\bar n_1 \cdot \bar n_2)-(n_1\cdot \bar{n}_2 )(\bar n_1 \cdot n_2)}{|x_1-x_2|^{2(d-2)}}\,,\label{sis2}
\end{equation}
where $\kappa_{(d)}$ is such (positive) constant  and 
$\bar n_1$ and $ n_1$ are unit vectors orthogonal to the surface and to each other, $n_1\cdot \bar n_1=0$. For fixed time slices, one can choose $n_1= n_2=\hat t$ and the integrand reduces to $-(\bar n_1\cdot \bar n_2)/|x_1-x_2|^{2(d-2)}$. 

In spite of its simplicity, the EMI model produces physically reasonable results in general dimensions \cite{Casini:2015woa,Bueno:2015rda,Witczak-Krempa:2016jhc,Bueno:2019mex,Estienne:2021lxh,Bueno:2021fxb}. Such results are rather similar to the free fermion ones and, in fact,  the EMI model  exactly coincides with a free  massless fermion in  $d=2$ \cite{Casini:2005rm}.\footnote{Besides, as shown in \cite{Agon:2021zvp}, the result for the mutual information of two arbitrarily boosted spheres in the EMI model exactly matches the contribution from the free fermion leading conformal block in general dimensions. However, it does not contain all the rest of conformal blocks present in the free fermion theory.} On the other hand, using the long-distance expansion of the mutual information to bootstrap its putative operatorial content, it has been shown that the EMI model cannot describe the mutual information of any actual CFT in higher dimensions \cite{Agon:2021zvp}.\footnote{This implies that the set of known axioms satisfied by EE in QFT is incomplete.} Still, as we have said, the EMI model produces reasonable  results which tend to resemble the free fermion ones in most situations, so we find it useful to include its predictions in the present cases.
In particular, in general dimensions the long-distance expression for the EMI model for two regions in the same time slice takes the universal form
\begin{equation}\label{Iemilong}
I_{\rm \ssc EMI}(A_1,A_2) \overset{r\gg R_{A_1,A_2}}{=} 4 \kappa_{(d)} (d-1)(d-2) \frac{{\rm vol}(A_1)\cdot {\rm vol}(A_2)}{r^{2(d-1)}}+\dots
\end{equation}
which is the same scaling as the one corresponding to a free fermion.




{\bf Disk regions.} The result for the mutual information of two identical disks of diameters $R$ and separated a distance $r$ in the three-dimensional EMI model can be extracted from the result presented in \cite{Agon:2021zvp}, where it was computed in the more general case of boosted spheres in general dimensions. Defining $r$ as the distance between disks such that $r\rightarrow 0$ when the disks touch at a point and $x\equiv r/R$, one finds
\begin{equation}\label{diskEMI}
I_{\rm \ssc EMI}(A_1,A_2)= 2\pi^2 \kappa_{(3)}  \frac{\left[\sqrt{x(2+x)}-2x(2+x)(1+x-\sqrt{x(2+x)})\right]}{x(1+x)(2+x)}\, .
\end{equation}
In the almost-touching regime, as well as in the long-distance one, one finds
\begin{align}
I_{\rm \ssc EMI} &\overset{x\rightarrow 0}{=}  \frac{\sqrt{2}\pi^2 \kappa_{(3)}}{\sqrt{x}} -4 \pi^2 \kappa_{(3)} + \mathcal{O}(\sqrt{x})  \, , \\
I_{\rm \ssc EMI} &\overset{x\rightarrow \infty} {=}  \frac{\pi^2 \kappa_{(3)} }{2x^4} - \frac{2\pi^2\kappa_{(3)} }{x^5}+\mathcal{O}(1/x^6) \, .
\end{align}
On the other hand, the EE disk coefficient $F_0$, which we use to normalize our curves in Section \ref{i2arbi}, is given for the EMI model by $F_0^{\rm \ssc EMI}=2\pi^2 \kappa_{(3)}$.


{\bf Rectangle regions.} Let us now consider the other case of interest for us, which corresponds to two identical rectangles of sides $ \xi R \times R$ separated along the $X$ direction a distance $r$ ---measured between the rightmost edge of the left rectangle and the leftmost edge of the right one. In the $r \rightarrow 0$ limit, the two rectangles would touch each other. The boundary of the leftmost rectangle is defined as $\partial A = \{ [X,0], X\in [0, \xi R]; [X,R],X\in [0, \xi R]; [0,Y],Y\in [0,R]; [\xi R, Y],Y\in [0,R]\}$. The boundary of $B$ is obtained by shifting the resulting figure along the $X$ axis  a distance $\xi R+r $, $\partial B = \{ [X,0], X\in [\xi R+r , 2\xi R+r]; [X,R],X\in [\xi R+r ,2 \xi R+r]; [\xi R+r ,Y],Y\in [0,R]; [2\xi R+r, Y],Y\in [0,R]\}$.

In principle there would be 16 contributions to the mutual information coming from the $4\times 4$ pairings of segments from the two rectangles. However, observe that 8 of them vanish, corresponding to segments perpendicular to each other.  It is not difficult to see that the remaining contributions can be written as
\begin{equation}
I_{\rm \ssc EMI}(A_1,A_2)=-2\kappa_{(3)} \left[ 2\Delta_1+\Delta_2+\Delta_3+2\Delta_4+2\Delta_5\right]\, ,
\end{equation}
where
\begin{align}
\Delta_1&=\Delta[R,0,R,r+\xi R]\, , \quad  \Delta_2=-\Delta[R,0,R,r+2\xi R]\, , \quad  \Delta_3=-\Delta[R,0,R,r]\, , \\
\Delta_4&=\Delta[\xi R, \xi R+r,2\xi R+r,0]\, , \quad \Delta_5=-\Delta[\xi R, \xi R+r,2\xi R+r,R]\, ,
\end{align}
and we defined
\begin{equation}
\Delta[\alpha,\beta,\gamma,\chi]\equiv \int_0^\alpha \int_{\beta}^\gamma \frac{d x_1 d x_2}{\chi^2+(x_1-x_2)^2}\, .
\end{equation}
Performing the integrals and simplifying a bit, one finds the final result
\begin{align}\notag 
I_{\rm \ssc EMI}=4\kappa_{(3)}\left[ \frac{\arccot(x)}{x}-\frac{2 \arccot(x+\xi)}{x+\xi}+\frac{\arccot(x+2\xi)}{x+2\xi} -2(x+\xi) \arctan(x+\xi) \right. \\ \left. +x \arctan(x) + (x+2\xi) \arctan(x+2\xi)  -\log \left[\frac{(x+\xi)^4(1+x^2)(1+(x+2\xi)^2)}{x^2(x+2\xi)^2(1+(x+\xi)^2)^2} \right]\right]\, , \label{rectEMI}
\end{align}
where we defined $x\equiv r/R$. For almost-touching rectangles and in the long-distance regime one finds, respectively,
\begin{align}
I_{\rm \ssc EMI} &\overset{x\rightarrow 0}{=}  \frac{k_{\rm \ssc EMI}^{(3)} } {x}+8\kappa_{(3)} \log x + g(\xi) + \mathcal{O}(x)  \, , \\
I_{\rm \ssc EMI} &\overset{x\rightarrow \infty} {=} \frac{8\kappa_{(3)}\xi^2}{x^4} - \frac{32 \kappa_{(3)} \xi^3}{x^5}+ \mathcal{O}(x^{-6})  \, , 
\end{align}
where $k_{\rm \ssc EMI}^{(3)} =2\pi \kappa_{(3)}$ is the coefficient appearing in the entanglement entropy of a strip region, $g(\xi)$ is a complicated function of $\xi$ and $\xi^2/x^4$ in the second expression is the product of the areas of the two rectangles $A,B$ divided by their separation to the fourth power, $\xi^2/x^4={\rm area}(A) {\rm area}(B)/r^4$, as expected from the general formula, \req{Iemilong}.

\subsection{MEMI model}
A simple modified version of the EMI model can be constructed such that the long-distance scaling reproduces the one corresponding to a free scalar field. This ``Modified'' EMI (MEMI) model is defined, in the case of a fixed time slice, by a mutual information of the form
\begin{equation}
I_{\rm \ssc MEMI}(A_1,A_2)=2\mu_{(d)}\,\int_{\partial A_1} d \sigma_1  \, \int_{\partial A_2} d \sigma_2 \, \frac{1}{|x_1-x_2|^{2(d-2)}}\,,\label{sis2}
\end{equation}
where $\mu_{(d)}$ is a positive constant. It is easy to see that, for long separations, this behaves as
 \begin{equation}\label{memii}
I_{\rm \ssc MEMI}(A_1,A_2)\overset{r\gg R_{A_1,A_2}}{=}2\mu_{(d)}\,\frac{{\rm area}(\partial A_1) \cdot {\rm area} (\partial A_2)}{r^{2(d-2)}}\, ,
\end{equation}
which is indeed the same scaling as for the free scalar. Note that this formula seems problematic in the case of regions $A_1$, $A_2$ which have very small volumes but are bounded by finite area surfaces. In that situation, one would expect the mutual information to vanish, a behavior which is not reproduced by \req{memii}, since  $\mu_{(d)}$ is a fixed constant. In other words, the MEMI model does not obey the monotonicity property of the mutual information and thus, it can not describe an actual mutual information for general geometric regions. 

On the other hand, interestingly,  we have verified that the MEMI model for the case of disjoint spheres is proportional to the conformal block of an intermediate scalar operator with scaling dimension $d-2$. In those cases the MEMI model can thus be interpreted as the resumed contribution of the replica primary operator $\phi^i \phi^j$ and its descendants to the mutual information of a theory with a free scalar $\phi$ in its spectrum. A similar statement holds for the EMI model, where the associated replica operator is the conserved current $\bar{\psi}_i \gamma_\mu \psi_j$ associated to a free fermion \cite{Agon:2021zvp}. It would be interesting to explore the extent to which similar statements might exist for higher spin generalizations or interacting fields.  Another unusual feature of the MEMI model is that the universal term in the EE of a disk region turns out to vanish for this model, \ie $F_0^{\rm\ssc MEMI}=0$. Hence, when presenting the results for this model we will normalize the mutual information in a way such that at short distances the result tends to the free scalar one (in the main text, we denote the model with such normalization by MEMI$^\star$).

{\bf Disk regions.} In the case of two identical disks of diameters $R$ separated a distance $r$ defined as in the EMI case above, we have for the MEMI model
\begin{equation}\label{diskMEMI}
I_{\rm \ssc MEMI}=  \frac{2\pi^2\mu_{(3)}}{(x+1)\sqrt{x(x+2)}}\, .
\end{equation}
For small and large separations one finds, respectively,
\begin{align}
I_{\rm \ssc MEMI} &\overset{x\rightarrow 0}{=}  \frac{\sqrt{2}\pi^2 \mu_{(3)}}{\sqrt{x}} -\frac{5\pi^2 \mu_{(3)} \sqrt{x}}{2\sqrt{2}}  + \mathcal{O}(x^{3/2})  \, , \\
I_{\rm \ssc MEMI} &\overset{x\rightarrow \infty} {=}  \frac{2\pi^2 \mu_{(3)} }{x^2} - \frac{4\pi^2\mu_{(3)} }{x^3}+\mathcal{O}(1/x^4) \, .
\end{align}

{\bf Rectangle regions.} Following analogous steps to the EMI model ones, one can  also obtain the mutual information corresponding to two rectangle regions of sides $\xi R \times R$ in the MEMI model. The final result reads
\begin{align}\notag 
I_{\rm \ssc MEMI}=4\mu_{(3)} \Big[ &+\frac{\arccot(x)}{x}+\frac{2 \arccot(x+\xi)}{x+\xi}+\frac{\arccot(x+2\xi)}{x+2\xi} -2(x+\xi) \arctan(x+\xi) \\ &+x \arctan(x) + (x+2\xi) \arctan(x+2\xi)  +\log \left[\frac{(x+\xi)^4}{ (1+x^2)(1+(x+2\xi)^2)} \right] \\  \notag &+2\int_0^{\infty} d t \,\, t e^t \left[\frac{x}{e^{2t} x^2+1}-\frac{2\xi+x}{e^{2t}(x+2\xi)^2+1} \right]  \Big]\, . \label{rectMEMI}
\end{align}
This behaves respectively, in the almost-touching and long-distance regimes as
\begin{align}
I_{\rm \ssc MEMI} &\overset{x\rightarrow 0}{=}  \frac{k_{\rm \ssc MEMI}^{(3)} } {x}+\frac{2\pi \mu_{(3)}}{\xi} \log x+\tilde g(\xi) + \mathcal{O}(x)  \, , \\
I_{\rm \ssc MEMI} &\overset{x\rightarrow \infty} {=} \frac{8\mu_{(3)} (1+\xi)^2}{x^2} - \frac{16 \mu_{(3)} \xi (1+\xi)^2}{x^3}+ \mathcal{O}(x^{-4})  \, , 
\end{align}
where $\tilde g(\xi) $ is some complicated function of $\xi$.

\bibliographystyle{ucsd}
\bibliography{N-partite-info-v3}

\end{document}